\documentclass[allclo,onecollarge,natbib]{svjour2}
\bibpunct{[}{]}{;}{n}{}{,} % to get "[numbered]" references from natbib
\smartqed  % flush right qed marks, e.g. at end of proof
\usepackage{amsfonts}
\usepackage{amssymb}
\usepackage{clpsref}
\usepackage{graphicx}
\usepackage{pzccal}

\journalname{Few Body Systems}

\usepackage{amsmath}

\newcommand{\lsim}{\mathrel{\rlap{\lower4pt\hbox{\hskip0pt$\sim$}}
\raise1pt\hbox{$<$}}}           %less than or approx. symbol
\newcommand{\gsim}{\mathrel{\rlap{\lower4pt\hbox{\hskip0pt$\sim$}}
\raise1pt\hbox{$>$}}}           %greater than or approx. symbol

\usepackage{color}
\definecolor{purple}{rgb}{0.5,0,0.5}
\definecolor{blue}{rgb}{0.0,0,0.9}

\begin{document}

\title{Spectrum of hadrons with strangeness}

\authorrunning{Chen Chen \emph{et al}.}
\titlerunning{Spectrum of hadrons with strangeness}        % if too long for running head
\sloppy

\author{Chen Chen
\and
        Lei Chang
\and
        Craig D.~Roberts
\and
        Shaolong Wan
\and
        David J.~Wilson
}


\institute{Chen Chen \and Shaolong Wan
\at
Institute for Theoretical Physics and Department of Modern Physics,\\
University of Science and Technology of China, Hefei 230026, P. R. China
\and
Lei Chang
\at
Institut f\"ur Kernphysik, Forschungszentrum J\"ulich, D-52425 J\"ulich, Germany
\and
Chen Chen  \and Craig D.~Roberts  %(\email{cdroberts@anl.gov})
\and David J.~Wilson
\at
Physics Division, Argonne National Laboratory, Argonne
Illinois 60439, USA %[http://www.phy.anl.gov/theory]
\and
Chen Chen \and Craig D.~Roberts
\at
Department of Physics, Illinois Institute of Technology, Chicago Illinois 60616, USA
}

\date{5 April 2012}
%\date{27 February 2012}
%\date{9 February 2012}
%\date{Version 1.15: 27 January 2012}

\maketitle

\begin{abstract}
We describe a calculation of the spectrum of strange and nonstrange hadrons that simultaneously correlates the dressed-quark-core masses of meson and baryon ground- and excited-states within a single framework.
The foundation for this analysis is a symmetry-preserving Dyson-Schwinger equation treatment of a vector$\times$vector contact interaction.
Our results exemplify and highlight the deep impact of dynamical chiral symmetry breaking on the hadron spectrum: an accurate description of the meson spectrum entails a similarly successful prediction of the spectrum of baryons, including those with strangeness.
The analysis also provides numerous insights into baryon structure.  For example, that baryon structure is largely flavour-blind, the first radial excitation of ground-state baryons is constituted almost entirely from axial-vector diquark correlations, and DCSB is the foundation for the ordering of low-lying baryon levels; viz., $(1/2)^+$, $(1/2)^+$, $(1/2)^-$.

\end{abstract}

%\tableofcontents

\section{Introduction}
\label{intro}
Quantum chromodynamics (QCD) is a relativistic quantum field theory that is generally believed to describe the strongly interacting part of the Standard Model.  If so, it is Nature's only known example of an essentially nonperturbative fundamental theory.  This is the difficulty: in attempting to understand QCD one must immediately confront a unique nonperturbative problem.  Never before have we been confronted by a theory whose elementary excitations are not those degrees-of-freedom readily accessible via experiment; i.e.,
whose elementary excitations are \emph{confined}.  Moreover, it appears that QCD generates forces which are so strong that less-than 2\% of a nucleon's mass can be attributed to the current-quark masses that appear in the QCD Lagrangian; viz., forces that generate mass from (almost) nothing, a phenomenon known as dynamical chiral symmetry breaking (DCSB).  It follows that the Higgs mechanism is largely irrelevant to the bulk of normal matter in the Universe.  Instead the most important mass generating mechanism is the strong interaction effect of DCSB.
%\emph{a priori}, no one has any idea as to what such a theory is nonperturbatively capable of producing.  In order to decide whether QCD
Neither confinement nor DCSB is apparent in QCD's Lagrangian and yet they play the dominant role in determining the observable characteristics of real-world QCD.  The physics of hadrons is ruled by emergent phenomena, such as these, which can only be elucidated through the employment of nonperturbative methods in quantum field theory.\footnote{
In connection with these phenomena, it is important to appreciate that the static potential measured in numerical simulations of quenched lattice-regularised QCD is not related in any known way to the question of light-quark confinement.  It is a basic feature of QCD that light-quark creation and annihilation effects are essentially nonperturbative
and therefore it is impossible in principle to compute a potential between two light quarks.  These points are elucidated, e.g., in Sect.\,2.4 of Ref.\,\protect\cite{Roberts:2012sv}.}
This is both the greatest novelty and the biggest challenge within the Standard Model.

One method by which to validate QCD is computation of its hadron spectrum and subsequent comparison with modern experiment.  Indeed, this is an integral part of the international effort in nuclear physics.  The $N^\ast$ programme \cite{Aznauryan:2009da,Aznauryan:2011ub} and the search for hybrid and exotic mesons \cite{Carman:2005ps,Crede:2008vw} together address the questions: which hadron states and resonances are produced by QCD, and how are they constituted?  Herein, motivated by this intense effort in hadron spectroscopy, we extend Ref.\,\cite{Roberts:2011cf} and treat ground- and excited-state hadrons with $s$-quark content.  Furthermore, as Ref.\,\cite{Roberts:2011cf} was a precursor to a wide-ranging study of nucleon elastic and transition form factors \cite{Roberts:2011wy,Wilson:2011aa}, the study we describe herein is also a necessary step toward a comprehensive analysis of form factors that involve hadrons with strangeness.

We bring to these tasks a continuum perspective based on QCD's Dyson-Schwinger equations (DSEs) \cite{Roberts:2012sv,Chang:2011vu,Bashir:2012fs} and within this framework we use a symmetry-preserving treatment of a vector$\times$vector contact interaction because that has proven to be a reliable tool in spectrum calculations.  It is appropriate to remark that this interaction produces form factors which are too hard \cite{Roberts:2011wy,Wilson:2011aa,GutierrezGuerrero:2010md,Roberts:2010rn} but, when interpreted carefully, they, too, can be used to draw valuable insights.

To explain our choice of interaction we note by contrast that DSE kernels with a closer connection to perturbative QCD; namely, which preserve QCD's one-loop renormalisation group behaviour, have long been employed in studies of the spectrum and interactions of mesons \cite{Jain:1993qh,Maris:1997tm,Maris:2003vk}.  Such kernels are developed in the rainbow-ladder approximation, which is the leading-order in a systematic and symmetry-preserving truncation scheme \cite{Munczek:1994zz,Bender:1996bb}; and their model input is expressed via a statement about the nature of the gap equation's kernel at infrared momenta.  With a single parameter that expresses a confinement length-scale or strength \cite{Maris:2002mt,Eichmann:2008ae}, they have successfully described and predicted numerous properties of vector \cite{Eichmann:2008ae,Maris:1999nt,Bhagwat:2006pu,Qin:2011dd,Qin:2011xq} and pseudoscalar mesons \cite{Eichmann:2008ae,Qin:2011dd,Qin:2011xq,Maris:1998hc,Maris:2000sk,Maris:2002mz,Bhagwat:2007ha} with masses less than 1\,GeV, and ground-state baryons \cite{Eichmann:2008ef,Eichmann:2011vu,Eichmann:2011ej,Eichmann:2011pv}.  Such kernels are also reliable for ground-state heavy-heavy mesons \cite{Bhagwat:2006xi}.

On the other hand, whilst model-independent results for properties of pseudoscalar meson excited states have been established \cite{Holl:2004fr,Holl:2005vu}, the rainbow-ladder truncation is quantitatively inaccurate for the spectrum of light-quark mesons with masses greater than 1\,GeV, for reasons which are understood \cite{Qin:2011dd,Qin:2011xq}.  In fact, an explanation of the spectrum of such states requires that the kernels used in formulating the associated bound-state problems are essentially nonperturbative, incorporating effects of DCSB which it has only recently become possible to express \cite{Chang:2009zb,Chang:2010hb,Chang:2011ei,Bashir:2011dp}.  An equivalent formulation of the baryon bound-state problem is not yet available.  Furthermore, technical difficulties associated with the analytic structure of rainbow-ladder kernels constructed using realistic interactions \cite{Maris:1997tm,Maris:2000sk,Bhagwat:2002tx} have so far prevented computation of the spectrum of meson excited states, and the excited states and parity partners of ground-state baryons.

Key elements in a successful spectrum computation are: symmetries and the pattern by which they are broken; the mass-scale associated with confinement and DCSB; and full knowledge of the physical content of bound-state kernels.  These features are present in the informed use of a symmetry-preserving treatment of a vector$\times$vector contact interaction.  This underlies the success of that approach in Ref.\,\cite{Roberts:2011cf}, which produced the first unified DSE description of the spectrum of light-quark meson and baryon ground- and excited-states, and is promising to provide a bridge between QCD and dynamical coupled-channels reaction models \cite{Wilson:2011aa}.  We undertake the current study in the expectation of similar reward in connection with strange hadrons.  This is critical because contemporary hadron structure calculations, which typically omit meson-cloud effects, should not directly be compared with experiment but instead with the bare-masses, -couplings, etc., determined via coupled-channels analyses \cite{Gasparyan:2003fp,Matsuyama:2006rp,Suzuki:2009nj,Doring:2010ap}.

In Sect.\,\ref{sec:Faddeev} we explain our interaction, and its application to mesons and colour-antitriplet quark-quark correlations.  The latter bear no relation to the pointlike diquark degrees-of-freedom employed in some models of the constituent-quark type.  Instead, they are dynamical elements that arise naturally in solving a Faddeev equation with QCD-based interactions; and, as we shall make clear, they are crucial in understanding the baryon spectrum.  Section~\ref{sec:BaryonFEs} describes the general structure of the Faddeev equations and solution amplitudes, and explains the impact of omitting resonant (meson cloud) contributions when constructing the Faddeev kernel.  Our results are presented and discussed in Sect.\,\ref{sec:results}; and we provide a summary and perspective in Sect.\,\ref{sec:epilogue}.

\section{Elements in the Faddeev Equation}
\label{sec:Faddeev}
\begin{figure}[t]
\centerline{%
\includegraphics[clip,width=0.50\textwidth]{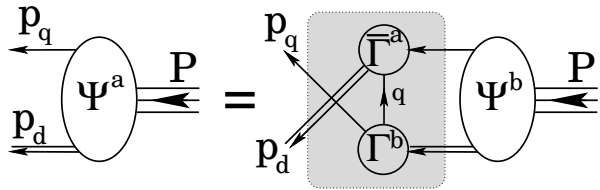}}
\caption{\label{fig:FaddeevI} Poincar\'e covariant Faddeev equation employed herein to calculate baryon properties.  $\Psi$ in Eq.\,(\protect\ref{PsiBaryon}) is the Faddeev amplitude for a baryon of total momentum $P= p_q + p_d$.  It expresses the relative momentum correlation between the dressed-quark and -diquarks within the baryon.  The shaded region demarcates the kernel of the Faddeev equation, Sect.\,\protect\ref{app:FEgeneral}, in which: the \emph{single line} denotes the dressed-quark propagator, Sect.\,\protect\ref{sec:gap};
$\Gamma$ is the diquark Bethe-Salpeter amplitude, Sect.\,\protect\ref{sec:qqBSA};
and the \emph{double line} is the diquark propagator, Eqs.\,(\ref{scalarqqprop}), (\ref{qqavprop}).  Quarks within a diquark are correlated via gluon exchange but the kernel in this Faddeev equation expresses binding within the baryon through diquark breakup and reformation, which is mediated by exchange of a dressed-quark with momentum $q$.}
\end{figure}

We base our description of baryon bound-states on a Poincar\'e-covariant Faddeev equation, which is illustrated in Fig.\,\ref{fig:FaddeevI}.  Introduced in Ref.\,\cite{Cahill:1988dx}, its key elements are the dressed-quark and -diquark propagators, and the diquark Bethe-Salpeter amplitudes.  All are completely determined once the quark-quark interaction kernel is specified and, as explained in the Introduction, we use
\begin{equation}
\label{njlgluon}
%g^2 D_{\mu \nu}(p-q) = \delta_{\mu \nu} \frac{1}{m_G^2}\,,
g^2 D_{\mu \nu}(p-q) = \delta_{\mu \nu} \frac{4 \pi \alpha_{\rm IR}}{m_G^2}\,,
\end{equation}
%where $m_G$ is a gluon mass-scale,
where $m_G=0.8\,$GeV is a gluon mass-scale typical of the one-loop renormalisation-group-improved interaction detailed in Ref.\,\cite{Qin:2011dd}, and the fitted parameter $\alpha_{\rm IR} = 0.93 \pi$ is commensurate with contemporary estimates of the zero-momentum value of a running-coupling in QCD \cite{Aguilar:2009nf,Oliveira:2010xc,Aguilar:2010gm,Boucaud:2010gr,Pennington:2011xs,Wilson:2012em}.  We embed Eq.\,\eqref{njlgluon} in a rainbow-ladder truncation of the DSEs, which is the leading-order in the most widely used, global-symmetry-preserving truncation scheme \cite{Bender:1996bb}.  This means
\begin{equation}
\label{RLvertex}
\Gamma_{\nu}(p,q) =\gamma_{\nu}
\end{equation}
in the gap equation and in the subsequent construction of the Bethe-Salpeter kernels.

One may view the interaction in Eq.\,(\ref{njlgluon}) as being inspired by models of the Nambu--Jona-Lasinio type \cite{Nambu:1961tp} but our treatment is atypical.  Used to build a rainbow-ladder truncation of the DSEs, Eqs.\,\eqref{njlgluon}, (\ref{RLvertex}) produce results for low-momentum-transfer observables that are practically indistinguishable from those produced by more sophisticated interactions \cite{Roberts:2011wy,Wilson:2011aa,GutierrezGuerrero:2010md,Roberts:2010rn}.

\subsection{Dressed-quark propagator}
\label{sec:gap}
Using Eqs.\,(\ref{njlgluon}), (\ref{RLvertex}), the gap equation for a quark of flavour $f$ becomes
\begin{equation}
 S_f^{-1}(p) =  i \gamma \cdot p + m_f +  \frac{16\pi}{3}\frac{\alpha_{\rm IR}}{m_G^2} \int\!\frac{d^4 q}{(2\pi)^4} \,
\gamma_{\mu} \, S_f(q) \, \gamma_{\mu}\,,   \label{gap-1}
\end{equation}
where $m_f$ is the quark's current-mass.  (Our Euclidean metric conventions are detailed in App.\,\ref{App:EM}.)  Equation~\eqref{gap-1} possesses a quadratic divergence, even in the chiral limit.  When the divergence is regularised in a Poincar\'e covariant manner, the solution is
\begin{equation}
\label{genS}
S_f(p)^{-1} = i \gamma\cdot p + M_f\,,
\end{equation}
where $M_f$ is momentum-independent and determined by
\begin{equation}
M_f = m_f + M_f\frac{4\alpha_{\rm IR}}{3\pi m_G^2} \int_0^\infty \!ds \, s\, \frac{1}{s+M_f^2}\,.
\end{equation}

\begin{table}[t]
\caption{\label{Tab:DressedQuarks}
Computed dressed-quark properties, required as input for the Bethe-Salpeter and Faddeev equations, and computed values for in-hadron condensates \protect\cite{Brodsky:2010xf,Chang:2011mu,Brodsky:2012ku}.  All results obtained with $\alpha_{\rm IR} =0.93 \pi$ and (in GeV) $\Lambda_{\rm ir} = 0.24\,$, $\Lambda_{\rm uv}=0.905$.  N.B.\ These parameters take the values determined in the spectrum calculation of Ref.\,\protect\cite{Roberts:2011cf}; and we assume isospin symmetry throughout.
(All dimensioned quantities are listed in GeV.)}
\begin{center}
\begin{tabular*}%{|c|c|c|c|c|c|c|}\hline
{\hsize}
{
c@{\extracolsep{0ptplus1fil}}
c@{\extracolsep{0ptplus1fil}}
c@{\extracolsep{0ptplus1fil}}
c@{\extracolsep{0ptplus1fil}}
c@{\extracolsep{0ptplus1fil}}
c@{\extracolsep{0ptplus1fil}}
c@{\extracolsep{0ptplus1fil}}
c@{\extracolsep{0ptplus1fil}}
c@{\extracolsep{0ptplus1fil}}
c@{\extracolsep{0ptplus1fil}}}\hline
$m_u$ & $m_s$ & $m_s/m_u$ & $M_0$ &   $M_u$ & $M_s$ & $M_s/M_u$ & $\kappa_0^{1/3}$ & $\kappa_\pi^{1/3}$ & $\kappa_K^{1/3}$ \\\hline
0.007  & 0.17 & 24.3 & 0.36 & 0.37 & 0.53 & 1.43  & 0.241 & 0.243 & 0.246
\\\hline
\end{tabular*}
\end{center}
\end{table}

Our regularisation procedure follows Ref.\,\cite{Ebert:1996vx}; i.e., we write
\begin{eqnarray}
\frac{1}{s+M^2} & = & \int_0^\infty d\tau\,{\rm e}^{-\tau (s+M^2)}  \rightarrow  \int_{\tau_{\rm uv}^2}^{\tau_{\rm ir}^2} d\tau\,{\rm e}^{-\tau (s+M^2)}
%\label{RegC}\\
%
 =
\frac{{\rm e}^{- (s+M^2)\tau_{\rm uv}^2}-{\rm e}^{-(s+M^2) \tau_{\rm ir}^2}}{s+M^2} \,, \label{ExplicitRS}
\end{eqnarray}
where $\tau_{\rm ir,uv}$ are, respectively, infrared and ultraviolet regulators.  It is apparent from the rightmost expression in Eq.\,(\ref{ExplicitRS}) that a finite value of $\tau_{\rm ir}=:1/\Lambda_{\rm ir}$ implements confinement by ensuring the absence of quark production thresholds \cite{Roberts:2012sv,Chang:2011vu,Krein:1990sf}.  Since Eq.\,(\ref{njlgluon}) does not define a renormalisable theory, then $\Lambda_{\rm uv}:=1/\tau_{\rm uv}$ cannot be removed but instead plays a dynamical role, setting the scale of all dimensioned quantities.  Using Eq.\,\eqref{ExplicitRS}, the gap equation becomes
\begin{equation}
%M = m +  \frac{M}{3\pi^2 m_G^2} \,{\cal C}(M^2;\tau_{\rm ir},\tau_{\rm uv})\,,
%M = m +  \frac{M}{3\pi^2 m_G^2} \,{\cal C}^{\rm iu}(M^2)\,,
M_f = m_f + M_f\frac{4\alpha_{\rm IR}}{3\pi m_G^2}\,\,{\cal C}^{\rm iu}(M_f^2)\,,
\label{gapactual}
\end{equation}
where ${\cal C}^{\rm iu}(\sigma)/\sigma = \overline{\cal C}^{\rm iu}(\sigma) = \Gamma(-1,\sigma \tau_{\rm uv}^2) - \Gamma(-1,\sigma \tau_{\rm ir}^2)$, with $\Gamma(\alpha,y)$ being the incomplete gamma-function.

In Table~\ref{Tab:DressedQuarks} we report values of $u$- and $s$-quark properties, computed from Eq.\,\eqref{gapactual}, that will subsequently be used in bound-state calculations: the input ratio $m_s/\bar m$, where $\bar m = (m_u+m_d)/2$, is consistent with contemporary estimates \cite{Leutwyler:2009jg}.
N.B.\ It is a feature of Eq.\,\eqref{gapactual} that in the chiral limit, $m_f=m_0=0$, a nonzero solution for $M_0:= \lim_{m_f\to 0} M_f$ is obtained so long as $\alpha_{\rm IR}$ exceeds a minimum value.  With $\Lambda_{\rm ir,uv}$ as specified in the Table, that value is $\alpha_{\rm IR}^c\approx 0.4\pi$.  In the Table we also include chiral-limit and physical-mass values of the in-pseudoscalar-meson condensate \cite{Brodsky:2010xf,Chang:2011mu,Brodsky:2012ku}, $\kappa_H$, which is the dynamically generated mass-scale that characterises DCSB.  A growth with current-quark mass is anticipated in QCD \cite{Maris:1997tm,Roberts:2011ea}.

\subsection{Mesons and diquark correlations}
\label{sec:qqBSA}
\subsubsection{Mesons}
\label{sec:qqBSAMeson}
The rainbow-ladder truncation of the gap and Bethe-Salpeter equations provides a good approximation for ground-state vector- and charged-pseudoscalar-mesons \cite{Chang:2011vu,Chang:2009zb,Bender:2002as,Bhagwat:2004hn}.  We therefore employ it herein, in which case the homogeneous Bethe-Salpeter equation (BSE) for a meson comprised of quarks with flavours $f$, $\bar g$ is
\begin{equation}
\Gamma_{f\bar g}(k;P) =  - \frac{16 \pi}{3} \frac{\alpha_{\rm IR}}{m_G^2}
\int \! \frac{d^4q}{(2\pi)^4} \gamma_\mu S_f(q+P) \Gamma_{f\bar g}(q;P)S_g(q) \gamma_\mu \,,
\label{LBSEI}
\end{equation}
where $P$ is the total momentum of the bound-state.  This equation has a solution for $P^2=-m_{f\bar g}^2$, where $m_{f\bar g}$ is the bound-state's mass.

Here we illustrate the nature of the BSE via two relevant examples; viz., the negatively charged kaon and the kindred $K^\ast$ vector meson, both of which possess $s \bar u$ flavour structure.  The interaction in Eq.\,\eqref{njlgluon} supports a kaon Bethe-Salpeter amplitude of the form
\begin{equation}
\label{KaonBSA}
\Gamma_K(P) = i \gamma_5 \,E_K(P) + \frac{1}{2 M_R} \gamma_5 \gamma\cdot P \, F_K(P)\,,
\end{equation}
where\footnote{The choice one makes for the mass-dimensioned constant, $M_R$, has no effect on any result.}
$M_R = M_s M_u/[M_s + M_u]$.  If one inserts Eq.\,\eqref{KaonBSA} into Eq.\,\eqref{LBSEI} and employs the symmetry-preserving regularisation of the contact interaction explained, e.g., in Ref.\,\cite{Wilson:2011aa}, which requires
\begin{equation}
0 = \int_0^1d\alpha \,
\left[ {\cal C}^{\rm iu}(\omega(M_u^2,M_s^2,\alpha,P^2))
+ \, {\cal C}^{\rm iu}_1(\omega(M_u^2,M_s^2,\alpha,P^2))\right], \label{avwtiP}
\end{equation}
where
\begin{eqnarray}
\label{eq:omega}
\omega(M_u^2,M_s^2,\alpha,P^2) &=& M_u^2 (1-\alpha) + \alpha M_s^2 + \alpha(1-\alpha) P^2\,,\\
{\cal C}^{\rm iu}_1(z) &=& - z (d/dz){\cal C}^{\rm iu}(z) = z\left[ \Gamma(0,M^2 \tau_{\rm uv}^2)-\Gamma(0,M^2 \tau_{\rm ir}^2)\right] ,\rule{2em}{0ex}
\label{eq:C1}
\end{eqnarray}
then the explicit form of the kaon BSE is
\begin{equation}
\label{bsefinalE}
\left[
\begin{array}{c}
E_{K}(P)\\
F_{K}(P)
\end{array}
\right]
= \frac{4 \alpha_{\rm IR}}{3\pi m_G^2}
\left[
\begin{array}{cc}
{\cal K}_{EE}^K & {\cal K}_{EF}^K \\
{\cal K}_{FE}^K & {\cal K}_{FF}^K
\end{array}\right]
\left[\begin{array}{c}
E_{K}(P)\\
F_{K}(P)
\end{array}
\right],
\end{equation}
with
\begin{subequations}
\label{pionKernel}
\begin{eqnarray}
\nonumber
{\cal K}_{EE}^K &=&
\int_0^1d\alpha \bigg\{
{\cal C}^{\rm iu}(\omega(M_u^2, M_s^2, \alpha, P^2))  \\
&&+ \bigg[ M_u M_s-\alpha (1-\alpha) P^2 - \omega(M_u^2, M_s^2, \alpha, P^2)\bigg]
\, \overline{\cal C}^{\rm iu}_1(\omega(M_u^2, M_s^2, \alpha, P^2))\bigg\},\\
{\cal K}_{EF}^K &=& \frac{P^2}{2 M_R} \int_0^1d\alpha\, \bigg[(1-\alpha)M_u+\alpha M_s\bigg]\overline{\cal C}^{\rm iu}_1(\omega(M_u^2, M_s^2, \alpha, P^2)),\\
{\cal K}_{FE}^K &=& \frac{2 M_R^2}{P^2} {\cal K}_{EF}^K ,\\
{\cal K}_{FF}^K &=& - \frac{1}{2} \int_0^1d\alpha\, \bigg[ M_u M_s+(1-\alpha) M_u^2+\alpha M_s^2\bigg] \overline{\cal C}^{\rm iu}_1(\omega(M_u^2, M_s^2, \alpha, P^2))\,.
\end{eqnarray}
\end{subequations}

Equation~\eqref{bsefinalE} is an eigenvalue problem, which has a solution for $P^2=-m_K^2$. The eigenvector is the kaon's Bethe-Salpeter amplitude, and in the computation of observables one must employ the canonically normalised amplitude; viz., the amplitude rescaled such that
\begin{equation}
\label{normcan}
1=\left. \frac{d}{d P^2}\Pi_K(Q,P)\right|_{Q=P},
\end{equation}
where
\begin{equation}
\Pi_K(Q,P)= 6 {\rm tr}_{\rm D} \int\! \frac{d^4q}{(2\pi)^4}\Gamma_{K}(-Q)
 \frac{\partial}{\partial P_\mu} S_s(q+P) \, \Gamma_{K}(Q)\, S_u(q)\,.
\end{equation}

Since Eq.\,\eqref{njlgluon} can only support a vector meson Bethe-Salpeter amplitude of the form
\begin{equation}
\label{KastBSA}
\Gamma_{K^\ast} = \gamma_\mu^\perp E_{K^\ast}(P)\,,
\end{equation}
where $P_\mu \gamma_\mu^\perp = 0$, the $K^\ast$ BSE is simpler; viz.,
\begin{equation}
1 - {\cal K}^{K^{\ast}}(-m_{K^\ast}^2) = 0\,,
\end{equation}
with
\begin{equation}
\label{KastKernel}
{\cal K}^{K^{\ast}}(P^2)=\frac{2\alpha_{\rm IR}}{3\pi m_G^2} \int_0^1d\alpha\,
\bigg[ M_u M_s - (1-\alpha)M_u^2-\alpha M_s^2-2\alpha(1-\alpha)P^2\bigg]
\overline{\cal C}_1^{\rm iu}(\omega(M_u^2, M_s^2, \alpha, P^2))\,,
\end{equation}
where we have used Eq.\,\eqref{avwtiP}.  In this case the canonical normalisation  condition can be written
\begin{equation}
\frac{1}{E_{K^\ast}^2} = 9 \mathpzc{m}_G^2 \left. \frac{d}{d z} {\cal K}^{K^{\ast}}(z)\right|_{z=-m_{K^\ast}^2} ,
\; \frac{1}{\mathpzc{m}_G^2} = \frac{4\pi \alpha_{\rm IR}}{m_G^2}\,.
\end{equation}

It should be plain that the analogous set of equations for the $\rho$-meson is obtained simply by replacing the $s$-quark by a $d$-quark throughout; and that for the $\phi$ by replacing the $\bar u$-quark by a $\bar s$-quark.  Other states are discussed in App.\,\ref{app:groundBSEs}.

\subsubsection{Diquark correlations}
\label{sec:qqBSADiquark}
The relevance of the rainbow-ladder meson BSE to the baryon Faddeev equation is explained, e.g., in Sect.\,2.1 of Ref.\,\cite{Roberts:2011cf}; namely, in this truncation one may obtain the mass and Bethe-Salpeter amplitude for a colour-antitriplet quark-quark correlation (diquark) with spin-parity $J^P$ from the equation for a $J^{-P}$-meson in which the only change is a halving of the interaction strength \cite{Cahill:1987qr}.  The flipping of the sign in parity occurs because it is opposite for fermions and antifermions.

At this point it is appropriate to remark that the rainbow-ladder truncation generates asymptotic diquark states.  Such states are not observed and their appearance is an artefact of the truncation.  Higher-order terms in the quark-quark scattering kernel, whose analogue in the quark-antiquark channel do not materially affect the properties of vector and flavour non-singlet pseudoscalar mesons, ensure that QCD's quark-quark scattering matrix does not exhibit singularities which correspond to asymptotic diquark states \cite{Bender:1996bb,Bender:2002as,Bhagwat:2004hn}.   Studies with kernels that exclude diquark bound states nevertheless support a physical interpretation of the masses, $m_{(qq)_{\!J^P}}$, obtained using the rainbow-ladder truncation; viz., the quantity $\ell_{(qq)^{\!J^P}}:=1/m_{(qq)_{\!J^P}}$ may be interpreted as a range over which the diquark correlation can propagate before fragmentation.

This caveat expressed, one may write the contact-interaction rainbow-ladder BSE for a colour-antitriplet diquark constituted from quarks with flavour $f$, $g$:
\begin{equation}
\Gamma^C_{f g}(k;P):=
\Gamma_{f g}(k;P)C^\dagger =  - \frac{8 \pi}{3} \frac{\alpha_{\rm IR}}{m_G^2}
\int \! \frac{d^4q}{(2\pi)^4} \gamma_\mu S_f(q+P) \Gamma^C_{fg}(q;P) S_g(q) \gamma_\nu \,,
\label{LBSEqq}
\end{equation}
where $C$ is the charge-conjugation matrix, Eq.\,\eqref{chargec}.  %(This connection is valid for an arbitrary interaction treated in rainbow-ladder truncation \cite{Cahill:1987qr}.)

Capitalising further on the connection between the meson and diquark sectors, one may readily write explicit forms of the BSEs and canonical normalisation conditions for scalar ($[fg]$) and axial-vector ($ \{ff\}$, $\{fg\}$) diquark correlations.  For example, the Bethe-Salpeter amplitude for a $J^P=0^+$ $[s,u]$-diquark is
\begin{equation}
\label{BSA[su]}
\Gamma^C_{[su]_{0^+}}(P) = i \gamma_5 \,E_{[su]_{0^+}}(P) + \frac{1}{2 M_R} \gamma_5 \gamma\cdot P \, F_{[su]_{0^+}}(P)\,,
\end{equation}
which satisfies the following BSE
\begin{equation}
\label{bse[su]E}
\left[
\begin{array}{c}
E_{[su]_{0^+}}(P)\\
F_{[su]_{0^+}}(P)
\end{array}
\right]
= \frac{2 \alpha_{\rm IR}}{3\pi m_G^2}
\left[
\begin{array}{cc}
{\cal K}_{EE}^K & {\cal K}_{EF}^K \\
{\cal K}_{FE}^K & {\cal K}_{FF}^K
\end{array}\right]
\left[\begin{array}{c}
E_{[su]_{0^+}}(P)\\
F_{[su]_{0^+}}(P)
\end{array}
\right].
\end{equation}
In this case the canonical normalisation condition is
\begin{equation}
1=\left. \frac{d}{d P^2}\Pi_{[su]_{0^+}}(Q,P)\right|_{Q=P},
\end{equation}
where
\begin{equation}
\Pi_{[su]_{0^+}}(Q,P)= 4 {\rm tr}_{\rm D} \int\! \frac{d^4q}{(2\pi)^4}\Gamma_{[su]_{0^+}}(-Q)
 \frac{\partial}{\partial P_\mu} S_s(q+P) \, \Gamma_{[su]_{0^+}}(Q)\, S_u(q)\,.
\end{equation}
Compared with Eq.\,\eqref{normcan}, the colour factor is different owing to the fact that diquarks are colour-antitriplets not singlets.

Following this pattern one may immediately write the BSE for $J^P=1^+$ $\{su\}$ diquark correlations; viz.,
\begin{equation}
1 - \frac{1}{2} K^{K^{\ast}}(-m_{\{su\}}^2) = 0\,,
\end{equation}
and the canonical normalisation condition:
\begin{equation}
\label{canonicalavqq}
\frac{1}{E_{\{su\}}^2} = 6 \mathpzc{m}_G^2 \left.\frac{d}{d z} K^{K^{\ast}}(z)\right|_{z=-m_{\{su\}}^2}.
\end{equation}

The analogous set of equations for axial-vector $\{uu\}$- and $\{ud\}$-diquarks are obtained by replacing the $s$-quark by either a $u$- or $d$-quark throughout; and that for the $\{ss\}$-diquark by replacing the $u$-quark by a $s$-quark.  Other correlations are discussed in App.\,\ref{app:groundBSEs}.

\begin{table}[t]
\caption{\label{Mesonmasses}
Row~1: Quark-core masses of ground-state mesons computed using our symmetry-preserving regularisation of the vector$\times$vector contact interaction, with the input from Table~\protect\ref{Tab:DressedQuarks}.   Row~2: Except for scalar mesons, values drawn from Ref.\,\protect\cite{Nakamura:2010zzi}, with weighted averages of mass-squared values reported, where appropriate.  For the isoscalar-scalar meson we list an estimate for the state's dressed-quark core \protect\cite{Pelaez:2006nj,RuizdeElvira:2010cs}.  Nothing is known about this value for the $I=1/2$ scalar.
Rows~3 and 4 repeat this pattern for the mesons' first radial excitation.  The theory error in Row~3 displays the outcome of varying the location of the node in the radial excitation's Bethe-Salpeter amplitude: $1/d_{\cal F} = 2 M^2 ( 1\pm 0.2)$.  An asterisk-marked mass in Row~4 indicates a state whose properties are poorly determined.
(All dimensioned quantities are listed in GeV.)
% rho ... K1
}
\begin{center}
\begin{tabular*}%{|c|c|c|c|c|c|c|}\hline
{\hsize}
{
l@{\extracolsep{0ptplus1fil}}
l@{\extracolsep{0ptplus1fil}}
|l@{\extracolsep{0ptplus1fil}}
l@{\extracolsep{0ptplus1fil}}
l@{\extracolsep{0ptplus1fil}}
l@{\extracolsep{0ptplus1fil}}
l@{\extracolsep{0ptplus1fil}}
l@{\extracolsep{0ptplus1fil}}
l@{\extracolsep{0ptplus1fil}}
l@{\extracolsep{0ptplus1fil}}
l@{\extracolsep{0ptplus1fil}}
l@{\extracolsep{0ptplus1fil}}}\hline
& &
$m_\pi$ & $m_K$ &
$m_\rho$ & $m_{K^\ast}$ & $m_\phi$ &
$m_\sigma$ & $m_\kappa$ &
$m_{a_1}$ & $m_{K_1}$ & $m_{f_1}$ \\\hline
n=0 & DSE \rule{0em}{3ex}&
0.14 & 0.50 &
0.93 & 1.03 & 1.13 &
1.29 & 1.40 &
1.38 & 1.48 & 1.59 \\
%...sep
%0, 0, 21, 17, 17, ... scalar comparison makes no sense
%
& expt. \rule{0em}{3ex}&
0.14 & 0.50 &
0.78 & 0.89 & 1.02 &
1.0 - 1.2 &  &
1.23 & 1.34 & 1.42 \\\hline
n=1 & DSE \rule{0em}{3ex}&
$1.33_{\pm 0.06}$ & $1.33_{\pm 0.07}$  &
$1.29_{\pm 0.05}$ & $1.40_{\pm 0.05}$ & $1.51_{\pm 0.05}$ &
$1.42_{\pm 0.02}$ & $1.53_{\pm 0.02}$ &
$1.47_{\pm 0.02}$ & $1.57_{\pm 0.01}$ & $1.67_{\pm 0.02}$ \\
& expt. \rule{0em}{3ex}&
$1.3_{\pm 0.1}$ & $1.46^\ast$ &
$1.46_{\pm 0.03}$ & $1.68^\ast$ & $1.68_{\pm 0.02}$ &
 &  &
$1.65_{\pm 0.02}$ &  &  \\\hline
%
%   & $m_{\pi}$ & $m_K$ & $m_{\rho}$ & $m_{K^\ast}$
%   & $m_{\sigma}$ & $m_{K_0^\ast}$ & $m_{a_1}$ & $m_{K_1}$ \\\hline
%
%DSE & 0.14 & 0.50 & 0.93 & 1.03 & 1.29 & 1.40 & 1.38 & 1.48\rule{0em}{2.5ex} \\\hline
%
%PDG & 0.140 & 0.50 & 0.78 &  0.89 & 1.0-1.2 & 1.43 & 1.24 & 1.27\\\hline
\end{tabular*}
\end{center}
\end{table}

\subsubsection{Mesons: computed masses}
Before discussing the results presented in Table~\ref{Mesonmasses}, it is necessary to recapitulate on an important modification of the rainbow-ladder Bethe-Salpeter kernel that one should implement before prediction and comparison with experiment are meaningful.  It has long been known that the rainbow-ladder truncation describes vector meson and flavour-nonsinglet pseudoscalar-meson ground-states very well but fails for their parity partners \cite{Qin:2011dd,Qin:2011xq,Maris:2006ea,Cloet:2007pi,Watson:2004kd,Fischer:2009jm}.  The origin and solution of this longstanding puzzle are now available following a novel reformulation of the BSE \cite{Chang:2009zb}, which is valid and tractable when the quark-gluon vertex is fully dressed.  In employing this approach to study the meson spectrum it was found that DCSB generates a large dressed-quark anomalous chromomagnetic moment and consequently that spin-orbit splitting between ground-state mesons is dramatically enhanced \cite{Chang:2010hb,Chang:2011ei,Chang:2010jq}.  This is the mechanism responsible for a magnified splitting between parity partners; namely, there are essentially nonperturbative DCSB corrections to the rainbow-ladder kernels, which largely-cancel in the pseudoscalar and vector channels but add constructively in the scalar and axial-vector channels.

With this in mind, we follow Ref.\,\cite{Roberts:2011cf} and introduce spin-orbit repulsion into the scalar- and pseudovector-meson channels through the artifice of a phenomenological coupling $g^2_{\rm SO}\leq 1$, introduced as a single, common factor multiplying the kernels defined in Eqs.\,(\ref{avKernelE}), (\ref{scalarKernelE}).  The value\footnote{NB.\, $g_{\rm SO}=1$ means no spin-orbit repulsion.  The mass changes slowly with diminishing $g_{\rm SO}$; e.g., $g_{\rm SO}=0.50$ yields $m_{a_1}=1.23\,$GeV.}
\begin{equation}
\label{gSO}
g_{\rm SO} = 0.24
\end{equation}
is chosen so as to obtain the experimental value for the $a_1$-$\rho$ mass-splitting, which we know to be achieved by the corrections described above \cite{Chang:2009zb,Chang:2010hb,Chang:2010jq}.  It is noteworthy that the shift in $m_{a_1}$ is accompanied by an increase of $m_\sigma$ and that the new value matches an estimate for the $\bar q q$-component of the $\sigma$-meson obtained using unitarised chiral perturbation theory \cite{Pelaez:2006nj,RuizdeElvira:2010cs}.

\begin{figure}[t]
\leftline{%
\includegraphics[clip,width=0.48\textwidth]{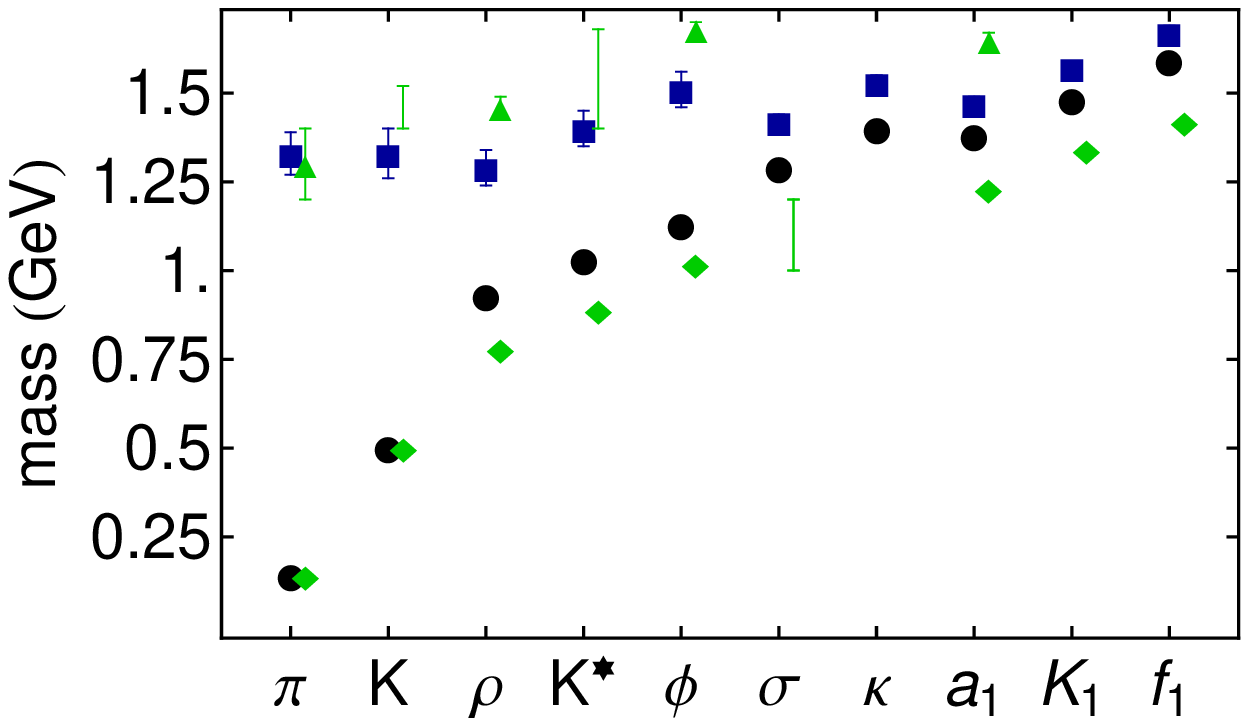}}\vspace*{-39ex}

\rightline{%
\includegraphics[clip,width=0.48\textwidth]{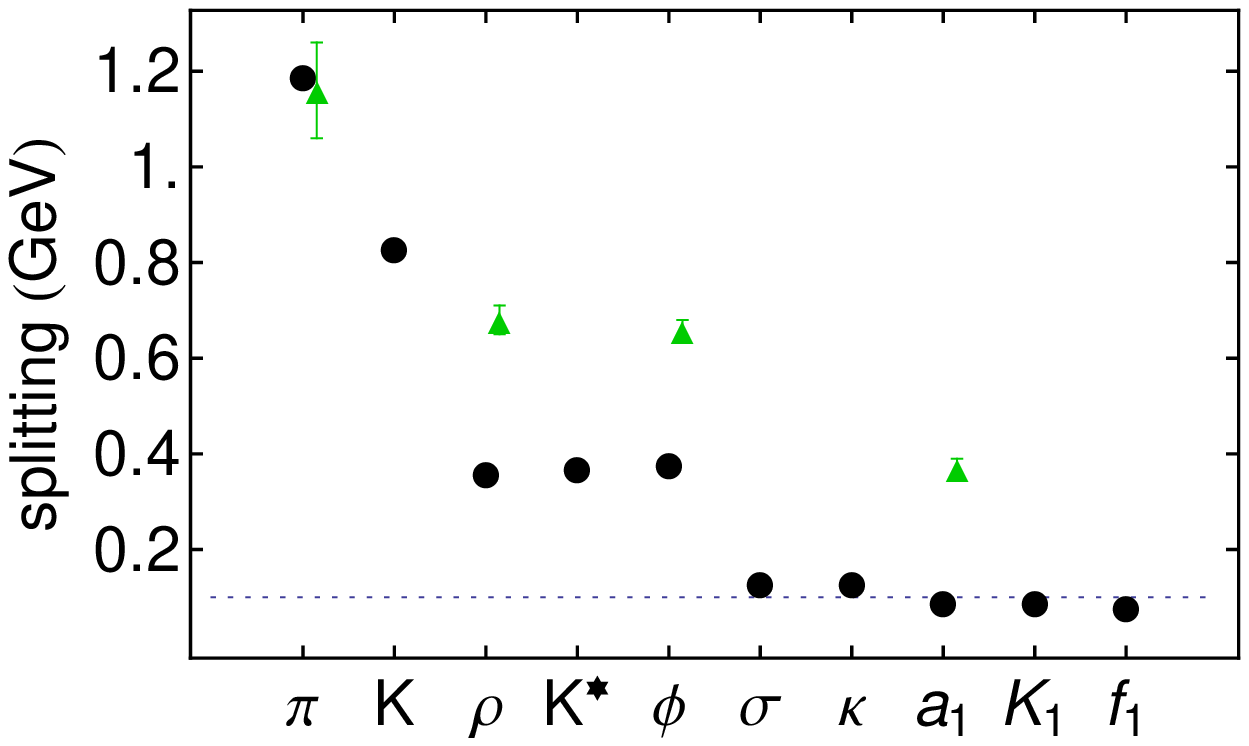}}

\caption{\label{fig:MesonMasses}
\underline{Left panel}: Pictorial representation of Table~\protect\ref{Mesonmasses}.  \emph{Circles} -- computed ground-state masses; \emph{squares} -- computed masses of radial excitations; \emph{diamonds} -- empirical ground-state masses in Row~2; and \emph{triangles} -- empirical radial excitation masses in Row~4.
\underline{Right panel}: \emph{Circles} -- computed splitting between the first radial excitation and ground state in each channel; and \emph{triangles} -- empirical splittings, where they are known.  The \emph{dashed line} marks a splitting of 0.1\,GeV.}
\end{figure}

This expedient produces the results for scalar and axial-vector mesons in Rows~1,3 of Table~\ref{Mesonmasses}, which reports calculated results for meson masses and compares them with available empirical values.  This information is represented pictorially in the left panel of Fig.\,\ref{fig:MesonMasses}.  Owing to our choice for the current-quark masses, $m_\pi$ and $m_K$ agree with experiment.  All other computed values for ground-states are greater than the empirical masses, where they are known.  This is typical of DCSB-corrected kernels that nevertheless omit resonant contributions; i.e., do not contain effects that may phenomenologically be associated with a meson cloud.  In Table~\ref{MesonBSAs} we list the canonically normalised Bethe-Salpeter amplitude for each meson.  These are the quantities used in calculating observable properties of mesons; and comparison with the kindred diquark amplitudes, listed in Table~\ref{DiquarkBSAs}, will subsequently be instructive.

\begin{table}[b]
\caption{\label{MesonBSAs}
The structure of meson Bethe-Salpeter amplitudes is described in Sect.\,\ref{sec:qqBSAMeson} and App.\,\ref{app:groundBSEs}.  Here we list the canonically normalised amplitude associated with each of the BSE eigenstates in Table~\protect\ref{Mesonmasses}.  Only pseudoscalar mesons involve two independent amplitudes when a vector$\times$vector contact interaction is treated systematically in rainbow-ladder truncation.
%
%Rows~1, 2: ground states; and Rows~3, 4: first radial excitation.
}
\begin{center}
\begin{tabular*}%{|c|c|c|c|c|c|c|}\hline
{\hsize}
{
l@{\extracolsep{0ptplus1fil}}
l@{\extracolsep{0ptplus1fil}}
|l@{\extracolsep{0ptplus1fil}}
l@{\extracolsep{0ptplus1fil}}
l@{\extracolsep{0ptplus1fil}}
l@{\extracolsep{0ptplus1fil}}
l@{\extracolsep{0ptplus1fil}}
l@{\extracolsep{0ptplus1fil}}
l@{\extracolsep{0ptplus1fil}}
l@{\extracolsep{0ptplus1fil}}
l@{\extracolsep{0ptplus1fil}}
l@{\extracolsep{0ptplus1fil}}}\hline
    &   &
$m_\pi$ & $m_K$ &
$m_\rho$ & $m_{K^\ast}$ & $m_\phi$ &
$m_\sigma$ & $m_\kappa$ &
$m_{a_1}$ & $m_{K_1}$ & $m_{f_1}$ \\\hline
n=0 &$E_{q\bar q}$ \rule{0em}{3ex}&
3.60 & 3.86 &
1.53 & 1.62 & 1.74 &
0.47 & 0.47 &
0.31 & 0.31 & 0.31 \\
    & $F_{q\bar q}$ \rule{0em}{3ex}&
0.48 & 0.60 &
 &  &  &
 &  &
 &  &  \\\hline
n=1 &$E_{q\bar q}$ \rule{0em}{3ex}&
0.83 & 0.76 &
0.72 & 0.70 & 0.66 &
0.34 & 0.35 &
0.28 & 0.28 & 0.28 \\
    & $F_{q\bar q}$ \rule{0em}{3ex}&
 0.05 & 1.18 &
 &  &  &
 &  &
 &  &  \\\hline
\end{tabular*}
\end{center}
\end{table}

In practical calculations, meson-cloud effects divide into two distinct types.  \label{page:pionloops}
The first is within the gap equation, where pseudoscalar meson loop corrections to the dressed-quark-gluon vertex act to reduce uniformly the mass-function of a dressed-quark \cite{Eichmann:2008ae,Cloet:2008fw,Blaschke:1995gr,Fischer:2007ze,Chang:2009ae}.  This effect can be pictured as a single quark emitting and reabsorbing a pseudoscalar meson.  It can be mocked-up by simply choosing the parameters in the gap equation's kernel so as to obtain a dressed-quark mass-function that is characterised by a mass-scale of approximately $400\,$MeV.  Such an approach has implicitly been widely employed with phenomenological success \cite{Chang:2011vu,Maris:2003vk,Roberts:2000aa,Roberts:2007jh}.  We employ it herein.

The second type of correction arises in connection with bound-states and may be likened to adding pseudoscalar meson exchange \emph{between} dressed-quarks within the bound-state \cite{Roberts:1988yz,Hollenberg:1992nj,Alkofer:1993gu,Mitchell:1996dn,Ishii:1998tw,%
Pichowsky:1999mu,Hecht:2002ej}, as opposed to the first type of effect; i.e., emission and absorption of a meson by the same quark.  The type-2 contribution is that computed in typical evaluations of meson-loop corrections to hadron observables based on a point-hadron Lagrangian.  These are the corrections that should be added to the calculated results in Table~\ref{Mesonmasses}.  The most complete computation of this sort predicts that such effects reduce $m_\rho$ by $0.13\,$GeV \cite{Pichowsky:1999mu}.  Applied to our result, this would produce $m_\rho^{\rm loop-corrected} = 0.8\,$GeV, in good agreement with the empirical value of $0.78\,$GeV.

These observations underpin a view that bound-state kernels which omit type-2 meson-cloud corrections should produce dressed-quark-core masses for hadron ground-states that are larger than the empirical values.  As we shall see, this is uniformly true herein.  Moreover, this perspective also has implications for the description of elastic and transition form factors \cite{Wilson:2011aa,Eichmann:2008ef,Cloet:2008wg,Cloet:2008re}.

The situation for radially excited states is less clear.  This may be seen from the right panel of Fig.\,\ref{fig:MesonMasses}, which depicts the computed mass splitting between ground-states and the first radial excitation in each channel; and also provides a comparison with experiment, when that is available.  The comparison suggests that our formulation of the contact interaction kernels produces the correct trend but underestimates the splitting by $\sim 0.2\,$GeV.  We note that this mismatch is reduced if type-2 meson-cloud corrections to the masses of radial excitations are smaller than for ground-states.  On the other hand, it might simply be that this underestimate is an error arising from the expedient we employ in order to define radial excitations within the contact interaction framework, which is discussed in App.\,\ref{sec:radial}.  Given that possibility, one must allow that our predictions for the dressed-quark-core masses of hadron first radial excitations might be $\sim 0.2\,$GeV too small.

\subsubsection{Diquarks: computed masses}
\label{subsub:diquarks}
\begin{figure}[t]
\leftline{%
\includegraphics[clip,width=0.48\textwidth]{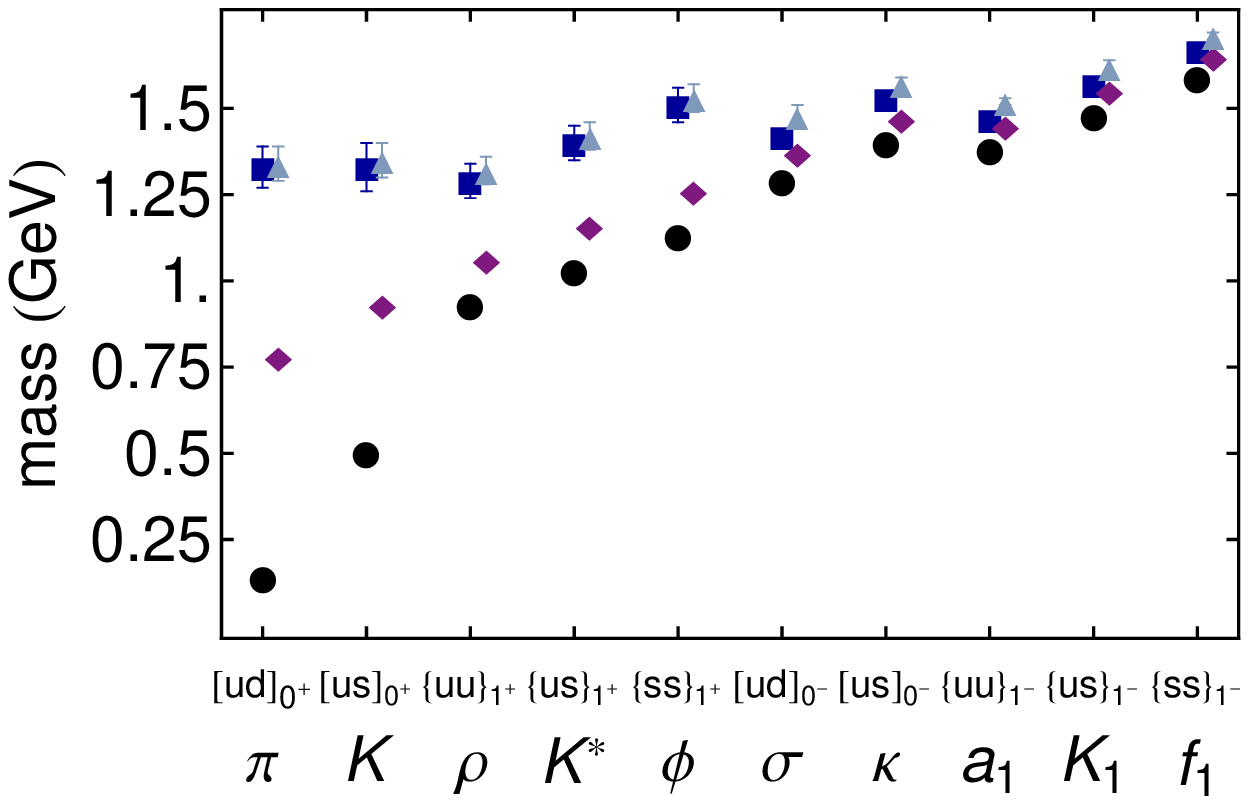}}\vspace*{-42.6ex}

\rightline{%
\includegraphics[clip,width=0.48\textwidth]{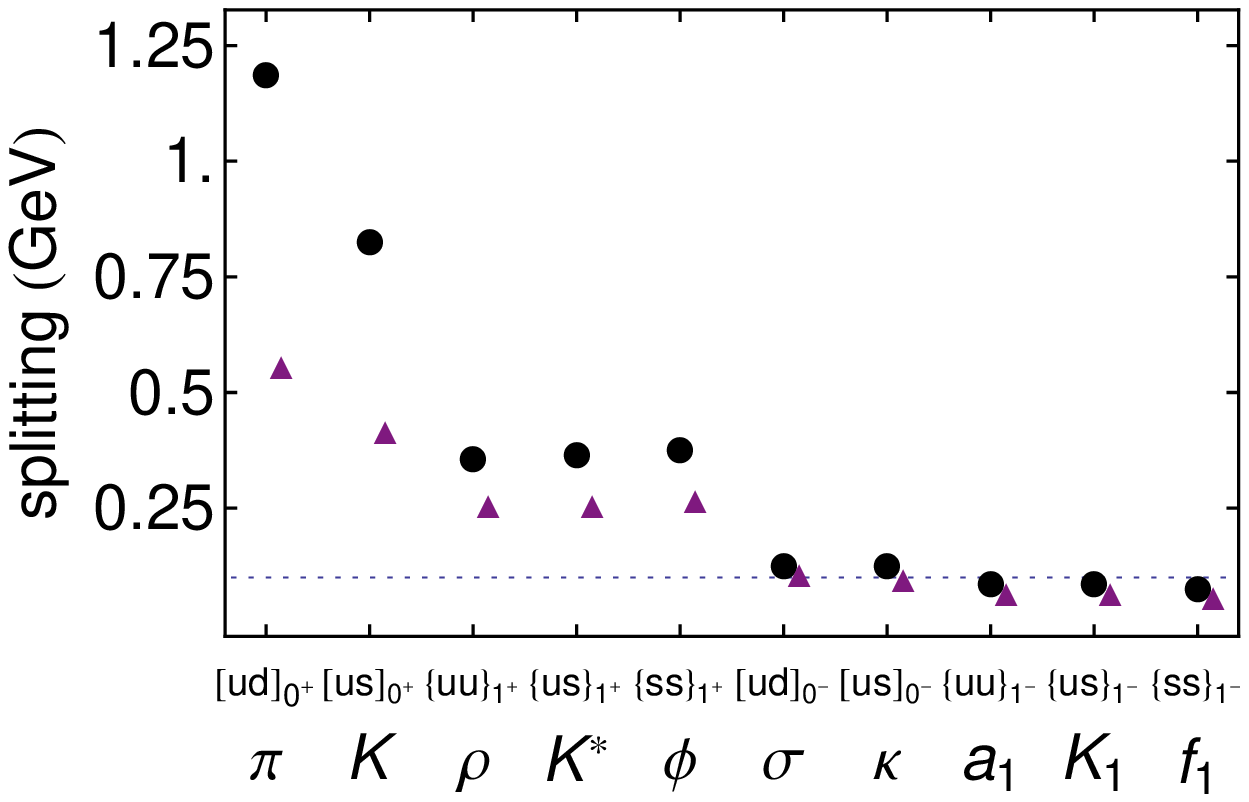}}

\caption{\label{fig:DiquarkMasses}
\underline{Left panel}: Pictorial representation of Table~\protect\ref{Diquarkmasses}.  \emph{Diamonds} -- ground-state diquark masses in Row~1;
\emph{circles} -- ground-state meson masses in Row~2;
\emph{triangles} -- masses of diquark first radial excitations in Row~3;
and
\emph{squares} -- masses of meson radial excitations in Row~4.
\underline{Right panel}:
\emph{Diamonds} -- for diquarks, computed splittings between first radial excitation and ground state; and \emph{circles} -- for mesons, computed splitting between the first radial excitation and ground state in each channel.  The \emph{dashed line} marks a splitting of 0.1\,GeV.}
\end{figure}

The preceding discussion of systematic trends within our predictions for meson masses is important to understanding the results of our Faddeev equation studies because of the connection between the meson and diquark BSEs, outlined in Sect.\,\ref{sec:qqBSADiquark}: predictions for mesons masses determine the diquark spectrum and hence impact heavily on the baryon spectrum.  In Table~\ref{Diquarkmasses} we therefore present results for the masses of diquark correlations and compare them with the meson masses in Table~\ref{Mesonmasses}.  This information is also depicted in the left panel of Fig.\,\ref{fig:DiquarkMasses}.

%-- Irrelevant comparison because these studies were pre-meson.cloud understanding.
%Before discussing relationships between our consistently computed meson and diquark masses, however, we compare our diquark masses with those obtained in two other DSE-based studies of diquark correlations: Ref.\,\cite{Burden:1996nh} in Row~3, which uses a separable interaction whose parameters are determined by fitting gap equation solutions; and Ref.\,\cite{Maris:2002yu} in Row~4, which employs the most widely-used rainbow-ladder-truncation interaction-model \cite{Maris:1999nt}.  Where states

\begin{table}[b]
\caption{\label{Diquarkmasses}
Row~1: Quark-core masses of diquark correlations that play a role in the octet and decuplet spectra of baryons, computed using our symmetry-preserving regularisation of the vector$\times$vector contact interaction, with the input from Table~\protect\ref{Tab:DressedQuarks}.
Row~2: Ground-state meson masses from Row~1 of Table~\protect\ref{Mesonmasses}.
%
%Row~3: Diquark masses computed in Ref.\,\protect\cite{Burden:1996nh}, wherein a $[u,s]_{0^-}$ is not bound; and Row~4: relevant diquark masses computed in Ref.\,\protect\cite{Maris:2002yu}.
%--- These studies are pre-meson.cloud understanding and hence not relevant.
%
Rows~3, 4 repeat the pattern of Rows~1, 2 for the diquarks' first radial excitations.  The theory error in these rows displays the outcome of varying the location of the node in the radial excitation's Bethe-Salpeter amplitude: $1/d_{\cal F} = 2 M^2 ( 1\pm 0.2)$.
(All dimensioned quantities are listed in GeV.)
}
\begin{center}
\begin{tabular*}%{|c|c|c|c|c|c|c|}\hline
{\hsize}
{
l@{\extracolsep{0ptplus1fil}}
l@{\extracolsep{0ptplus1fil}}
|l@{\extracolsep{0ptplus1fil}}
l@{\extracolsep{0ptplus1fil}}
l@{\extracolsep{0ptplus1fil}}
l@{\extracolsep{0ptplus1fil}}
l@{\extracolsep{0ptplus1fil}}
l@{\extracolsep{0ptplus1fil}}
l@{\extracolsep{0ptplus1fil}}
l@{\extracolsep{0ptplus1fil}}
l@{\extracolsep{0ptplus1fil}}
l@{\extracolsep{0ptplus1fil}}}\hline
& &
$[u,d]_{0^+}$ & $[s,u]_{0^+}$ &
$\{u,u\}_{1^+}$ & $\{s,u\}_{1^+}$ & $\{s,s\}_{1^+}$ &
$[u,d]_{0^-}$ & $[s,u]_{0^-}$ &
$\{u,u\}_{1^-}$ & $\{s,u\}_{1^-}$ & $\{s,s\}_{1^-}$ \\\hline
n=0 &
$qq$ \rule{0em}{3ex}&
0.78 & 0.93 &
1.06 & 1.16 & 1.26 &
1.37 & 1.47 &
1.45 & 1.55 & 1.65 \\
&
$q\bar q$ \rule{0em}{3ex}&
0.14 & 0.50 &
0.93 & 1.03 & 1.13 &
1.29 & 1.40 &
1.38 & 1.48 & 1.59 \\\hline
%
% & Ref.\,\protect\cite{Burden:1996nh} \rule{0em}{3ex}&
% 0.74 & 0.82 &
% 0.95 & $1.05$ & $1.13$ &
% 1.50 &   &
% 1.47 & 1.53 & 1.64  \\
% 5, 12, 10, 9, 10, -9, -1.4, 1.3, 0.7 ... rms.rel.err.=8%
%
% & Ref.\,\protect\cite{Maris:2002yu} \rule{0em}{3ex}&
% 0.82 & 1.10 &
% 1.02 & $1.30_{\pm 0.06}$ & $1.44_{\pm 0.04}$ &
%   &   &
%   &   &   \\\hline
% -5, -18, 3.8, -12, -14 ... rms.rel.err.=12%
%
n=1   &
$qq$ \rule{0em}{3ex}&
$1.34_{\pm 0.05}$ & $1.35_{\pm 0.05}$  &
$1.32_{\pm 0.04}$ & $1.42_{\pm 0.04}$ & $1.53_{\pm 0.04}$ &
$1.48_{\pm 0.03}$ & $1.57_{\pm 0.02}$ &
$1.52_{\pm 0.01}$ & $1.62_{\pm 0.02}$ & $1.71_{\pm 0.01}$ \\
&
$q\bar q$ \rule{0em}{3ex}&
$1.33_{\pm 0.06}$ & $1.33_{\pm 0.07}$  &
$1.29_{\pm 0.05}$ & $1.40_{\pm 0.05}$ & $1.51_{\pm 0.05}$ &
$1.42_{\pm 0.02}$ & $1.53_{\pm 0.02}$ &
$1.47_{\pm 0.02}$ & $1.57_{\pm 0.01}$ & $1.67_{\pm 0.02}$ \\\hline
\end{tabular*}
\end{center}
\end{table}

It is plain from Fig.\,\ref{fig:DiquarkMasses} that the level ordering of diquark correlations is precisely the same as that for mesons.  Moreover, in all diquark channels, except the scalar, the mass of the diquark's partner meson is a fair guide to the diquark's mass: the meson mass bounds the diquark's mass from below; and the splitting is always less than 0.13\,GeV and decreases with increasing meson mass.

The scalar diquark channels are particular owing to DCSB and the Goldstone boson character of the partner pseudoscalar mesons.  We note that in a two-color version of QCD, the scalar diquark is also a Goldstone mode \cite{Roberts:1996jx,Bloch:1999vk}, a long-known result of Pauli-G\"ursey symmetry \cite{Pauli:1957,Gursey:1958}.  A memory of the symmetry persists in the three-color theory and is evident here in low masses for the scalar diquarks.  That they are low is highlighted by the right panel of Fig.\,\ref{fig:DiquarkMasses}, which shows large splittings between the ground and excited states in the scalar diquark channel.  Notwithstanding this, the scalar diquark correlations are split widely from the true Goldstone mode mesons.

In constructing baryon Faddeev equations, the canonically normalised diquark Bethe-Salpeter amplitudes are critical because they determine the strength of the correlation within a given baryon.  We list them in Table~\ref{DiquarkBSAs}.  Notably, the amplitudes of the positive-parity states are much larger than those of states with negative-parity: for like-flavour content, the ratio always exceeds $5$.  This pattern repeats that established by the ground-state mesons, see Table~\ref{MesonBSAs}.

\begin{table}[t]
\caption{\label{DiquarkBSAs}
The structure of diquark Bethe-Salpeter amplitudes is described in Sect.\,\ref{sec:qqBSADiquark} and App.\,\ref{app:groundBSEs}.  Here we list all canonically normalised amplitudes that are relevant to the baryons we consider.  Only scalar diquarks involve two independent amplitudes.
%
%Rows~1, 2: ground states; and Rows~3, 4: first radial excitation.
}
\begin{center}
\begin{tabular*}%{|c|c|c|c|c|c|c|}\hline
{\hsize}
{
l@{\extracolsep{0ptplus1fil}}
|l@{\extracolsep{0ptplus1fil}}
l@{\extracolsep{0ptplus1fil}}
l@{\extracolsep{0ptplus1fil}}
l@{\extracolsep{0ptplus1fil}}
l@{\extracolsep{0ptplus1fil}}
l@{\extracolsep{0ptplus1fil}}
l@{\extracolsep{0ptplus1fil}}
l@{\extracolsep{0ptplus1fil}}
l@{\extracolsep{0ptplus1fil}}
l@{\extracolsep{0ptplus1fil}}}\hline
    &
$[u,d]_{0^+}$ & $[s,u]_{0^+}$ &
$\{u,u\}_{1^+}$ & $\{s,u\}_{1^+}$ & $\{s,s\}_{1^+}$ &
$[u,d]_{0^-}$ & $[s,u]_{0^-}$ &
$\{u,u\}_{1^-}$ & $\{s,u\}_{1^-}$ & $\{s,s\}_{1^-}$ \\\hline
$E_{qq}$ \rule{0em}{3ex}&
2.74 & 2.91 &
1.30 & 1.36 & 1.42 &
0.40 & 0.39 &
0.27 & 0.27 & 0.26 \\
$F_{qq}$ \rule{0em}{3ex}&
0.31 & 0.40 &
 &  &  &
 &  &
 &  &  \\\hline
\end{tabular*}
\end{center}
\end{table}

\section{Baryon Faddeev Equations}
\label{sec:BaryonFEs}
\subsection{General structure of the Faddeev amplitudes}
\label{app:FEgeneral}
A spin-$1/2$ baryon is represented by a Faddeev amplitude \cite{Cahill:1988dx}
\begin{equation}
\label{PsiBaryon}
\Psi = \Psi_1 + \Psi_2 + \Psi_3  \,,
\end{equation}
where the subscript identifies the bystander quark and, e.g., $\Psi_{1,2}$ are obtained from $\Psi_3$ by a cyclic permutation of all the quark labels.  We employ the simplest realistic representation of $\Psi$, so that an octet baryon is composed from a sum of scalar and axial-vector diquark correlations:
\begin{equation}
\label{Psi} \Psi_3(p_j,\alpha_j,\varphi_j) = {\cal N}_{\;\Psi_3}^{0^+} + {\cal N}_{\;\Psi_3}^{1^+},
\end{equation}
with $(p_j,\alpha_j,\varphi_j)$ the momentum, spin and flavour labels of the
quarks constituting the bound state, and $P=p_1+p_2+p_3$ the system's total momentum.

It is conceivable that pseudoscalar and vector diquarks could play a role in the Faddeev amplitudes of ground-state $J^P=(1/2)^+$ baryons.  However, such correlations have opposite parity and hence can only appear in concert with nonzero quark angular momentum.  Since one expects ground-states to possess the minimum possible amount of quark orbital angular momentum and these diquark correlations are significantly more massive than the scalar and axial-vector (see Fig.\,\ref{fig:DiquarkMasses}), they can safely be ignored in computing properties of the ground state.

In order to assist in explicating the structure of the diquark pieces in Eq.\,(\ref{Psi}), we define a set of flavour matrices
\begin{equation}
\label{flavourarrays}
\begin{array}{ccc}
{\tt t}^{1=[ud]} = \left[\begin{array}{ccc}
                    0 & 1 & 0 \\
                    -1 & 0 & 0 \\
                    0 & 0 & 0
                    \end{array}\right],
&
{\tt t}^{2=[us]} = \left[\begin{array}{ccc}
                    0 & 0 & 1 \\
                    0 & 0 & 0 \\
                    -1 & 0 & 0
                    \end{array}\right],
&
{\tt t}^{3=[ds]} = \left[\begin{array}{ccc}
                    0 & 0 & 0 \\
                    0 & 0 & 1 \\
                    0 & -1 & 0
                    \end{array}\right],\\[4ex]
{\tt t}^{4=\{uu\}} = \left[\begin{array}{ccc}
                    \surd 2 & 0 & 0 \\
                    0 & 0 & 0 \\
                    0 & 0 & 0
                    \end{array}\right],
&
{\tt t}^{5=\{ud\}} = \left[\begin{array}{ccc}
                    0 & 1 & 0 \\
                    1 & 0 & 0 \\
                    0 & 0 & 0
                    \end{array}\right],
&
{\tt t}^{6=\{us\}} = \left[\begin{array}{ccc}
                    0 & 0 & 1 \\
                    0 & 0 & 0 \\
                    1 & 0 & 0
                    \end{array}\right],\\[4ex]
{\tt t}^{7=\{dd\}} = \left[\begin{array}{ccc}
                    0 & 0 & 0 \\
                    0 & \surd 2 & 0 \\
                    0 & 0 & 0
                    \end{array}\right],
&
{\tt t}^{8=\{ds\}} = \left[\begin{array}{ccc}
                    0 & 0 & 0 \\
                    0 & 0 & 1 \\
                    0 & 1 & 0
                    \end{array}\right],
&
{\tt t}^{9=\{ss\}} = \left[\begin{array}{ccc}
                    0 & 0 & 1 \\
                    0 & 0 & 0 \\
                    0 & 0 & \surd 2
                    \end{array}\right].
\end{array}
\end{equation}
Employing these matrices, the scalar diquark piece in Eq.\,(\ref{Psi}) can be written
\begin{equation}
{\cal N}_{\;\Psi_3}^{0^+}(p_j,\alpha_j,\varphi_j)=
\sum_{[\varphi_1\varphi_2]\varphi_3\in \Psi} \bigg[{\tt t}^{[\varphi_1\varphi_2]}\,
\Gamma_{[\varphi_1\varphi_2]}^{0^+}(\frac{1}{2}p_{[12]};K)\bigg]_{\alpha_1
\alpha_2}^{\varphi_1 \varphi_2}\,
\Delta_{[\varphi_1\varphi_2]}^{0^+}(K) \,[{\cal S}^{\Psi}(\ell;P) u^\Psi(P)]_{\alpha_3}^{\varphi_3} ,
\label{calS}
\end{equation}
where:
$K= p_1+p_2=: p_{\{12\}}$,
$p_{[12]}= p_1 - p_2$, $\ell := (-p_{\{12\}} + 2 p_3)/3$;
\begin{equation}
\label{scalarqqprop}
\Delta_{[\varphi_1\varphi_2]}^{0^+}(K) = \frac{1}{K^2+m_{[\varphi_1\varphi_2]_{0^+}}^2}
\end{equation}
is a propagator for the scalar diquark formed from quarks $1$ and $2$, with $m_{[\varphi_1\varphi_2]_{0^+}}$ the mass-scale associated with this $[\varphi_1 \varphi_2]_{0^+}$ diquark; $\Gamma_{[\varphi_1\varphi_2]}^{0^+}\!$ is the canonically-normalised Bethe-Salpeter amplitude describing the relative momentum correlation between the quarks; ${\cal S}$, a $4\times 4$ Dirac matrix, describes the relative quark-diquark momentum correlation within the baryon; and the spinor satisfies
\begin{equation}
(i\gamma\cdot P + M_\Psi)\, u^\Psi(P) =0= \bar u^\Psi(P)\, (i\gamma\cdot P + M_\Psi)\,,
\end{equation}
with $M_\Psi$ the baryon mass obtained by solving the Faddeev equation.  We note that $u^\Psi$ also possesses another column-vector degree of freedom; viz.,
\begin{equation}
\begin{array}{ccc}
u_p= \left[
\begin{array}{c}
[ud] u \\
\{uu\} d \\
\{ud\} u
\end{array} \right], &
u_{\Sigma^+}=
\left[ \begin{array}{c}
[us] u \\
\{uu\} s \\
\{us\} u
\end{array} \right],
u_{\Xi^0}=
\left[
\begin{array}{c}
[us] s \\
\{us\} s \\
\{ss\} u
\end{array}\right],&
u_\Lambda= \frac{1}{\surd 2}
\left[
\begin{array}{c}
\surd 2 [ud] s \\
\, [ud] s - [ds] u \\
\{us\} d - \{d s\} u
\end{array}
\right].
\end{array}
\label{uColumn}
\end{equation}
Owing to our assumption of isospin symmetry, the unlisted octet charge states are degenerate with their listed partners.

The axial-vector part of Eq.\,(\ref{Psi}) is
\begin{equation}
{\cal N}_{\;\Psi_3}^{1^+}(p_j,\alpha_j,\varphi_j)=
\sum_{\{\varphi_1\varphi_2\}\varphi_3\in \Psi} \bigg[{\tt t}^{\{\varphi_1\varphi_2\}}\,
\Gamma_{\mu\{\varphi_1\varphi_2\}}^{1^+}(\frac{1}{2}p_{[12]};K)\bigg]_{\alpha_1
\alpha_2}^{\varphi_1 \varphi_2}\,
\Delta_{1^+\mu\nu}^{\{\varphi_1\varphi_2\}}(K) \,[{\cal A}_\nu^{\Psi}(\ell;P) u^\Psi(P)]_{\alpha_3}^{\varphi_3} ,
\label{calA}
\end{equation}
where
\begin{equation}
\Delta_{1^+ \mu\nu}^{\{\varphi_1\varphi_2\}}(K) = \frac{1}{K^2+m_{m^2_{\{\varphi_1 \varphi_2\}_{1^+}}}} \, \left(\delta_{\mu\nu} + \frac{K_\mu K_\nu}{m^2_{\{\varphi_1 \varphi_2\}_{1^+}}}\right) ,
\label{qqavprop}
\end{equation}
is a propagator for the axial-vector diquark formed from quarks $1$ and $2$ and the other elements in Eq.\,\eqref{calA} are obvious analogues of those in Eq.\,\eqref{calS}.

In connection with decuplet baryons we note that it is not possible to combine an isospin-zero diquark with an isospin-1/2 quark to obtain isospin-3/2 and hence the $\Delta$ is comprised solely from axial-vector correlations.  This sets the pattern for the remaining decuplet baryons, which may therefore be expressed via
\begin{equation}
\label{DecupletFA}
\Psi_3^{10}(p_i,\alpha_i,\varphi_i) = {\cal D}_{\Psi_3^{10}}^{1+}(p_j,\alpha_j,\varphi_j),
\end{equation}
with
\begin{equation}
{\cal D}_{\,\Psi_3^{10}}^{1^+}(p_j,\alpha_j,\varphi_j)=
\! \sum_{\{\varphi_1\varphi_2\}\varphi_3\in \Psi^{10}} \bigg[{\tt t}^{\{\varphi_1\varphi_2\}}\,
\Gamma_{\{\varphi_1\varphi_2\}}^{1^+}(\frac{1}{2}p_{[12]};K)\bigg]_{\alpha_1
\alpha_2}^{\varphi_1 \varphi_2} \rule{-1ex}{0ex}
\Delta_{1^+\mu\nu}^{\{\varphi_1\varphi_2\}}(K) \,[{\cal D}_{\nu\rho}^{\Psi^{10}}(\ell;P) u_\rho^{\Psi^{10}}(P)]_{\alpha_3}^{\varphi_3} ,
\label{calD}
\end{equation}
where $u_\rho^{\Psi^{10}}(P)$ is a Rarita-Schwinger spinor and, as with octet baryons, in constructing the Faddeev equations we focus on that member of each isospin multiplet which has maximum electric charge; viz.,
\begin{equation}
\begin{array}{ccc}
u_\Delta= \left[
\begin{array}{c}
\{uu\} u \\
\end{array} \right], &
u_{\Sigma^\ast}=
\left[ \begin{array}{c}
\{uu\} s \\
\{us\} u
\end{array} \right],
u_{\Xi^\ast}=
\left[
\begin{array}{c}
\{us\} s \\
\{ss\} u
\end{array}\right],&
u_\Omega=
\left[
\begin{array}{c}
\{ss\}s
\end{array}
\right].
\end{array}
\end{equation}

The general forms of the matrices ${\cal S}^{\Psi}(\ell;P)$, ${\cal A}^{\Psi}_\nu(\ell;P)$ and ${\cal D}^{\Psi^{10}}_{\nu\rho}(\ell;P)$, which describe the momentum-space correlation between the quark and diquark in the octet and decuplet baryons, respectively, are described in Refs.\,\cite{Cloet:2007pi,Oettel:1998bk}.  The requirement that ${\cal S}^{\Psi}(\ell;P)$ represent a positive energy baryon entails
\begin{equation}
\label{SexpI}
{\cal S}^{\Psi}(\ell;P) = s^{\Psi}_1(\ell;P)\,\mathbf{I}_{\rm D} + \left(i\gamma\cdot \hat\ell - \hat\ell \cdot \hat P\, \mathbf{I}_{\rm D}\right)\,s^{\Psi}_2(\ell;P)\,,
\end{equation}
where $(\mathbf{I}_{\rm D})_{rs}= \delta_{rs}$, $\hat \ell^2=1$, $\hat P^2= - 1$.  In the baryon rest frame, $s^{\Psi}_{1,2}$ describe, respectively, the upper, lower component of the bound-state baryon's spinor.  Placing the same constraint on the axial-vector component, one has
\begin{equation}
\label{AexpI}
 {\cal A}^{\Psi}_\nu(\ell;P) = \sum_{n=1}^6 \, p_n^\Psi(\ell;P)\,\gamma_5\,A^n_{\nu}(\ell;P)\,,
\end{equation}
where ($ \hat \ell^\perp_\nu = \hat \ell_\nu + \hat \ell\cdot\hat P\, \hat P_\nu$, $ \gamma^\perp_\nu = \gamma_\nu + \gamma\cdot\hat P\, \hat P_\nu$)
\begin{equation}
\label{AfunctionsI}
\begin{array}{lll}
A^1_\nu= \gamma\cdot \hat \ell^\perp\, \hat P_\nu \,,\; &
A^2_\nu= -i \hat P_\nu \,,\; &
A^3_\nu= \gamma\cdot\hat \ell^\perp\,\hat \ell^\perp\,,\\
A^4_\nu= i \,\hat \ell_\mu^\perp\,,\; &
A^5_\nu= \gamma^\perp_\nu - A^3_\nu \,,\; &
A^6_\nu= i \gamma^\perp_\nu \gamma\cdot\hat \ell^\perp - A^4_\nu\,.
\end{array}
\end{equation}

Finally, requiring also that ${\cal D}^{\Psi^{10}}_{\nu\rho}(\ell;P)$ be an eigenfunction of $\Lambda_+(P)$, one obtains
\begin{equation}
\label{DeltaFA}
{\cal D}^{\Psi^{10}}_{\nu\rho}(\ell;P) = {\cal S}^{\Psi^{10}}(\ell;P) \, \delta_{\nu\rho} + \gamma_5{\cal A}^{\Psi^{10}}_\nu(\ell;P) \,\ell^\perp_\rho \,,
\end{equation}
with ${\cal S}^{\Psi^{10}}$ and ${\cal A}^{\Psi^{10}}_\nu$ given by obvious analogues of Eqs.\,(\ref{SexpI}) and (\ref{AexpI}), respectively.

Having in hand detailed forms for the dressed-quark propagators, the diquark Bethe-Salpeter amplitudes and the diquark propagators, it is now possible to write Faddeev equations for the baryons.  As apparent in Fig.\,\ref{fig:FaddeevI}, the kernels of those equations involve diquark breakup and reformation via exchange of a dressed-quark.  In proceeding we follow Ref.\,\cite{Roberts:2011cf} and make a drastic simplification; namely, in the Faddeev equation for a baryon of type $B$, the quark exchanged between the diquarks is represented as
\begin{equation}
S^{\rm T}(k) \to \frac{g_B^2}{M_f}\,,
\label{staticexchange}
\end{equation}
where the superscript ``T'' indicates matrix transpose, $f$ is the quark's flavour and $g_B$ is discussed below.  This is a variant of the so-called ``static approximation,'' which itself was introduced in Ref.\,\cite{Buck:1992wz} and has subsequently been used in studies of a range of nucleon properties \cite{Wilson:2011aa,Bentz:2007zs}.  In combination with diquark correlations generated by Eq.\,(\ref{njlgluon}), whose Bethe-Salpeter amplitudes are momentum-independent, Eq.\,(\ref{staticexchange}) generates Faddeev equation kernels which themselves are momentum-independent.  It follows that Eqs.\,\eqref{SexpI}, \eqref{AexpI} simplify dramatically, with only those terms that are independent of the relative momentum surviving:
\begin{subequations}
\label{calSAcontact}
\begin{eqnarray}
{\cal S}^{\Psi}(\ell;P)  \to {\cal S}^{\Psi}(P) &=& s^{\Psi}(P)\,\mathbf{I}_{\rm D} \,,\\
{\cal A}^{\Psi}_\nu(\ell;P) \to {\cal A}^{\Psi}_\mu(P) &=& a_1^\Psi(P)\,i\gamma_5\gamma_\mu + a_2^{\Psi}(P) \gamma_5 \hat P_\mu\,. \label{calAcontact}
\end{eqnarray}
\end{subequations}

\subsection{Explicit example: $\Lambda$ baryon}
\label{sec:FELambda}
In App.\,\ref{app:FEDelta} we derive the Faddeev equation for the $\Delta^{++}$ resonance, whose simple spin-flavour structure makes it a useful illustrative example; and in App.\,\ref{app:FENucleon} the nucleon's Faddeev equation is reported, with indications of the minor changes in analysis that are necessary to complete its derivation.
Here, on the other hand, we illustrate different aspects of the construction of the Faddeev equation by considering the $\Lambda$ baryon, which is an isospin-zero, $J=(1/2)^+$ state constituted from a single quark of each flavour and hence has a complicated spin-flavour amplitude.
%\footnote{N.B.\ Derivations of the nucleon and $\Delta$ Faddeev equations are reported in Sect.\,4.1 and App.\,C.2 of Ref.\,\cite{Roberts:2011cf}.}

Recalling comments made when opening Sect.\,\ref{app:FEgeneral}, five possible diquark combinations are possible for the ground-state $\Lambda$:
\begin{equation}
s[ud]_{0^+},\; d[us]_{0^+}, \;u[ds]_{0^+},\; d\{us\}_{1^+},\; u\{ds\}_{1^+} .
\end{equation}
Of these, $s[ud]_{0^+}$ has $I=0$, whilst the others do not possess good isospin.  This lies behind a mixing that obscures the $\Lambda$ and $\Sigma^0$ octet baryon isospin-eigenstates and entails that building the flavour structure of the $\Lambda$ Faddeev kernel is quite involved.  States of good isospin can be constructed as follows: with
\begin{equation}
\begin{array}{cc}
 V=\left(
  \begin{array}{c}
   [ud]s\\[0.7ex]
   [us]d\\[0.7ex]
   [ds]u\\[0.7ex]
   \{us\}d\\[0.7ex]
   \{ds\}u
  \end{array}
  \right), &
 O=\left(
  \begin{array}{ccccc}
   1 & 0 & 0 & 0 & 0\\[0.7ex]
   0 & \frac{1}{\surd 2} & -\frac{1}{\surd 2} & 0 & 0\\[0.7ex]
   0 & \frac{1}{\surd 2} & \frac{1}{\surd 2} & 0 & 0\\[0.7ex]
   0 & 0 & 0 & \frac{1}{\surd 2} & -\frac{1}{\surd 2}\\[0.7ex]
   0 & 0 & 0 & \frac{1}{\surd 2} & \frac{1}{\surd 2}
  \end{array}
  \right) ,
 \end{array}
\label{VOmatrices}
\end{equation}
then each of the entries in the new column vector
\begin{equation}
\label{goodV}
 \tilde{V}=OV=\frac{1}{\surd 2}
 \left(
  \begin{array}{cc}
   \surd 2 [ud]s\,, & I=0\\[0.7ex]
   [us]d-[ds]u\,, & I=0\\[0.7ex]
   [us]d+[ds]u\,, & I=1\\[0.7ex]
   \{us\}d-\{ds\}u\,, & I=0 \\[0.7ex]
   \{us\}d+\{ds\}u\,, & I=1
  \end{array}
  \right)
\end{equation}
is an eigenvectors of isospin, with the eigenvalue indicated.

A consideration of Fig.\,\ref{fig:FaddeevI} reveals that the column vector $V$ satisfies a Faddeev equation of the form $V = K_{uds} V$, which it is helpful to write explicitly:
\begin{equation}
\left(
  \begin{array}{c}
   [ud]s\\[0.7ex]
   [us]d\\[0.7ex]
   [ds]u\\[0.7ex]
   \{us\}d\\[0.7ex]
   \{ds\}u
  \end{array}
  \right)
=
\left(
\begin{array}{ccccc}
0 & K_{[ud],[us]} & K_{[ud],[ds]} & K_{[ud],\{us\}} & K_{[ud],\{ds\}} \\
K_{[us],[ud]} & 0 & K_{[us],[ds]} & 0 & K_{[us],\{ds\}} \\
K_{[ds],[ud]} & K_{[ds],[us]} & 0 & K_{[ds],\{us\}} & 0 \\
K_{\{us\},[ud]} & 0 & K_{\{us\},[ds]}  & 0 & K_{\{us\},\{ds\}} \\
K_{\{ds\},[ud]} & K_{\{ds\},[us]} & 0 & K_{\{ds\},\{us\}} & 0
\end{array}
\right)
\left(
  \begin{array}{c}
   [ud]s\\[0.7ex]
   [us]d\\[0.7ex]
   [ds]u\\[0.7ex]
   \{us\}d\\[0.7ex]
   \{ds\}u
  \end{array}
  \right),
\end{equation}
where, e.g., $K_{[ud],[us]}$ describes the breakup of a $[us]$ scalar diquark through emission of a dressed $u$-quark, which then joins the $d$-quark to form a $[ud]$ scalar diquark, leaving the $s$-quark as a bystander.  In the kernel, the repeated flavour label always indicates the exchanged quark.  Using the notation connected with Eq.\,\eqref{flavourarrays}, one may write
\begin{equation}
K_{uds} = \left(
\begin{array}{ccccc}
0 & K_{12} & K_{13} & K_{16} & K_{18} \\
K_{21} & 0 & K_{23} & 0 & K_{28} \\
K_{31} & K_{32} & 0 & K_{36} & 0 \\
K_{61} & 0 & K_{63}  & 0 & K_{68} \\
K_{81} & K_{82} & 0 & K_{86} & 0
\end{array}
\right)\,.
\end{equation}

We now make two entries explicit:
\begin{subequations}
\begin{eqnarray}
K_{12} & = & U_F^{\rm T} {\tt t}^2 \Gamma_{[us]}(l_{ds}) S_u^{\rm T} {\tt t}^{1{\rm T}} \bar\Gamma_{[ud]}(-k_{ud}) S_d(l_d) \Delta_{[us]}(l_{us}) \, U_F \,, \\
& = & \Gamma_{[us]}(l_{ds}) S_u^{\rm T} \bar\Gamma_{[ud]}(-k_{ud}) S_d(l_d) \Delta_{[us]}(l_{us}) =: {\cal K}_{12}\,,   \\
K_{13} & = & U_F^{\rm T} {\tt t}^3 \Gamma_{[ds]}(l_{ds}) S_d^{\rm T} {\tt t}^{1 {\rm T}} \bar\Gamma_{[ud]}(-k_{qq}) S_u(l_u) \Delta_{[ds]}(l_{ds}) U_F,\\
& = & - \Gamma_{[ds]}(l_{ds}) S_d^{\rm T}  \bar\Gamma_{[ud]}(-k_{qq}) S_u(l_u) \Delta_{[ds]}(l_{ds}) =: -{\cal K}_{13}\,,
\end{eqnarray}
\end{subequations}
where $U_F^{\rm T}=(1,1,1)$ is a device that we use to collapse the flavour structure.  Repeating this procedure for all entries and using isospin symmetry, one arrives at
\begin{equation}
K_{uds} = \left(
\begin{array}{ccccc}
0 & {\cal K}_{\;12} & -{\cal K}_{\;13} & - {\cal K}_{\;16} & {\cal K}_{\;16} \\
{\cal K}_{\;21} & 0 & {\cal K}_{\;23} & 0 & {\cal K}_{\;28} \\
-{\cal K}_{\;21} & {\cal K}_{\;23} & 0 & {\cal K}_{\;28} & 0 \\
-{\cal K}_{\;61} & 0 & {\cal K}_{\;62}  & 0 & {\cal K}_{\;68} \\
{\cal K}_{\;61} & {\cal K}_{\;62} & 0 & {\cal K}_{\;68} & 0
\end{array}
\right)\,.
\end{equation}

As the kernel $K_{uds}$ contains mixing that obscures the $\Lambda$ and $\Sigma^0$ isospin-eigenstate baryons, we now employ the matrix $O$ in Eq.\,\eqref{VOmatrices} and construct the non-mixing kernel:
\begin{equation}
{\cal K}_{uds} = O \, K_{uds} \, O^{\rm T} =
\left(
  \begin{array}{ccccc}
   0 & \surd 2 {\cal K}_{\;12} & 0 & -\surd 2 {\cal K}_{\;16} & 0\\[0.7ex]
   \surd 2 {\cal K}_{\;21} & -{\cal K}_{\;23} & 0 & -{\cal K}_{\;28} & 0\\[0.7ex]
   0 & 0 & {\cal K}_{\;23} & 0 & {\cal K}_{\;28}\\[0.7ex]
   -\surd 2 {\cal K}_{\;61} & -{\cal K}_{\;63} & 0 & -{\cal K}_{\;68} & 0\\[0.7ex]
   0 & 0 & {\cal K}_{\;63} & 0 & {\cal K}_{\;68}
  \end{array}
  \right).
\end{equation}
Rows 1, 2, 4 map $I=0$ into itself, whereas rows 3, 5 do the same for $(I,I_z) = (1,0)$.  Focusing on the $I=0$ sector, one arrives at the following Faddeev equation for the $\Lambda$ baryon: $\tilde{V}_\Lambda = {\cal K}^\Lambda_{\,uds}\tilde{V}_\Lambda$; i.e., explicitly,
\begin{equation}
\tilde{V}_\Lambda=\frac{1}{\surd 2}
\left(
  \begin{array}{c}
   \surd 2 [ud]s \\[0.7ex]
   [us]d-[ds]u\\[0.7ex]
   \{us\}d-\{ds\}u\\[0.7ex]
  \end{array}
  \right)
=
\left(
  \begin{array}{ccc}
   0 & \surd 2 {\cal K}_{\;12} &  -\surd 2 {\cal K}_{\;16} \\[0.7ex]
   \surd 2 {\cal K}_{\;21} & -{\cal K}_{\;23} & -{\cal K}_{\;28} \\[0.7ex]
   -\surd 2 {\cal K}_{\;61} & -{\cal K}_{\;63} &  -{\cal K}_{\;68}
  \end{array}
  \right)
\frac{1}{\surd 2}
\left(
  \begin{array}{c}
   \surd 2 [ud]s \\[0.7ex]
   [us]d-[ds]u\\[0.7ex]
   \{us\}d-\{ds\}u\\[0.7ex]
  \end{array}
  \right). \label{FELambda}
\end{equation}

Acknowledging that the axial-vector diquark components of all baryon amplitudes involve two scalar functions, as evident in Eq.\,\eqref{calAcontact}, we re-express ${\cal K}^\Lambda_{\,uds}$ as follows:
\begin{equation}
{\cal K}^\Lambda_{\;uds} = \left(
  \begin{array}{cccc}
   0 & \surd 2 {\cal K}^\Lambda_{\;12} &  -\surd 2 {\cal K}^\Lambda_{\;16_1} &  -\surd 2 {\cal K}^\Lambda_{\;16_2} \\[0.7ex]
   \surd 2 {\cal K}^\Lambda_{\;21} & -{\cal K}^\Lambda_{\;23} & -{\cal K}^\Lambda_{\;28_1} & -{\cal K}^\Lambda_{\;28_2}\\[0.7ex]
   -\surd 2 {\cal K}^\Lambda_{\;6_11} & -{\cal K}^\Lambda_{\;6_13} &  -{\cal K}^\Lambda_{\;6_18_1} &  -{\cal K}^\Lambda_{\;6_18_2}\\[0.7ex]
   -\surd 2 {\cal K}^\Lambda_{\;6_2 1} & -{\cal K}^\Lambda_{\;6_2 3} &  -{\cal K}^\Lambda_{\;6_2 8_1} &  -{\cal K}^\Lambda_{\;6_2 8_2}\\[0.7ex]
  \end{array}
  \right);
  \label{KernelLambda}
\end{equation}
i.e., adding subscripts associated with the diquark labels $6,8$.  Having arrived at this form, one can write compact algebraic expressions for each of the entries, which we list in App.\,\ref{App:Lambda}.

Equation~\eqref{FELambda} is an eigenvalue problem whose solution yields the mass for the dressed-quark-core of the $\Lambda$-resonance.  If one sets $g_\Lambda=1$, then the Faddeev equation yields $m_\Lambda = 1.39\,$GeV and the unit-normalised eigenvector
\begin{equation}
\begin{array}{cccc}
s_\Lambda^{1} & s_\Lambda^{[2,3]} & a_{\Lambda 1}^{[6,8]} &a_{\Lambda 2}^{[6,8]} \\
0.65 & 0.59 & -0.47 & -0.020
%{0.654631, 0.591301, -0.470556, -0.019947}
\end{array}\,.
\end{equation}
The eigenvector corresponds to the following probabilities: $P_{\Lambda}^1= 43$\%, $P_\Lambda^{[2,3]}=35$\%, $P_\Lambda^{[6,8]}=22$\%; i.e., the ground-state $\Lambda$ is 22\% axial vector diquark and 78\% scalar, with a roughly equal probability of both $I=0$ scalar diquark configurations.  Comparison with Eq.\,\eqref{nucleonqqratio} reveals an interesting result; viz., $P_\Lambda^{[6,8]}=P_N^4+P_N^5$.  So, despite the differences in flavour structure, the net axial-vector diquark content of the nucleon and $\Lambda$ is the same.

\subsection{Pion-loops and baryon masses}
We discussed the nature of resonance contributions to meson bound-state kernels on page~\pageref{page:pionloops}.  Similar remarks may be made in connection with baryon Faddeev equations and this provides us with a means by which to fix $g_B$ in Eq.\,\eqref{staticexchange}; viz., formulae such as those in Ref.\,\cite{Young:2002cj} can be used to estimate the size of type-2 meson-loop corrections to baryon masses computed using the Faddeev equations herein.  The straightforward application of such expressions, using a common meson-baryon form-factor mass-scale of $0.8\,$GeV, yields a shift of $(-300\,$MeV$)$ in $m_N$ and $(-270\,$MeV$)$ in $m_\Delta$, from which one may infer that our type-2 uncorrected Faddeev equations should produce $m_N=1.24\,$GeV and $m_\Delta=1.50\,$GeV, values which Apps.\,\ref{app:FEDelta},\ref{app:FENucleon} show to be of the appropriate size.  For the $\Delta$-resonance there is another estimate, which is arguably more sophisticated.  Namely, that produced by the Excited Baryon Analysis Center \cite{Suzuki:2009nj}, which used a realistic coupled-channels model to remove meson dressing from the $\Delta$ and expose a dressed-quark-core bare-mass of $1.39\,$GeV.  Following these observations, which are condensed from Ref.\,\cite{Roberts:2011cf}, we return to Eq.\,(\ref{staticexchange}) and choose
\begin{equation}
\label{gNgDelta}
g_8:=g_{B=N,\Lambda,\Sigma,\Xi} = 1.18\,,\;
g_{10}:= g_{\Delta,\Sigma^\ast,\Xi^\ast,\Omega} = 1.56\, ,
\end{equation}
so that the Faddeev equations in Apps.\,\ref{app:FEDelta},\ref{app:FENucleon} produce $m_N=1.14\,$GeV, $m_\Delta = 1.39\,$GeV, $\delta m = m_\Delta - m_N = 0.25\,$GeV because these outcomes are consistent with the information presented above and Refs.\,\cite{Ishii:1998tw,Hecht:2002ej,Cloet:2008re}.  N.B.\ Notwithstanding this common choice for the couplings within the octet and separately within the decuplet, the kernels in different strangeness sectors are still distinct owing to the appearance of the dressed-quark mass in the denominator of Eq.\,\eqref{staticexchange}.

%We remark that it is most appropriate to view Eq.\,\eqref{staticexchange} as implementing an approximation to the Faddeev equation kernel in a given channel instead of merely as a simplification of the propagator for the exchanged quark.  From that perspective one might argue that the Bethe-Salpeter and Faddeev amplitudes could influence the integrand's support within the channel under consideration and hence that $g_B$ should rightly depend on isospin and strangeness, contrary to the assumptions expressed in Eq.\,\eqref{gNgDelta}.

\begin{table}[t]
\caption{\label{OctetDecupletMasses}
Dressed-quark-core masses of ground-state octet and decuplet baryons, those of their radial excitations and of all their parity partners, computed using our symmetry-preserving formulation of the vector$\times$vector contact interaction supplemented by Eqs.\,\eqref{gNgDelta} and the material in Sect.\,\ref{sec:Faddeev}.
Row~1: Baryon ground-states.  The lowest mass state in each channel has positive parity.
Row~3: First radial excitations of the ground-states, which provide the second level in each channel.  The theory error in this row displays the outcome of varying the location of the node in the radial excitation's Faddeev amplitude: $1/d_{\cal F} = 2 M^2 ( 1\pm 0.2)$.
Row~5: Parity partners of the baryon ground-states, which provide the third level in each channel.
Row~7: First radial excitations of the parity partner to each of the baryon ground-states, which provide the fourth level in each channel.
Masses in the rows labelled ``expt.'' are taken from Ref.\,\protect\cite{Nakamura:2010zzi}.  To explain the notation we observe that the entry $1.44_{P_{11}}$ in the ``N'' column is read as ``$N(1440)P_{11}$'', and a hyphen in any position indicates that no empirically known resonance can confidently be associated with the associated state.
(All dimensioned quantities are listed in GeV.)
}
\begin{center}
\begin{tabular*}%{|c|c|c|c|c|c|c|}\hline
{\hsize}
{
l@{\extracolsep{0ptplus1fil}}
l@{\extracolsep{0ptplus1fil}}
l@{\extracolsep{0ptplus1fil}}
|l@{\extracolsep{0ptplus1fil}}
l@{\extracolsep{0ptplus1fil}}
l@{\extracolsep{0ptplus1fil}}
l@{\extracolsep{0ptplus1fil}}
|l@{\extracolsep{0ptplus1fil}}
l@{\extracolsep{0ptplus1fil}}
l@{\extracolsep{0ptplus1fil}}
l@{\extracolsep{0ptplus1fil}}}\hline
& & & \rule{0ex}{2.5ex}
$N$ & $\Lambda$ & $\Sigma$ & $\Xi$
   & $\Delta$ & ${\Sigma^\ast}$ & ${\Xi^\ast}$ & $\Omega$ \\\hline
$P=+$ & n=0 & \rule{0ex}{2.5ex}
DSE & 1.14 & 1.26 & 1.35 & 1.51 & 1.39 & 1.51 & 1.63 & 1.76\rule{0em}{2.5ex} \\
& & \rule{0ex}{2.5ex}
expt. & 0.94 & 1.12 & 1.19 &  1.31 & $1.23_{P_{33}}$ & $1.39_{P_{13}}$ & $1.53_{P_{13}}$ & 1.67\\\hline
$P=+$ & n=1 & \rule{0ex}{2.5ex}
DSE & $1.82_{\pm 0.07}$ & $1.89_{\pm 0.07}$ & $2.09_{\pm 0.01}$ & $2.16_{\pm 0.01}$ &
$1.84_{\pm 0.04}$ & $1.94_{\pm 0.04}$ & $2.04_{\pm 0.04}$ & $2.14_{\pm 0.04}$\rule{0em}{2.5ex} \\
& & \rule{0ex}{2.5ex}
expt. & $1.44_{P_{11}}$ & $1.60_{P_{01}}$ & $1.66_{P_{11}}$ &  - & $1.60_{P_{33}}$ & - & - & -\\\hline
$P=-$ & n=0 & \rule{0ex}{2.5ex}
%DSE-old & 2.22 & 2.31 & 2.33 & 2.43 & 2.25 & 2.35 & 2.46 & 2.56\rule{0em}{2.5ex} \\
DSE & 2.30 & 2.40 & 2.41 & 2.51 & 2.33 & 2.44 & 2.55 & 2.65\rule{0em}{2.5ex} \\
& & \rule{0ex}{2.5ex}
expt. & $1.54_{S_{11}}$ & $1.67_{S_{01}}$ & $1.75_{S_{11}}$ & - & $1.70_{D_{33}}$ & $1.67_{D_{13}}$ & $1.82_{D_{13}}$ & - \\\hline
$P=-$ & n=1 & \rule{0ex}{2.5ex}
%DSE-old & $2.29_{\pm 0.01}$ & $2.39_{\pm 0.02}$ & $2.40_{\pm 0.01}$ & $2.49_{\pm 0.01}$ & $2.33_{\pm 0.02}$ & $2.42_{\pm 0.01}$ & $2.52_{\pm 0.01}$ & $2.62_{\pm 0.02}$\rule{0em}{2.5ex} \\
DSE & $2.35_{\pm 0.01}$ & $2.45_{\pm 0.01}$ & $2.46_{\pm 0.01}$ & $2.55_{\pm 0.01}$ & $2.39_{\pm 0.01}$ & $2.49_{\pm 0.01}$ & $2.59_{\pm 0.01}$ & $2.70_{\pm 0.01}$\rule{0em}{2.5ex} \\
& & \rule{0ex}{2.5ex}
expt. & $1.65_{S_{11}}$ & $1.80_{S_{01}}$ & - &  - & - & $1.94_{D_{13}}$ & - & -\\\hline
\end{tabular*}
\end{center}
\end{table}

\section{Baryon spectrum}
\label{sec:results}
%\subsection{Ground states}
\subsection{Faddeev equations}
The Faddeev equations for all ground-state baryons in our formulation of the contact interaction are presented in App.\,\ref{App:AllFEs}.

Turning to radial excitations we note that, in analogy with mesons, the leading Tchebychev moment of the bound-state amplitude for a baryon's first radial excitation should possess a single zero.  Hence, as advocated in Ref.\,\cite{Roberts:2011cf}, it is possible to estimate masses for these states by employing the expedient described in App.\,\ref{sec:radial}.  With the zero located as prescribed in Eq.\,(\ref{locatezero}), no new parameters are introduced.  Following this reasoning, the Faddeev equation for the first radial excitation of each baryon is simply obtained by making the following replacement throughout the equations in App.\,\ref{App:AllFEs}:
\begin{equation}
\label{groundtoradial}
\overline{\cal C}_1(\sigma) \to \overline{\cal F}_1(\sigma)\,.
\end{equation}

In a more general setting one might imagine that a baryon's first radial excitation could be an admixture of two components: one with a zero in the Faddeev amplitude, describing a radial excitation of the quark-diquark system; and the other with a zero in the diquark's Bethe-Salpeter amplitude, which represents an internal excitation of the diquark.  The procedure in App.\,\ref{sec:radial} can conceivably distinguish between these components via a mixing term whose strength is $\propto E_{(fg)_{J^P}}E_{(fg)^\ast_{J^P}}$, where the latter is the excited diquark's amplitude.  Owing to orthogonality of the two-body ground- and first-radially-excited states, we anticipate that this mixing term is negligible.  Under this assumption, a baryon's first radial excitation is predominantly a radial excitation of the quark-diquark system.  Should a state constituted from a radially-excited diquark exist, then its mass will be larger because $[E_{(fg)^\ast_{J^P}}^2/m_{(fg)^\ast_{J^P}}^2]\ll [E_{(fg)_{J^P}}^2/m_{(fg)_{J^P}}^2]$.

The case of parity partners is more complicated.  One must typically reanalyse each isospin/flavour channel, beginning with the equation represented by Fig.\,\ref{fig:FaddeevI}, because of the altered Dirac structure of the diquark correlations that are involved.  For example, the nucleon's parity partner is composed of pseudoscalar and vector diquark correlations and its Faddeev amplitude must change sign under a parity transformation.  These alterations lead to changes in the locations of the $\gamma_5$ matrices in the Faddeev equation and also a reduction in the number of terms because the pseudoscalar diquark does not possess a $F(P)$-component [Eq.\,\eqref{BSA[su]} cf.\ Eq.\,\eqref{sBSA}].
With the exception of the $\Delta(\frac{3}{2}^-)$, whose Faddeev equation is written in Eq.\,\eqref{DeltaPPartnerFE}, we do not report parity-partner Faddeev equations herein.  As will be seen from Eq.\,\eqref{DeltaPPartnerFE}, however, the Faddeev kernels that one obtains via this reanalysis have similar forms to those for the positive parity states, listed in App.\,\ref{App:AllFEs}, but sign changes are introduced\footnote{
These changes were inadvertently overlooked in Ref.\,\protect\cite{Roberts:2011cf}.  They lead to increases in the mass of parity partners but the effect is always smaller than 4\%.}
and, throughout, one has the replacements
\begin{equation}
E_{(fg)_{J^P}} \to E_{(fg)_{J^{-P}}}\,, \; m_{(fg)_{J^P}} \to m_{(fg)_{J^{-P}}}.
\end{equation}
Equations for the radial excitations of the parity partners are readily obtained by using Eq.\,(\ref{groundtoradial}).

\subsection{Computed Results}
\begin{figure}[t]
\leftline{%
\includegraphics[clip,width=0.48\textwidth]{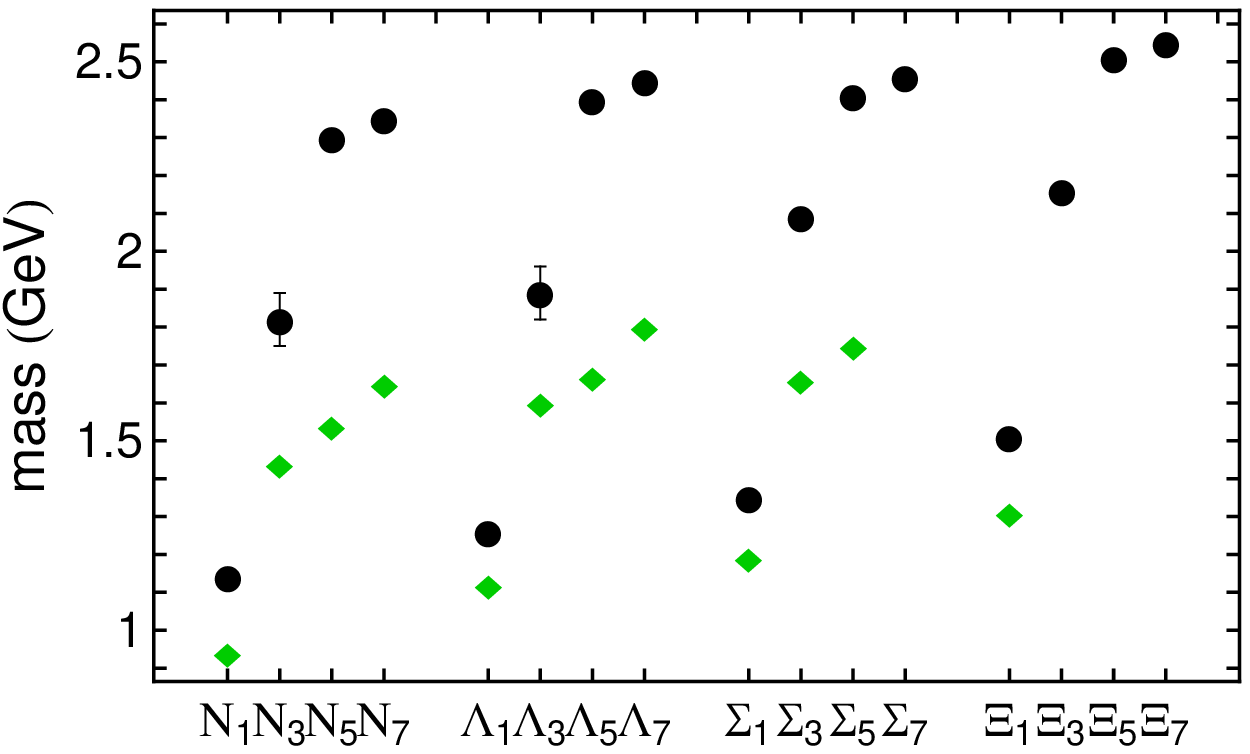}}\vspace*{-39ex}

\rightline{%
\includegraphics[clip,width=0.48\textwidth]{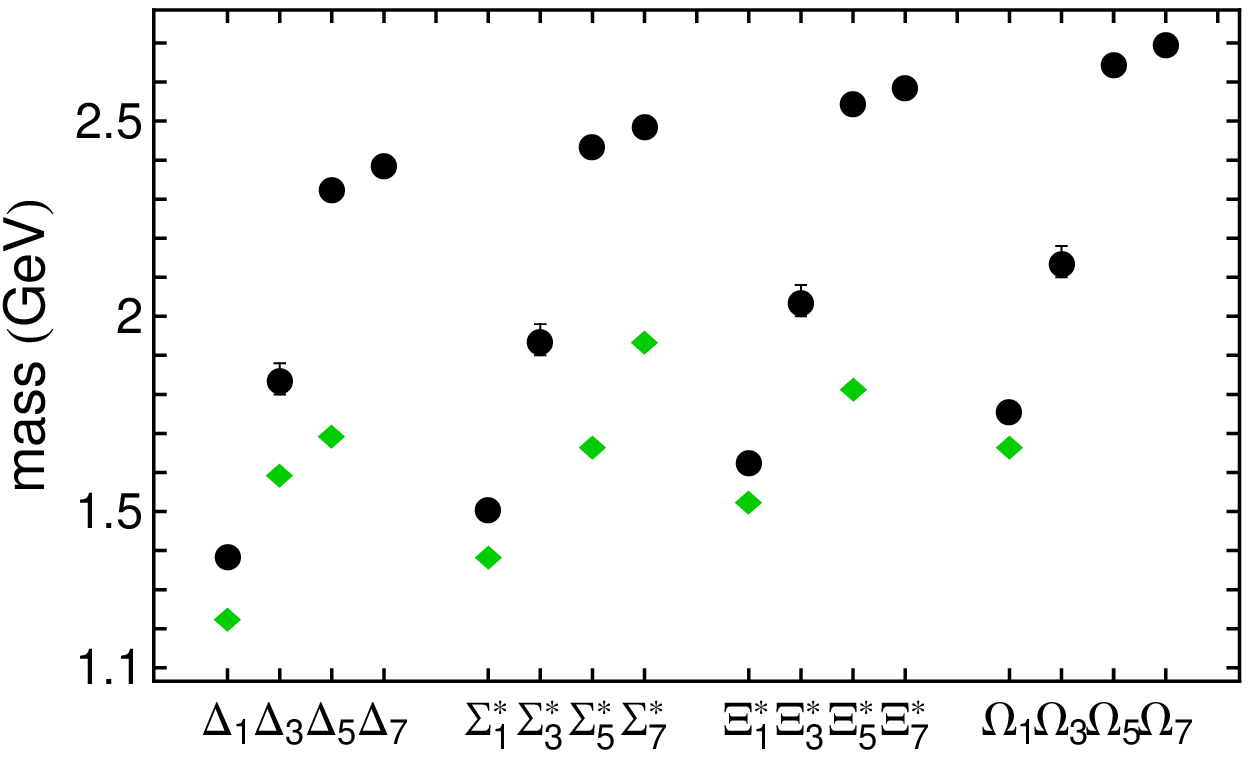}}

\caption{\label{fig:OctetMasses}
\underline{Left panel}: Pictorial representation of octet masses in Table~\protect\ref{OctetDecupletMasses}.  \emph{Circles} -- computed masses; and \emph{diamonds} -- empirical masses.  On the horizontal axis we list a particle name with a subscript that indicates its row in the table; e.g., $N_1$ means nucleon column, row 1.  In this way the labels step through ground-state, radial excitation, parity partner, parity partner's radial excitation.
\underline{Right panel}: Analogous plot for the decuplet masses in Table~\protect\ref{OctetDecupletMasses}.}
\end{figure}

We have solved all the Faddeev equations, supplemented by Eqs.\,\eqref{gNgDelta} and the material in Sect.\,\ref{sec:Faddeev}, and obtained the masses and eigenvectors of the ground-state octet and decuplet baryons.  The masses are listed in Table~\ref{OctetDecupletMasses}, with this information also depicted in Fig.\,\ref{fig:OctetMasses}.

It is evident in Fig.\,\ref{fig:OctetMasses} that, as with mesons, our computed baryon masses lie uniformly above the empirical values.  We view this as a success because our results are those for the baryons' dressed-quark-cores, whereas the empirical values include effects associated with type-2 meson-cloud effects, which typically produce sizable reductions \cite{Gasparyan:2003fp,Suzuki:2009nj}.  Our values may reasonably be viewed as bare-mass inputs appropriate for dynamical coupled-channels analyses of the hadron spectrum \cite{Aznauryan:2011ub}.  This was explained and illustrated for the nucleon and $\Delta$-resonance in Sect.\,4.5 of Ref.\,\cite{Roberts:2011cf} and in particular for the Roper resonance in Ref.\,\cite{Wilson:2011aa}.  Here we reiterate those instances in which a comparison can be made:
\begin{equation}
\begin{array}{l|llllll}
    & N_{940} P_{11} & N_{1440} P_{11} & N_{1535}S_{11} & N_{1650} S_{11}  & \Delta_{1232} P_{33} & \Delta_{1700} D_{33} \\\hline
{\rm Table}\;\ref{OctetDecupletMasses}\; ({\rm DSE}) & 1.14 & 1.82_{0.07} & 2.30 & 2.35_{0.01} & 1.39 & 2.33
\\
M_{B}^0 \; (\mbox{Ref.\,\protect\cite{Gasparyan:2003fp}}) & 1.24 &  & 2.05 & 1.92 & 1.46 & 2.25\\
M_{B}^0 \; (\mbox{Ref.\,\protect\cite{Suzuki:2009nj}})  & & 1.76 & 1.80 & 1.88 & 1.39 & 1.98
\end{array}
\end{equation}
where $M_B^0$, when it appears, is the relevant bare mass inferred in the associated coupled-channels analysis.  These bare masses have hitherto been uncertain and dependent on model details.  However, as we made no attempt to fit them, their proximity to our results suggests that it might now be possible to place these bare masses on firmer ground, investing them with meaning within the context of hadron structure calculations that have a traceable connection with QCD.

\begin{figure}[t]
\centerline{%
\includegraphics[clip,width=0.60\textwidth]{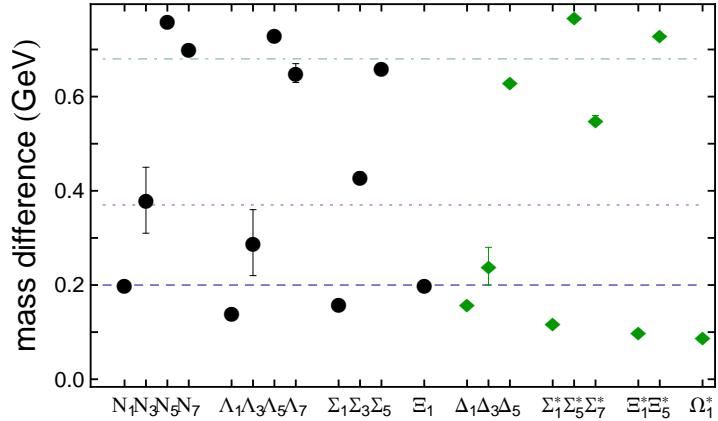}}

\caption{\label{fig:ODMassDifferences}
Theory-experiment mass differences computed, where possible, from Table~\protect\ref{OctetDecupletMasses}.
The dashed line is a difference of 0.2\,GeV; the dotted line, 0.37\,GeV; and the dot-dashed line, 0.68\,GeV.
Horizontal axis: particle name with a subscript that indicates its row in the table; e.g., $N_1$ means nucleon column, row 1.  In this way the labels step through ground-state, radial excitation, parity partner, parity partner's radial excitation.}
\end{figure}

In Fig.\,\ref{fig:ODMassDifferences} we plot the theory-experiment mass differences computed, where possible, from Table~\protect\ref{OctetDecupletMasses}.
This difference is uniformly less-than 0.2\,GeV for ground-states.  Moreover, for decuplet ground-states it decreases uniformly with increasing strangeness, a result that owes to the simplicity of the decuplet Faddeev amplitudes.
For octet radial excitations the difference is roughly $0.4\,$GeV; and for higher excitations it is around $0.7\,$GeV.  This identification of band clustering for the excited states is, perhaps, arguable but the scatter can be attributed, at least in part, to experimental uncertainty.
Notwithstanding this, these features suggest a regularity in the strength of type-2 meson-loop corrections within excitation bands, with the case being strongest for ground-states.

In this connection we remark that it is most appropriate to view Eq.\,\eqref{staticexchange} as implementing an approximation to the Faddeev equation kernel in a given channel, instead of merely as a simplification of the propagator for the exchanged quark.  From such a perspective one might argue that the Bethe-Salpeter and Faddeev amplitudes could influence the integrand's support within the channel under consideration and hence that $g_B$ should rightly depend on strangeness, contrary to the assumptions expressed in Eqs.\,\eqref{gNgDelta}.  Such a dependence may potentially alter our expectations regarding the size of meson-loop corrections to the mass of baryons with different strangeness content.
Notwithstanding this, we note that a strangeness-dependent $g_B$ is a second order effect within the rainbow-ladder framework because the propagator of the exchanged quark cannot itself exhibit a dependence on the environment in which it is subsequently embedded.  Any effect should therefore be small.

A full analysis of the impact of Eqs.\,\eqref{gNgDelta} can only be achieved by comparing our results with those obtained when Eq.\,\eqref{staticexchange} is not used.
Notably, however, the pattern of results in the decuplet, reported here and in Ref.\,\cite{Roberts:2011cf}, is consistent with the estimates of meson-loop corrections in Ref.\,\cite{Young:2002cj}; and hence a strangeness-dependent $g_{10}$ is not demanded.
In relation to the octet some might argue, following the decuplet pattern or Ref.\,\cite{Young:2009zb}, that the magnitude of the meson-loop correction to an octet baryon's mass should fall as the number of its $s$-quark constituents increases.  We do not find this.  However, we can obtain $m_\Xi^{\rm th} - m_\Xi^{\rm emp} = m_\Omega^{\rm th} - m_\Omega^{\rm emp}$ by choosing $g_\Xi = 1.15\,g_N$.  This small modification is well within bounds that may reasonably be applied to a first study and does not impact upon our discussion of band clustering, reducing masses of other $\Xi$-baryons in Table~\ref{OctetDecupletMasses} by less-than 1\%.
%: it reduces masses of by less-than 1\%.
%RE 2.16->2.14, PP 2.51->2.49 PPRE 2.55->2.54

It is also interesting to consider the baryons' diquark content, which is revealed by the Faddeev amplitude; i.e., the eigenvector produced by the Faddeev equation, and listed in Table~\ref{OctetEigenvectors}.  It is apparent that the relative strength of spin-zero cf.\  spin-one diquarks is fixed within a spectral level, almost independent of strangeness: the spin-zero-diquark content is 78\% in ground-states; 0\% in the positive-parity radial excitations; $52\pm 10$\% in the ground-state baryon parity-partners; and $46\pm 10$\% in  radial excitations of the latter.  This outcome is not surprising but, instead, reassuring because it indicates that baryon structure is largely flavour-blind within our framework, something one should also expect in QCD.

This discussion highlights a striking feature of Table~\ref{OctetEigenvectors}; namely, in possessing zero probability for $J=0$ diquark content, the Faddeev amplitudes corresponding to the first radial excitation of each ground-state are vastly different from the amplitudes for every other state.  This suggests a fascinating new possibility for the structure of these first radial excitations of baryon dressed-quark cores, which was first noted in connection with the Roper resonance \cite{Roberts:2011ym}.

To explain this remark, we focus first on the octet ground-states, whose Faddeev amplitudes describe states that are dominated by their scalar diquark components (78\%).  The axial-vector components are significantly smaller but nevertheless important.  For the nucleon, this heavy weighting of the scalar diquark component persists in solutions obtained with more sophisticated Faddeev equation kernels (see, e.g., Table~2 in Ref.\,\cite{Cloet:2008re}).  From a perspective provided by the octet states' parity partners and the first radial excitation of these states, in which the scalar and axial-vector diquark probabilities are 52\%-48\% and 46\%-54\%, respectively, the scalar diquark content of the octet ground-states actually appears to be unnaturally large.

\begin{table}[t]
\caption{\label{OctetEigenvectors}
Contact interaction Faddeev amplitudes for each of the octet baryons and their low-lying excitations.  The superscript in the expression $s^i$ or $a^i$ is a diquark enumeration label associated with Eq.\,\protect\eqref{flavourarrays}, except for $[2,3]$ and $[6,8]$, which are the $I=0$ combinations in Eq.\,\protect\eqref{goodV}.}
\begin{center}
\begin{tabular*}%{|c|c|c|c|c|c|c|}\hline
{\hsize}
{
l@{\extracolsep{0ptplus1fil}}
l@{\extracolsep{0ptplus1fil}}
|r@{\extracolsep{0ptplus1fil}}
r@{\extracolsep{0ptplus1fil}}
r@{\extracolsep{0ptplus1fil}}
r@{\extracolsep{0ptplus1fil}}
r@{\extracolsep{0ptplus1fil}}
r@{\extracolsep{0ptplus1fil}}
r@{\extracolsep{0ptplus1fil}}
r@{\extracolsep{0ptplus1fil}}
r@{\extracolsep{0ptplus1fil}}
r@{\extracolsep{0ptplus1fil}}
r@{\extracolsep{0ptplus1fil}}
r@{\extracolsep{0ptplus1fil}}
r@{\extracolsep{0ptplus1fil}}
|r@{\extracolsep{0ptplus1fil}}
}\hline
& & $s^1$ & $s^2$ & $s^{[2,3]}$ & $a^4_1$ & $a^4_2$ & $a^5_1$ & $a^5_2$ &
   $a^6_1$ & $a^6_2$ & $a^{[6,8]}_1$ & $a^{[6,8]}_2$ & $a^9_1$ & $a^9_2$ & $P_{J=0} $ \\\hline
%
%{-0.88076, 0.381399, -0.26969, 0.0635812, -0.0449587
$(P=+,n=0)$ \rule{0ex}{2.5ex}
& $N$       & 0.88 & & & -0.38 & 0.27 & -0.06 & 0.04 & & & & & & & 78\%\\
\rule{0ex}{2.5ex} %0.665199, 0.592559, -0.446234, -0.0851997
& $\Lambda$ & 0.67 & & -0.27 & & & & & & & -0.45 & -0.09 & & & 79\%\\
\rule{0ex}{2.5ex} % 0.846007, -0.448581, 0.259438, 0.124556, 0.0150227
& $\Sigma$  & & 0.85 & & -0.45 & 0.26 & & & 0.12 & 0.02 & & & & & 72\%\\
\rule{0ex}{2.5ex} % 0.90756, 0.139085, 0.0771653, 0.388633, 0.000779965
& $\Xi$     & & 0.91 & & 0.14 & 0.08 & & & & & & & 0.39 & 0.00 & 82\%\\\hline
%
%{-0.0224215, 0.523515, -0.370181, -0.626309, 0.442868
$(P=+,n=1)$ \rule{0ex}{2.5ex}
& $N$       & -0.02 & & & 0.52 & -0.37 & -0.63 & 0.44 & & & & & & & 0\%\\
\rule{0ex}{2.5ex} %0.0289835, 0.0576172, -0.777198, 0.625942
& $\Lambda$ & 0.03 & & 0.06 & & & & & & & -0.78 & 0.63 & & & 0\%\\
\rule{0ex}{2.5ex} % 0.00110028, -0.0441436, 0.0234577, 0.832595, -0.55162
& $\Sigma$  & & 0.00 & & -0.04 & 0.02 & & & 0.83 & -0.55 & & & & & 0\%\\
\rule{0ex}{2.5ex} % {-0.0018074, 0.0128419, -0.997976, -0.0249166, 0.0570602
& $\Xi$     & & 0.00 & & 0.01 & -1.00 & & & & & & & -0.02 & 0.06 & 0\%\\\hline
%
%-0.716545, 0.393592, -0.278311, -0.411657, 0.291085 - old
%0.70786, -0.408572, 0.288904, 0.407052, -0.287829
$(P=-,n=0)$ \rule{0ex}{2.5ex}
& $N$       & 0.71 & & & -0.41 & 0.29 & 0.41 & -0.29 & & & & & & & 50\%\\
\rule{0ex}{2.5ex}
%0.591393, 0.47779, -0.468137, 0.450353 - old
%{0.642881, 0.441948, -0.466504, 0.416844}
& $\Lambda$ & 0.64 & & 0.44 & & & & & & & -0.47 & 0.42 & & & 61\%\\
\rule{0ex}{2.5ex}
% {-0.664162, 0.423174, -0.234943, -0.517578, 0.238176 - old
%0.612954, -0.473122, 0.229851, 0.550492, -0.211116
& $\Sigma$  & & 0.61 & & -0.47 & 0.23 & & & 0.55 & -0.21 & & & & & 38\%\\
\rule{0ex}{2.5ex}
% 0.744109, -0.0362036, 0.457358, 0.357132, -0.329047 - old
% 0.764606, -0.335287, 0.345178, 0.32869, -0.275273
& $\Xi$     & & 0.76 & & -0.34 & 0.35 & & & & & & & 0.33 & -0.28 & 58\%\\\hline
%
%0.654189, -0.409872, 0.289824, 0.461912, -0.326621 - old
%0.664177, -0.413713, 0.292539, 0.448799, -0.317349
$(P=-,n=1)$ \rule{0ex}{2.5ex}
& $N$       & 0.66 & & & -0.41 & 0.29 & 0.45 & -0.32 & & & & & & & 44\%\\
\rule{0ex}{2.5ex}
%.538365, 0.463659, -0.490803, 0.504278 - old
%0.602296, 0.43312, -0.479324, 0.468929
& $\Lambda$ & 0.60 & & 0.43 & & & & & & & -0.48 & 0.47 & & & 55\%\\
\rule{0ex}{2.5ex}
% {-0.60596, 0.438094, -0.234767, -0.564808, 0.258384 - old
% {-0.574633, 0.467243, -0.226295, -0.584772, 0.241482
& $\Sigma$  & & 0.57 & & -0.47 & 0.23 & & & 0.58 & -0.24 & & & & & 33\%\\
\rule{0ex}{2.5ex}
% 0.702784, -0.030075, 0.493372, 0.367585, -0.355887 - old
% 0.731716, -0.339988, 0.375419, 0.334042, -0.310606
& $\Xi$     & & 0.73 & & -0.34 & 0.37 & & & & & & & 0.33 & -0.31 & 54\%\\\hline
\end{tabular*}
\end{center}
\end{table}

One can nevertheless understand the structure of the octet ground-states.  As with so much else in hadron physics, the composition of these flavour octet states is intimately connected with DCSB.  In Sect.\,\ref{subsub:diquarks} we observed that in a two-color version of QCD, scalar diquarks are Goldstone modes \cite{Roberts:1996jx,Bloch:1999vk} and that memories of this persist in the three-color theory.  For example, in the low masses of scalar diquark correlations (Table~\ref{Diquarkmasses}); and in the large values of their canonically normalised Bethe-Salpeter amplitudes (Table~\ref{DiquarkBSAs}) and hence strong quark$+$quark$-$diquark coupling within the octet ground-states.  (A qualitatively identical effect explains the large value of the $\pi N$ coupling constant and its analogues involving other pseudoscalar-mesons and octet-baryons.)  There is no commensurate enhancement mechanism associated with the axial-vector diquark correlations.  Therefore the scalar diquark correlations dominate within octet ground-states.

With the Faddeev equation treatment described herein, the effect on the first radial excitations is dramatic: orthogonality of the ground- and excited-states forces the radial excitations to be constituted almost entirely from axial-vector diquark correlations.  It is critical to check whether this outcome survives with Faddeev equation kernels built from a momentum-dependent interaction.

This brings us to a final, very significant observation; namely, the match between our computed level ordering and that of experiment, something which has historically been difficult for models to obtain (see, e.g., the discussion in Ref.\,\cite{Capstick:2000qj}) and is not achieved in contemporary numerical simulations of lattice-regularised QCD (see, e.g., Ref.\,\cite{Edwards:2011jj}).  In particular, our calculations produce a parity-partner for each ground-state that is always more massive than its first radial excitation so that, in the nucleon channel, e.g., the first $J^P=\frac{1}{2}^-$ state lies above the second $J^P=\frac{1}{2}^+$ state.
%Important to this success is incorporation of DCSB effects into the Bethe-Salpeter kernels associated with the scalar and axial-vector channels, as discussed in connection with Eq.\eqref{gSO}.
%--Not sure about the claim above. I set gSO=1 and mass of parity partners dropped by only 1GeV. ... NO: this was radial excitation of nucleon parity partner.  Changing code to work for ground-state, the new mass is 2.11GeV; i.e., 110MeV less than Hannes' result.  Still above radial excitation for the nucleon.  Perhaps DCSB is important ... parity parter diquarks don't have the sort of strong q+q->qq coupling that the scalar diquark has.

A veracious expression of DCSB in the meson spectrum is critical to this success.  One might ask why and how?  It is DCSB that both ensures the dressed-quark-cores of pseudoscalar and vector mesons are far lighter than those of their parity partners and produces strong quark$+$antiquark$-$meson couplings, which are expressed in large values for the canonically normalised Bethe-Salpeter amplitudes (Table~\ref{MesonBSAs}).  The remnants of Pauli-G\"ursey symmetry described previously entail that these features are carried into the diquark sector: as evident in Fig.\,\ref{fig:DiquarkMasses} and Table~\ref{DiquarkBSAs} and their comparison with Fig.\,\ref{fig:MesonMasses} and Table~\ref{MesonBSAs}.  The inflated masses but, more importantly, the suppressed values of the Bethe-Salpeter amplitudes for negative-parity diquarks, in comparison with those of positive-parity diquarks, guarantee the level ordering we have computed: attraction in a given channel diminishes with the square of the Bethe-Salpeter amplitude (see App.\,\ref{App:AllFEs}).  Hence, an approach within which DCSB cannot be realised or a simulation whose parameters are such that the importance of DCSB is suppressed will both necessarily have difficulty reproducing the experimental ordering of levels.\footnote{A phenomenological meson mass formula is used in Ref.\,\protect\cite{deTeramond:2012rt} as the basis for the analytic inference of baryon mass formulae.  This transmits patterns in the meson spectrum into the baryon sector.  The baryon mass formulae inferred thereby are consistent with the empirical level ordering.  This pathway bears a qualitative resemblance to our finding.
}

\section{Summary and Perspective}
\label{sec:epilogue}
We described the first DSE-based calculation of the spectrum of strange and nonstrange hadrons that simultaneously correlates the dressed-quark-core masses of meson and baryon ground- and excited-states within a single symmetry-preserving framework.
The nonstrange sector was described previously \cite{Roberts:2011cf}.  Therein, the physical pion mass is obtained with $m_u=m_d=m=7\,$MeV; and five parameters are used to define the gap-, Bethe-Salpeter- and Faddeev-equations.  In a comparison with relevant quantities, the study records a value of 13\% for the overall root-mean-square-relative-error$/$degree-of freedom; %($\overline{\mbox{rms}}$);
and notable amongst the results is agreement between the computed masses of nonstrange baryon dressed-quark-cores and the bare masses employed in modern dynamical coupled-channels models of meson-baryon reactions.
We reproduce all results in Ref.\,\cite{Roberts:2011cf}; and in extending that work to hadrons with strangeness, a $s$-quark current-mass of $0.17\,$GeV produces the physical kaon mass, and we neither introduce new model parameters nor change the values of those determined previously.

In explaining mesons we capitalised on recent progress in understanding the far-reaching effects of dynamical chiral symmetry breaking (DCSB) within the Bethe-Salpeter kernel to improve upon the rainbow-ladder truncation in the scalar and axial-vector channels.  This enabled us to produce a spectrum of meson dressed-quark-core masses in which empirical values are typically overestimated by $\sim 10$\%.  This is a marked success since our DCSB-corrected kernel omits resonant contributions; i.e., effects that may phenomenologically be associated with a meson cloud.  Indeed, since meson cloud contributions typically induce a material reduction in hadron dressed-quark-core masses,
%no approach that omits such effects, whether deliberately or inadvertently, can be viewed as providing reliable insights or understanding unless it overestimates masses by a similar magnitude.
any approach that omits such effects, whether deliberately or inadvertently, should be viewed with caution unless it overestimates masses by a similar magnitude.  Failing this, extreme care must be exercised before drawing insights or claiming understanding therefrom.

Our description of meson dressed-quark-cores enabled a prediction of the spectrum of baryons, including those with strangeness, that is similarly successful.  Deconstructing the masses, we arrived at numerous insights into baryon structure.
For example, we found that the diquark content of baryons is largely independent of strangeness; namely, that baryon structure is flavour-blind within our framework, something one should also expect in QCD.
Another noteworthy result concerns the first radial excitation of each ground-state; viz., they possess negligible probability for $J=0$ diquark content.  Thus, in striking contrast to the Faddeev amplitudes of every other state, the radial excitations are constituted almost entirely from axial-vector diquark correlations.  This possibility was first noted in connection with the Roper resonance \cite{Roberts:2011ym}.
Lastly and significantly, our computed level ordering matches that of experiment.  In particular, the parity-partner for each ground-state is always more massive than its first radial excitation; i.e., the first $J^P=\frac{1}{2}^-$ state always lies above the second $J^P=\frac{1}{2}^+$ state.  We showed that a veracious expression of DCSB in the meson spectrum is critical to this outcome, so that an approach within which DCSB cannot be realised or a simulation whose parameters are such that the importance of DCSB is suppressed will both necessarily have difficulty reproducing the experimental ordering of levels.
Furthermore, our results suggest that realistic simulation parameters, including near-empirical masses for dynamical light-quarks, must be achieved in order for lattice-regularised QCD to confirm the role of diquark correlations in baryon structure.

The analysis and results described herein further strengthen the claim that a symmetry-preserving treatment of a vector$\times$vector contact interaction is a useful tool for the study of phenomena characterised by probe momenta less-than the dressed-quark mass.  As remarked above, whilst this interaction produces form factors which are too hard, even they, when interpreted carefully, can be used to draw valuable insights, especially concerning relationships between different hadrons.  This foundation is therefore being used in the computation of elastic and transition form factors involving baryons with strangeness.

Studies employing a symmetry-preserving regularisation of the contact interaction serve as a useful surrogate, exploring domains which analyses using interactions that more closely resemble those of QCD are as yet unable to enter.  They are critical at present in attempts to use experimental data as a tool for charting the nature of the quark-quark interaction at long-range; i.e., for identifying distinct signals of the running of the couplings and masses in QCD.

\begin{acknowledgements}
We thank A.~Bashir, I.\,C.~Clo\"et, M.\,U.~D\"oring, T.-S.\,H.~Lee, V.\,I.~Mokeev, S.-x.~Qin, H.\,L.\,L.~Roberts and S.\,M.~Schmidt for helpful discussions.
Chen Chen acknowledges the support of the China Scholarship Council (file no.\ 2010634019).
This work was also supported by:
Forschungszentrum J\"ulich GmbH;
and U.\,S.\ Department of Energy, Office of Nuclear Physics, contract no.~DE-AC02-06CH11357.
\end{acknowledgements}

\appendix
\setcounter{equation}{0}
\renewcommand{\theequation}{\Alph{section}.\arabic{equation}}
\section{Euclidean Conventions}
\label{App:EM}
In our Euclidean formulation:
\begin{equation}
p\cdot q=\sum_{i=1}^4 p_i q_i\,;
\end{equation}
% \begin{equation}
\begin{equation}
\{\gamma_\mu,\gamma_\nu\}=2\,\delta_{\mu\nu}\,;\;
\gamma_\mu^\dagger = \gamma_\mu\,;\;
\sigma_{\mu\nu}= \frac{i}{2}[\gamma_\mu,\gamma_\nu]\,; \;
% \end{equation}
% and
% \begin{equation}
{\rm tr}\,[\gamma_5\gamma_\mu\gamma_\nu\gamma_\rho\gamma_\sigma]=
-4\,\epsilon_{\mu\nu\rho\sigma}\,, \epsilon_{1234}= 1\,.
% \end{equation}
\end{equation}

A positive energy spinor satisfies
\begin{equation}
\bar u(P,s)\, (i \gamma\cdot P + M) = 0 = (i\gamma\cdot P + M)\, u(P,s)\,,
\end{equation}
where $s=\pm$ is the spin label.  It is normalised:
\begin{equation}
\bar u(P,s) \, u(P,s) = 2 M \,,
\end{equation}
and may be expressed explicitly:
\begin{equation}
u(P,s) = \sqrt{M- i {\cal E}}
\left(
\begin{array}{l}
\chi_s\\
\displaystyle \frac{\vec{\sigma}\cdot \vec{P}}{M - i {\cal E}} \chi_s
\end{array}
\right)\,,
\end{equation}
with ${\cal E} = i \sqrt{\vec{P}^2 + M^2}$,
\begin{equation}
\chi_+ = \left( \begin{array}{c} 1 \\ 0  \end{array}\right)\,,\;
\chi_- = \left( \begin{array}{c} 0\\ 1  \end{array}\right)\,.
\end{equation}
For the free-particle spinor, $\bar u(P,s)= u(P,s)^\dagger \gamma_4$.

The spinor can be used to construct a positive energy projection operator:
\begin{equation}
\label{Lplus} \Lambda_+(P):= \frac{1}{2 M}\,\sum_{s=\pm} \, u(P,s) \, \bar
u(P,s) = \frac{1}{2M} \left( -i \gamma\cdot P + M\right).
\end{equation}

A negative energy spinor satisfies
\begin{equation}
\bar v(P,s)\,(i\gamma\cdot P - M) = 0 = (i\gamma\cdot P - M) \, v(P,s)\,,
\end{equation}
and possesses properties and satisfies constraints obtained via obvious analogy
with $u(P,s)$.

A charge-conjugated Bethe-Salpeter amplitude is obtained via
\begin{equation}
\label{chargec}
\bar\Gamma(k;P) = C^\dagger \, \Gamma(-k;P)^{\rm T}\,C\,,
\end{equation}
where ``T'' denotes a transposing of all matrix indices and
$C=\gamma_2\gamma_4$ is the charge conjugation matrix, $C^\dagger=-C$.  We note that
\begin{equation}
C^\dagger \gamma_\mu^{\rm T} \, C = - \gamma_\mu\,, \; [C,\gamma_5] = 0\,.
\end{equation}

In describing decuplet resonances we employ a Rarita-Schwinger spinor to represent a covariant spin-$3/2$ field.  The positive energy
spinor is defined by the following equations:
\begin{equation}
\label{rarita}
(i \gamma\cdot P + M)\, u_\mu(P;r) = 0\,,\;
\gamma_\mu u_\mu(P;r) = 0\,,\;
P_\mu u_\mu(P;r) = 0\,,
\end{equation}
where $r=-3/2,-1/2,1/2,3/2$.  It is normalised:
\begin{equation}
\bar u_{\mu}(P;r^\prime) \, u_\mu(P;r) = 2 M\,,
\end{equation}
and satisfies a completeness relation
\begin{equation}
\label{Deltacomplete}
\frac{1}{2 M}\sum_{r=-3/2}^{3/2} u_\mu(P;r)\,\bar u_\nu(P;r) =
\Lambda_+(P)\,R_{\mu\nu}\,,
\end{equation}
where
\begin{equation}
R_{\mu\nu} = \delta_{\mu\nu} \mbox{\boldmath $I$}_{\rm D} -\frac{1}{3} \gamma_\mu \gamma_\nu +
\frac{2}{3} \hat P_\mu \hat P_\nu \mbox{\boldmath $I$}_{\rm D} - i\frac{1}{3} [ \hat P_\mu
\gamma_\nu - \hat P_\nu \gamma_\mu]\,,
\end{equation}
with $\hat P^2 = -1$, which is very useful in simplifying the Faddeev equation for a positive energy decuplet state.

\setcounter{equation}{0}
\section{Bethe-Salpeter equations}
\label{app:groundBSEs}
We discussed the BSEs for ground-state charged-pseudoscalar and vector mesons in Sect.\,\ref{sec:qqBSAMeson}, and for ground-state scalar and axial-vector diquark correlations in Sect.\,\ref{sec:qqBSADiquark}.  Here we report formulae for other channels and also radial excitations in each channel.  In the latter context, some remarks are in order.

\subsection{Radial excitations of pseudoscalar mesons and scalar diquarks}
\label{sec:radial}
In quantum mechanics the radial wave function for a bound-state's first radial excitation possesses a single zero.  A similar feature is expressed in quantum field theory: namely, in a fully covariant approach a single zero is usually seen in the relative-momentum dependence of the leading Tchebychev moment of the dominant Dirac structure in the bound state amplitude for a meson's first radial excitation \cite{Qin:2011xq,Holl:2004fr,Holl:2005vu}.  The existence of radial excitations is therefore clear evidence against the possibility that the interaction between quarks is momentum-independent: a bound-state amplitude that is independent of the relative momentum cannot exhibit a single zero.  One may also express this differently; namely, if the location of the zero is at $k_0^2$, then a momentum-independent interaction can only produce reliable results for phenomena that probe momentum scales $k^2\ll k_0^2$.  In QCD, $k_0^2 \sim 2 M^2 \sim (0.5\,{\rm GeV})^2$ \cite{Qin:2011xq,Holl:2004fr,Holl:2005vu}.

In the phenomenological application of a contact interaction, however, this difficulty has been skirted by means of an expedient employed in Refs.\,\cite{Volkov:1996br,Volkov:1999xf}; i.e., one inserts a zero by hand into the kernels expressed in Eqs.\,\eqref{pionKernel}.  This means that one identifies the BSE for a radial excitation as the form of Eq.\,(\ref{LBSEI}) obtained with Eq.\,(\ref{njlgluon}) and insertion into the integrand of a factor
\begin{equation}
\label{locatezero}
1 - d_{\cal F} q^2 ,
\end{equation}
which forces a zero into the kernel at $q^2=1/d_{\cal F}$, where $d_{\cal F}$ is a parameter.  Plainly, the presence of this zero reduces the coupling in the BSE and hence increases the bound-state's mass.  Although this may not be as transparent with a more sophisticated interaction, a qualitatively equivalent mechanism is responsible for the elevated values of the masses of radial excitations.

The expedient may readily be illustrated in connection with pseudoscalar mesons.  First define the function
\begin{eqnarray}
{\cal F}^{\rm iu}(\omega(M_u^2,M_s^2,\alpha,P^2))
&=& {\cal C}^{\rm iu}(\omega(M_u^2,M_s^2,\alpha,P^2))
- d_{\cal F} {\cal D}^{\rm iu}(\omega(M_u^2,M_s^2,\alpha,P^2))\,,\\
{\cal D}^{\rm iu}(\omega(M_u^2,M_s^2,\alpha,P^2)) & = & \int_0^\infty ds\,s^2\,\frac{1}{s+\omega^2}
\to  \int_{r_{\rm uv}^2}^{r_{\rm ir}^2} d\tau\, \frac{2}{\tau^3} \,
\exp\left[-\tau \omega(M_u^2,M_s^2,\alpha,P^2)\right],
\end{eqnarray}
with, as usual, ${\cal F}^{\rm iu}_1(z) = - z (d/dz){\cal F}^{\rm iu}(z)$ and $\overline{\cal F}_1(z) = {\cal F}_1(z)/z$.  Then the BSE for the kaon's first radial excitation, denoted $K^1$, is obtained with the kernel
\begin{subequations}
\label{kaonKernelradial}
\begin{eqnarray}
\label{kaon1total}
{\cal K}^{K^1} &=&
\left[\begin{array}{cc}
{\cal K}_{EE}^{K^1} & {\cal K}_{EF}^{K^1}\\
\tilde{\cal K}_{FE}^{K^1} &
\tilde{\cal K}_{FF}^{K^1}
\end{array}\right],\\
\nonumber
{\cal K}_{EE}^{K^1} &=&
\int_0^1d\alpha \bigg\{
{\cal F}^{\rm iu}(\omega(M_u^2, M_s^2, \alpha, P^2))  \\
&&+ \bigg[ M_u M_s-\alpha (1-\alpha) P^2 - \omega(M_u^2, M_s^2, \alpha, P^2)\bigg]
\, \overline{\cal F}^{\rm iu}_1(\omega(M_u^2, M_s^2, \alpha, P^2))\bigg\},\\
{\cal K}_{EF}^{K^1} &=& \frac{P^2}{2 M_R} \int_0^1d\alpha\, \bigg[(1-\alpha)M_u+\alpha M_s\bigg]\overline{\cal F}^{\rm iu}_1(\omega(M_u^2, M_s^2, \alpha, P^2)),\\
{\cal K}_{FE}^{K^1} &=& \frac{2 M_R^2}{P^2} {\cal K}_{EF}^{K^1} = M_R \int_0^1d\alpha\, \bigg[(1-\alpha)M_u+\alpha M_s\bigg]\overline{\cal F}^{\rm iu}_1(\omega(M_u^2, M_s^2, \alpha, P^2)),\\
%\frac{2 M_R^2}{P^2} {\cal K}_{EF}^K ,\\
%
{\cal K}_{FF}^{K^1} &=& - \frac{1}{2} \int_0^1d\alpha\, \bigg[ M_u M_s+(1-\alpha) M_u^2+\alpha M_s^2\bigg] \overline{\cal F}^{\rm iu}_1(\omega(M_u^2, M_s^2, \alpha, P^2))\,.
\end{eqnarray}
\end{subequations}
In Eq.\,\eqref{kaon1total},
\begin{equation}
\tilde{\cal K}_{FE}^{K^1} = {\cal K}_{FE}^{K^1} - \left.{\cal K}_{FE}^{K^1} \right|_{m_{s,u}\to 0}\,, \;
\tilde{\cal K}_{FF}^{K^1} = {\cal K}_{FF}^{K^1} - \left.{\cal K}_{FF}^{K^1} \right|_{m_{s,u}\to 0}\,.
\end{equation}
These subtractions are implemented so as to ensure that the leptonic decay constant of the radially-excited pseudoscalar meson vanishes in the chiral limit, which is a consequence of the axial-vector Ward-Takahashi identity \cite{Holl:2004fr,Holl:2005vu}.  The kernel for the radially-excited scalar diquark is obtained through obvious analogy with Eq.\,(\ref{bse[su]E}).

\subsection{Vector mesons and axial-vector diquarks}
We emphasise that the vector-meson and axial-vector-diquark Bethe-Salpeter amplitudes only assume the simple form in Eq.\,\eqref{KastBSA} when rainbow-ladder truncation is employed. Even with a momentum-independent interaction, these amplitudes possess two Dirac covariants immediately upon inclusion of next-to-leading-order corrections to the quark-gluon vertex; viz.,
\begin{equation}
\Gamma_\mu^{1^-}(P) =  \gamma_\mu^\perp E_{1^-}(P) \;
\longrightarrow \;  \gamma_\mu^\perp E_{1^-}(P) + i\frac{1}{M} \sigma_{\mu\nu} P_\nu F_{1^-}(P)\,,\; \gamma^\perp_\mu P_\mu = 0.
\end{equation}
Similar observations hold for a $g^2 D(p-q) \sim \delta^4(p-q)$ interaction \cite{Bender:1996bb,Bender:2002as,Bhagwat:2004hn}.

In studying radial excitations, there are no complications in these channels: one simply implements the replacement ${\cal C} \to {\cal F}$ in Eq.\,\eqref{KastKernel} and then works with the BSEs and canonical normalisation conditions obtained therefrom.

\subsection{Axial-vector mesons and vector diquarks}
Again owing to the simplicity of the interaction and truncation, the Bethe-Salpeter amplitude for an axial-vector meson composed from a $f$- and $\bar g$-quark is
\begin{equation}
\label{avBSA}
\Gamma_{\mu {1^+}}^{f\bar g}(P) =  \gamma_5\gamma_\mu^\perp E^{1^+}_{f\bar g}(P)\,.
\end{equation}
Inserting Eq.\,(\ref{avBSA}) into Eq.\,(\ref{LBSEI}) yields the following BSE:
\begin{eqnarray}
\label{avKernel}
0 & = & 1 + {\cal K}_{f\bar g}^{1^+}(-(m_{f\bar g}^{1^+})^2)\,,\\
{\cal K}_{f\bar g}^{1^+}(z) &=&
\frac{2\alpha_{\rm IR}}{3\pi m_G^2} \int_0^1d\alpha\,\bigg[
{\cal C}_1^{\rm iu}(\omega(M_g^2,M_f^2,\alpha,z))
+ (M_f M_g+\alpha (1-\alpha) z )
\overline{\cal C}_1^{\rm iu}(\omega(M_g^2,M_f^2,\alpha,z))\bigg] .
\label{avKernelE}
\end{eqnarray}

Given the discussion in Sect.\,\ref{sec:qqBSADiquark}, the Bethe-Salpeter amplitude for a $J^P=1^-$ correlation of quarks with flavour $f$, $g$ is given by
\begin{equation}
\label{vdqBSA}
\Gamma_{\mu {1^-}}^{fg\,C}(P) =  \gamma_5\gamma_\mu^\perp E_{\{fg\}_{1^-}}(P)\,.
\end{equation}
In a three flavour theory the allowed correlations are $\{uu\}$, $\{ud\}$, $\{dd\}$, $\{su\}$,$\{sd\}$, $\{ss\}$ and their masses are determined from
\begin{equation}
1 + \frac{1}{2} {\cal K}_{f\bar g}^{1^+}(-m^2_{\{fg\}_{1^-}})\,.
\end{equation}
%The BSE for the axial-vector diquark again follows immediately; viz.,
%\begin{equation}
%1+ \frac{1}{2} K^\rho(-m_{qq_{1^+}}^2) = 0\,.
%\end{equation}

The canonical normalisation conditions are
\begin{equation}
\frac{1}{(E^{1^+}_{f\bar g})^2} = - \left. 9 \mathpzc{m}_G^2 \frac{d}{dz}
{\cal K}_{f\bar g}^{1^+}(z)\right|_{z=-(m_{f\bar g}^{1^+})^2},\;
\frac{1}{E^2_{\{fg\}_{1^-}}} = - \left. 6 \mathpzc{m}_G^2 \frac{d}{dz} {\cal K}_{f\bar g}^{1^+}(z)\right|_{z=-m^2_{\{fg\}_{1^-}}}\,.
\end{equation}

Once again, in studying radial excitations there are no complications in these channels: one simply implements the replacement ${\cal C} \to {\cal F}$ in Eq.\,\eqref{avKernelE} and then works with the BSEs and canonical normalisation conditions obtained therefrom.

\subsection{Scalar Mesons and Pseudoscalar Diquarks}
\label{app:scalar}
The Bethe-Salpeter amplitude for a $f\bar g$ scalar meson is
\begin{equation}
\label{sBSA}
\Gamma^{f\bar g}_{0^+}(P) = \mathbf{I}_{\rm D} \, E^{0^+}_{f\bar g}(P)\,.
\end{equation}
As with axial-vector mesons, corrections beyond rainbow-ladder truncation do not generate new covariants.  Inserting Eq.\,\eqref{sBSA} into Eq.\,\eqref{LBSEI} produces the BSE
\begin{eqnarray}
\label{BSEscalarmeson}
0 &=& 1 + {\cal K}^{0^+}_{f\bar g}(-(m^{0^+}_{f\bar g})^2)\,,\\
\nonumber
{\cal K}^{0^+}_{f\bar g}(z) & = &
- \frac{4\alpha_{\rm IR}}{3\pi m_G^2}
\int_0^1d\alpha\,\bigg[
\bigg({\cal C}^{\rm iu}(\omega(M_g^2,M_f^2,\alpha,z))
-{\cal C}_1^{\rm iu}(\omega(M_g^2,M_f^2,\alpha,z) )\bigg)\\
&& \rule{7em}{0ex} -(M_f M_g + \alpha (1-\alpha) z )
\overline{\cal C}_1^{\rm iu}(\omega(M_g^2,M_f^2,\alpha,z))\bigg]\,.
\label{scalarKernelE}
\end{eqnarray}

The canonical normalisation condition is:
\begin{equation}
\frac{1}{(E^{0^+}_{f\bar g})^2} =  \left. -  \frac{9}{2} \mathpzc{m}_G^2 \, \frac{d}{dz}{\cal K}^{0^+}_{f\bar g}(z)\right|_{z=-(m^{0^+}_{f\bar g})^2}.
\end{equation}

It is worth reiterating at this point that when applied in the chiral limit to the isoscalar-scalar channel, Eq.\,\eqref{BSEscalarmeson} yields \cite{Roberts:2010gh}
\begin{equation}
m^2_{(0)0^{++}} = 2 M_0\,.
\end{equation}
Whilst this algebraic result does not persist beyond rainbow-ladder truncation, it is a useful test of symmetry conservation within an application of the contact interaction and also provides for a reasonable definition of a single dressed-quark mass-scale when a more sophisticated interaction is employed.

Given the discussion in Sect.\,\ref{sec:qqBSADiquark}, the Bethe-Salpeter amplitude for a pseudoscalar diquark is readily inferred from Eq.\,\eqref{sBSA}:
\begin{equation}
\label{sBSAqq}
\Gamma_{0^-}^{C\,[fg]}(P) = \mathbf{I}_{\rm D} \, E_{[fg]_{0^-}}(P)\,.
\end{equation}
In a three-flavour theory, allowed combinations are $[ud]$, $[su]$, $[sd]$.  The associated BSE takes the simple form
\begin{equation}
0 = 1 + {\cal K}^{0^+}_{f\bar g}(-m^2_{[fg]_{0^-}})
\end{equation}
and the canonical normalisation condition is:
\begin{equation}
\frac{1}{E_{[fg]_{0^-}}^2} = \left. -3 \mathpzc{m}_G^2 \, \frac{d}{dz}{\cal K}^{0^+}_{f\bar g}(z)\right|_{z=-m^2_{[fg]_{0^-}}}.
\end{equation}

\section{Ground-state Faddeev equations}
\label{App:AllFEs}
\setcounter{equation}{0}
\subsection{$\Delta$-resonance}
\label{app:FEDelta}
As remarked when opening Sect.\,\ref{sec:FELambda}, owing to its inherent simplicity the $\Delta$ is an ideal system with which to illustrate steps in the derivation of Faddeev equations that involve Dirac algebra and analysis of momentum integrals.  One begins by observing that with a momentum-independent kernel, the Faddeev amplitude cannot depend on relative momentum and  hence for the $\Delta$-resonance Eq.\,(\ref{DecupletFA}) becomes
\begin{equation}
{\cal D}_{\nu\rho}(\ell;P) u_\rho^\Delta(P) = f^\Delta(P) \, \mathbf{I}_{\rm D} \, u_\nu^\Delta(P)\,.
\label{DnurhoI}
\end{equation}
N.B.\ Regarding Eq.\,(\ref{DecupletFA}) in general, one might naively suppose that isospin-one tensor diquarks could play a material role in the Faddeev amplitude for a ground state $\Delta$.  However, this notion can quickly be discarded because ground-states are distinguished by containing the smallest amount of quark orbital angular momentum, $L$, and a tensor diquark is characterised by $L\geq 1$.

Using Eqs.\,(\ref{DnurhoI}), \eqref{staticexchange}, Fig.\,\ref{fig:FaddeevI} is realised as
\begin{equation}
f^\Delta(P) \, u_\mu^\Delta(P)  = 4 \frac{g_{10}^2}{M_u} \int\frac{d^4\ell}{(2\pi)^4}\,{\cal M}^\Delta_{\mu\nu}(\ell;P)\,f^\Delta(P) \, u_\nu^\Delta(P)\,,
\end{equation}
with ($K=-\ell+P$, $P^2=-m_\Delta^2$)
\begin{equation}
{\cal M}^\Delta_{\mu\nu}(\ell;P) = 2\,i\Gamma_{\rho\{uu\}}^{1^+}(K) i\bar \Gamma_{\mu\{uu\}}^{1^+}(-P) S(\ell) \Delta^{\{uu\}}_{1^+\rho\nu}(K)\,,
\end{equation}
where the ``2'' has arisen through isospin contractions.

At this point, one post-multiplies by $\bar u_\beta^\Delta(P;r)$ and sums over the polarisation index to obtain, using Eq.\,(\ref{Deltacomplete}),
\begin{equation}
\Lambda_+^\Delta(P) R^\Delta_{\mu\beta}(P)  = 4 \frac{g_{10}^2}{M_u} \int\frac{d^4\ell}{(2\pi)^4}\,{\cal M}^\Delta_{\mu\nu}(\ell;P)\,
\Lambda_+^\Delta(P) R^\Delta_{\nu\beta}(P) \,,
\end{equation}
which, after contracting with $\delta_{\mu\beta}$, yields
\begin{eqnarray}
1  &=& \frac{g_{10}^2}{M_u} \, {\rm tr}_{\rm D}\int\frac{d^4\ell}{(2\pi)^4}\,{\cal M}^\Delta_{\mu\nu}(\ell;P)\,
\Lambda_+^\Delta(P) R^\Delta_{\nu\mu}(P) \\
\nonumber & = & \frac{8}{3} \frac{g_{10}^2}{M_u m_\Delta^3} \frac{E_{\{uu\}_{1^+}}^2}{m_{\{uu\}_{1^+}}^2} \!\int\frac{d^4\ell}{(2\pi)^4}
\frac{1}{(K^2+m_{\{uu\}_{1^+}}^2)(\ell^2+M_u^2)}
\left( -\ell\cdot P \, [3 \, m_{\{uu\}_{1^+}}^2 m_\Delta^2 + (K\cdot P)^2] \right.\\
&& \rule{12em}{0ex} \left. +m_\Delta[ 2 m_\Delta \ell\cdot K K\cdot P + 3 M (m_{\{uu\}_{1^+}}^2 m_\Delta^2 + (K\cdot P)^2)]\right),
\end{eqnarray}
where $E_{\{uu\}_{1^+}}\!(K)$ is the canonically-normalised axial-vector diquark Bethe-Salpeter amplitude, explained in Eq.\,(\ref{canonicalavqq}).  Now, with the aid of a Feynman parametrisation, the right hand side becomes
\begin{eqnarray}
\nonumber
&& \frac{8}{3} \frac{g_{10}^2}{M_u m_\Delta^3} \frac{E_{\{uu\}_{1^+}}^2}{m_{\{uu\}_{1^+}}^2}
\! \int\frac{d^4\ell}{(2\pi)^4} \int_0^1 d\alpha\,
\frac{1}{[(\ell-\alpha P)^2 + \sigma(\alpha,M_u^2,m^2_{\{uu\}_{1^+}},m^2_\Delta)]^2}
\left( -\ell\cdot P \, [3 \, m_{\{uu\}_{1^+}}^2 m_\Delta^2 \right.\\
&& \rule{13em}{0ex}\left.+ (K\cdot P)^2] +m_\Delta[ 2 m_\Delta \ell\cdot K K\cdot P + 3 M (m_{\{uu\}_{1^+}}^2 m_\Delta^2 + (K\cdot P)^2)]\right)
\end{eqnarray}
where
\begin{equation}
%\sigma_\Delta(\alpha,M_u,m_{\{uu\}_{1^+}},m_\Delta)
\sigma(\alpha,x,y,z)=(1-\alpha)\, x + \alpha \, y - \alpha (1-\alpha)\, z\,.
\label{definesigma}
\end{equation}

We employ a symmetry-preserving regularisation scheme. Hence the shift $\ell \to =\ell^\prime+\alpha P$ is permitted, whereafter $O(4)$-invariance entails $\ell^\prime\cdot P=0$ so that one may set
\begin{equation}
\label{replacementsDelta}
\ell \cdot P \to \alpha P^2\,,\;
K\cdot P = (1-\alpha) P^2\,,\;
\ell\cdot K \to \alpha (1-\alpha) P^2\,,
\end{equation}
and therewith obtain
\begin{eqnarray}
1 &=& 8 \frac{g_{10}^2}{M_u} \frac{E_{\{uu\}_{1^+}}^2}{m_{\{uu\}_{1^+}}^2}
\int\frac{d^4\ell^\prime}{(2\pi)^4} \int_0^1 d\alpha\,
\frac{(m_{\{uu\}_{1^+}}^2 + (1-\alpha)^2 m_\Delta^2)(\alpha m_\Delta + M_u)}
{[\ell^{^\prime 2} + \sigma(\alpha,M_u^2,m^2_{\{uu\}_{1^+}},m^2_\Delta)]^2}\\
&=& \frac{g_{10}^2}{M_u}\frac{E_{\{uu\}_{1^+}}^2}{m_{\{uu\}_{1^+}}^2}\frac{1}{2\pi^2}
\int_0^1 d\alpha\, [m_{\{uu\}_{1^+}}^2 + (1-\alpha)^2 m_\Delta^2][\alpha m_\Delta + M_u]\overline{\cal C}^{\rm iu}_1(\sigma(\alpha,M_u^2,m^2_{\{uu\}_{1^+}},m_\Delta^2))\,.
\end{eqnarray}
This is an eigenvalue problem whose solution yields the mass for the dressed-quark-core of the $\Delta$-resonance.  If one sets $g_{10}=1$, then $m_\Delta = 1.60\,$GeV.

For the purpose of illustration, here we present the Faddeev equation for the parity partner of the ground-state $\Delta$-resonance:
\begin{equation}
1=  \frac{g_{10}^2}{M_u}\frac{E_{\{uu\}_{1^-}}^2}{m_{\{uu\}_{1^-}}^2} \frac{1}{2\pi^2}
\int_0^1 d\alpha\, [m_{\{uu\}_{1^-}}^2 + (1-\alpha)^2 m_{\Delta^-}^2][\alpha m_{\Delta^-} - M_u]\overline{\cal C}^{\rm iu}_1(\sigma(\alpha,M_u^2,m^2_{\{uu\}_{1^-}},m_{\Delta^-}^2))\,.
\label{DeltaPPartnerFE}
\end{equation}

\subsection{Nucleon}
\label{app:FENucleon}
As we have already observed, using Eq.\,(\ref{staticexchange}) the nucleon's Faddeev amplitude simplifies to the form in Eqs.\,\eqref{calSAcontact}, which here we write as
\begin{subequations}
\begin{eqnarray}
{\cal S}^N(P) = s_N^1(P) \,\mbox{\boldmath $I$}_{\rm D}\,,\;
{\cal A}^{Ni}_{\mu}(P) =  a_{N1}^{i}(P) i\gamma_5\gamma_\mu + a_{N2}^{i}(P) \gamma_5 \hat P_\mu \,, \; i=4,5\,,
\end{eqnarray}
\end{subequations}
where the superscripts are diquark enumeration labels associated with Eq.\,\eqref{flavourarrays}.  In terms of the associated amplitude, the nucleon's Faddeev equation takes the form
\begin{equation}
\left(\begin{array}{c}
s^N_1(P)\\[0.7ex]
a_{N1}^{4}(P)\\[0.7ex]
a_{N1}^{5}(P)\\[0.7ex]
a_{N2}^{4}(P)\\[0.7ex]
a_{N2}^{5}(P)\end{array}\right)
=
\left( \begin{array}{ccccc}
{\cal K}^N_{\;11} & -\surd 2 \, {\cal K}^N_{\;14_1} & {\cal K}^N_{\;14_1} & -\surd 2\, {\cal K}^N_{\;14_2} & {\cal K}^N_{\;14_2}\\[0.7ex]
-\surd 2\, {\cal K}^N_{\;4_11} & 0 & \surd 2\, {\cal K}^N_{\;4_14_1} & 0 & \surd 2\, {\cal K}^N_{\;4_14_2}\\[0.7ex]
{\cal K}^N_{\;4_11} & \surd 2 \, {\cal K}^N_{\;4_14_1} & {\cal K}^N_{4_14_1} & \surd 2\,{\cal K}^N_{\;4_14_2} & {\cal K}^N_{4_14_2} \\[0.7ex]
-\surd 2\, {\cal K}^N_{\;4_21} & 0 & \surd 2\, {\cal K}^N_{\;4_24_1} & 0 & \surd 2\, {\cal K}^N_{\;4_24_2} \\[0.7ex]
{\cal K}^N_{\;4_21} & \surd 2\, {\cal K}^N_{\;4_24_1} & {\cal K}^N_{\;4_24_1} & \surd 2\, {\cal K}^N_{\;4_24_2} & {\cal K}^N_{\;4_24_2}
\end{array}
\right)
\left(\begin{array}{c}
s^N_1(P)\\[0.7ex]
a_{N1}^{4}(P)\\[0.7ex]
a_{N1}^{5}(P)\\[0.7ex]
a_{N2}^{4}(P)\\[0.7ex]
a_{N2}^{5}(P)\end{array}\right),
\end{equation}
where the kernel expresses the isospin symmetry we have assumed; i.e., that the Bethe-Salpeter amplitudes for the $\{uu\}_{1^+}$ and $\{ud\}_{1^+}$ correlations are identical.  Define now
\begin{equation}
c_N = \frac{g_8^2}{4 \pi^2 M_u}\,,\;
\sigma_N^i = (1-\alpha)\,M_u^2 + \alpha\,m_i^2 - \alpha (1-\alpha) m_N^2\,,
\end{equation}
where $i=1,4$ are diquark labels associated with Eq.\,\eqref{flavourarrays}, then the entries in the nucleon's Faddeev kernel may be expressed as follows:
\begin{subequations}
{\allowdisplaybreaks
\begin{eqnarray}
{\cal K}^N_{\;11} & = & K^N_{EE}+K^N_{EF}+K^N_{FF}\,,\\
K^N_{EE} & = & c_N E_1^2 \!
\int_0^1 d\alpha \,\overline{\cal C}_1(\sigma_N^1)
(\alpha m_N + M_u)\,,\\
K^N_{EF} & = & - 2 c_N E_1 F_1\frac{m_N}{M_u} \!
\int_0^1 d\alpha \,\overline{\cal C}_1(\sigma_N^1)
(1-\alpha) (\alpha m_N + M_u)\,,\\
K^N_{FF} & = & c_N F_1^2\frac{m_1^2}{M_u^2} \!
\int_0^1 d\alpha \,\overline{\cal C}_1(\sigma_N^1)(\alpha m_N + M_u)\,;\\
{\cal K}^N_{\;14_1} & = & K^N_{E4_1} + K^N_{F4_1}\,,\\
K^N_{E4_1} &=& c_N \frac{E_1 E_4}{m_4^2}\!
\int_0^1 d\alpha \,\overline{\cal C}_1(\sigma_N^4)
( m_4^2 (3 M_u + \alpha m_N) + 2 \alpha (1-\alpha)^2 m_N^3)\,,\\
K^N_{F4_1} &=& -c_N \frac{F_1 E_4}{m_4^2} \frac{m_N}{M_u}\!
\int_0^1 d\alpha \,\overline{\cal C}_1(\sigma_N^4)
(1-\alpha)
(m_4^2 (M_u + 3 \alpha m_N) + 2 (1-\alpha)^2 M_u m_N^2) \,;\\
{\cal K}^N_{\;14_2} & = & K^N_{E42} + K^N_{F42}\,,\\
K^N_{E4_2} & = & c_N \frac{E_1 E_4}{m_4^2}\!
\int_0^1 d\alpha \,\overline{\cal C}_1(\sigma_N^4)
(\alpha m_N - M_u) ((1-\alpha)^2 m_N^2-m_4^2)\,,\\
K^N_{F4_2} & = & c_N \frac{F_1 E_4}{m_4^2} \frac{m_N}{M_u} \!
\int_0^1 d\alpha \,\overline{\cal C}_1(\sigma_N^4)
(1-\alpha)(\alpha m_N - M_u) ((1-\alpha)^2 m_N^2-m_4^2)\,;\\
{\cal K}^N_{\;4_11} & = & K^N_{4_1E} + K^N_{4_1F}\,,\\
K^N_{4_1E} & = & \frac{c_N}{3}\frac{E_1E_4}{m_4^2} \!
\int_0^1 d\alpha \,\overline{\cal C}_1(\sigma_N^1)
(\alpha m_N + M_u) (2 m_4^2 + (1-\alpha)^2 m_N^2)\,,\\
K^N_{4_1F} & = & -\frac{c_N}{3}\frac{F_1E_4}{m_4^2} \frac{m_N}{M_u} \!
\int_0^1 d\alpha \,\overline{\cal C}_1(\sigma_N^1)
(1-\alpha)(2 m_4^2 + (1-\alpha)^2 m_N^2) (\alpha m_N + M_u)\,;\\
{\cal K}^N_{\;4_21} & = & K^N_{4_2E} + K^N_{4_2F}\,,\\
K^N_{4_2E} & = & \frac{c_N}{3} \frac{E_1E_4}{m_4^2} \!
\int_0^1 d\alpha \,\overline{\cal C}_1(\sigma_N^1)
(\alpha m_N + M_u) (m_4^2 - 4 (1-\alpha)^2 m_N^2),\\
K^N_{4_2F} & = & \frac{c_N}{3} \frac{F_1E_4}{m_4^2}\frac{m_N}{M_u} \!
\int_0^1 d\alpha \,\overline{\cal C}_1(\sigma_N^1)
(1-\alpha) (5 m_4^2-2(1-\alpha)^2 m_N^2)(\alpha m_N + M_u)\,;\\
{\cal K}^N_{\;4_14_1} & = & -\frac{c_N}{3}\frac{E_4^2}{m_4^2} \!
\int_0^1 d\alpha \,\overline{\cal C}_1(\sigma_N^4)
( 2 m_4^2 (M_u-\alpha m_N) + (1-\alpha)^2 m_N^2 (M_u+5 \alpha m_N))\,;\\
{\cal K}^N_{\;4_14_2} & = & -\frac{2 c_N}{3}\frac{E_4^2}{m_4^2} \!
\int_0^1 d\alpha \,\overline{\cal C}_1(\sigma_N^4)
(-m_4^2+(1-\alpha)^2 m_N^2) (\alpha m_N - M_u)\,;\\
{\cal K}^N_{\;4_24_1} & = & -\frac{c_N}{3}\frac{E_4^2}{m_4^2} \!
\int_0^1 d\alpha \,\overline{\cal C}_1(\sigma_N^4)
(m_4^2(11 \alpha  m_N + M_u) - 2(1-\alpha)^2 m_N^2 (7\alpha m_N + 2 M_u))\,;\\
{\cal K}^N_{\;4_24_2}  & = & - \frac{5 c_N}{3} \frac{E_4^2}{m_4^2} \!
\int_0^1 d\alpha \,\overline{\cal C}_1(\sigma_N^4)
(m_4^2 - (1-\alpha)^2 m_N^2) (\alpha m_N - M_u)\,,
\end{eqnarray}}
\end{subequations}

\hspace*{-\parindent}with $E_1$, $F_1$, $E_4$ being canonically normalised Bethe-Salpeter amplitudes for diquarks corresponding to enumeration labels $i=1,4$ in Eq.\,\eqref{flavourarrays}.
This kernel was computed following the procedure detailed for the $\Delta$-resonance in App.\,\ref{app:FEDelta}.  During this process we employed the replacements in Eq.\,(\ref{replacementsDelta}), their analogues involving the scalar-diquark's momentum, $K_{[ud]_{0^+}}$, and $K_{[ud]_{0^+}}\cdot K_{\{uu\}_{1^+}}\to (1-\alpha)^2 P^2$.  In the present context, of course, $P^2=-m_N^2$.
If one sets $g_8=1$, then $m_N=1.27$.

Given the structure of the kernel, it is not surprising that the eigenvectors exhibit the pattern
\begin{equation}
a_{Nj}^4 = -\surd 2\, a_{Nj}^5,\, j=1,2\,.
\end{equation}
For example, at the mass presented in Table~\ref{OctetDecupletMasses}, the nucleon's unit-normalised Faddeev amplitude is
\begin{equation}
\label{nucleonqqratio}
\begin{array}{ccccc}
s^1_N & a_{N1}^4 & a_{N1}^5 & a_{N2}^4 & a_{N2}^5 \\
%-0.885498, -0.375276, 0.26536, 0.0556518, -0.0393518
%-0.880961, 0.380803, -0.269269, 0.0652776, -0.0461583
0.88 & 0.38 & -0.27 & -0.065 & 0.046
\end{array}\,.
\end{equation}
The axial-vector-diquark correlation provides 22\% of the unit normalisation.  %This is discussed further in connection with Fig.\,\ref{Fig6}.

\subsection{Kernel for the $\Lambda$ baryon}
\label{App:Lambda}
Here we make explicit each entry in the Faddeev equation kernel for the $\Lambda$-baryon, Eq.\,\eqref{KernelLambda}.  Define
\begin{equation}
c_{\Lambda}^f = \frac{g_8^2}{4 \pi^2 M_f}\,,\;
\sigma_{\Lambda}^{f,i} =
\sigma(\alpha,M_f^2,m_i^2,m_\Lambda^2)\,,
%(1-\alpha)\,M_I^2 + \alpha\,m_i^2 - \alpha(1-\alpha)m_{\Lambda}^2\,,
%
\end{equation}
where $\sigma(\alpha,x,y,z)$ was introduced in Eq.\,\eqref{definesigma}, $f$ labels a quark flavour and $i$ is the diquark enumeration label in Eq.\,\eqref{flavourarrays}, so that $m_i$ is the mass of the associated correlation, then
\begin{subequations}
{\allowdisplaybreaks
\begin{eqnarray}
{\cal K}^\Lambda_{\;12} & = & \frac{c_\Lambda^u}{2 M_u M_R}
\int_0^1 d\alpha \,\overline{\cal C}_1(\sigma_{\Lambda}^{u,2})
[E_1 M_u - F_1 m_{\Lambda}(1-\alpha)]
[2 E_2 M_R - F_2 m_{\Lambda}(1-\alpha)][M_u+\alpha m_{\Lambda}] \,, \\
\nonumber
{\cal K}^{\Lambda}_{\;16_1} & = & \frac{c_{\Lambda}^u E_6 }{M_u m_6^2} \!
\int_0^1 d\alpha \,\overline{\cal C}_1(\sigma_{\Lambda}^{u,6})
\bigg[
E_1 M_u[3M_u m_6^2+m_{\Lambda}(m_6^2+2m_{\Lambda}^2(1-\alpha)^2)\alpha]\\
&& \rule{15ex}{0ex}
- F_1 m_{\Lambda}(1-\alpha)(M_u[m_6^2+2m_{\Lambda}^2(1-\alpha)^2]+3m_6^2 m_{\Lambda} \alpha)
\bigg]\,,\\
{\cal K}^{\Lambda}_{\;16_2} &=& \frac{c_{\Lambda}^u E_6}{M_u m_6^2} \!
\int_0^1 d\alpha \,\overline{\cal C}_1(\sigma_{\Lambda}^{u,6})
[E_1 M_u+F_1 m_{\Lambda}(1-\alpha)][m_6^2-m_{\Lambda}^2(1-\alpha)^2][M_u-\alpha m_{\lambda}]\,,\\
{\cal K}^{\Lambda}_{\;21} &=& \frac{c_{\Lambda}^s }{2M_u M_R} \!
\int_0^1 d\alpha \,\overline{\cal C}_1(\sigma_{\Lambda}^{s,1})
[E_1 M_u - F_1 m_{\Lambda}(1-\alpha)]
[2 E_2 M_R-F_2 m_{\Lambda}(1-\alpha)]
[M_s+\alpha m_{\Lambda}]\,,\\
{\cal K}^{\Lambda}_{\;23} &=&  \frac{c_{\Lambda}^s}{4M_R^2} \!
\int_0^1 d\alpha \,\overline{\cal C}_1(\sigma_{\Lambda}^{u,2})
[2 E_2 M_R - F_2 m_\Lambda (1-\alpha)]^2 [M_u+\alpha m_\Lambda]\,,\\
\nonumber
{\cal K}^{\Lambda}_{\;28_1} & = & \frac{c_{\Lambda}^s E_6}{2 M_R m_6^2} \!
\int_0^1 d\alpha \,\overline{\cal C}_1(\sigma_{\Lambda}^{u,6})
\bigg[
2 E_2 M_R [3M_u m_6^2+m_\Lambda[m_6^2+2m_{\Lambda}^2(1-\alpha)^2]\alpha]\,,\\
&& \rule{15ex}{0ex}
- F_2 m_\Lambda (1-\alpha)[M_u(m_6^2+2m_{\Lambda}^2 (1-\alpha)^2)+3m_6^2 m_\Lambda \alpha]\bigg]\,,\\
{\cal K}^{\Lambda}_{\;28_2} &=& \frac{c_{\Lambda}^s E_6}{2 M_R m_6^2} \!
\int_0^1 d\alpha \,\overline{\cal C}_1(\sigma_{\Lambda}^{u,6})
[2 E_2 M_R + F_2 m_\Lambda (1-\alpha)][m_6^2 - m_\Lambda^2(1-\alpha)^2][M_u-\alpha m_\Lambda]\,, \\
{\cal K}^{\Lambda}_{\;6_11} &=&  \frac{c_{\Lambda}^u E_6 }{3M_u m_6^2} \!
\int_0^1 d\alpha \,\overline{\cal C}_1(\sigma_{\Lambda}^{s,1})
[E_1 M_u - F_1 m_\Lambda(1-\alpha)][2m_6^2+m_\Lambda^2(1-\alpha)^2][M_s+\alpha m_\Lambda]\,,\\
\nonumber
{\cal K}^{\Lambda}_{6_21} &= & \frac{c_{\Lambda}^u E_6}{3M_u m_6^2} \!
\int_0^1 d\alpha \,\overline{\cal C}_1(\sigma_{\Lambda}^{s,1})
\bigg[E_1 M_u [m_6^2-4m_\Lambda^2(1-\alpha)^2]\\
&&\rule{15ex}{0ex}
+ F_1 m_\Lambda [5m_6^2 - 2m_\Lambda^2(1-\alpha)^2](1-\alpha)\bigg][M_u+\alpha m_\Lambda]\,,\\
{\cal K}^{\Lambda}_{\;6_13} &=&  \frac{c_{\Lambda}^s E_6}{6 M_R m_6^2} \!
\int_0^1 d\alpha \,\overline{\cal C}_1(\sigma_{\Lambda}^{u,2})
[2 E_2 M_R - F_2 m_\Lambda(1-\alpha)] [2m_6^2+m_\Lambda^2(1-\alpha)^2] [M_u+\alpha m_\Lambda]\,,\\
\nonumber
{\cal K}^{\Lambda}_{\;6_2 3}& = &\frac{c_{\Lambda}^s E_6}{6M_R m_6^2} \!
\int_0^1 d\alpha \,\overline{\cal C}_1(\sigma_{\Lambda}^{u,2})
\bigg[2 E_2 M_R [m_6^2-4m_\Lambda^2(1-\alpha)^2]\\
&&\rule{15ex}{0ex}
+ F_2 m_\Lambda [5m_6^2 - 2m_\Lambda^2(1-\alpha)^2](1-\alpha)\bigg][M_u+\alpha m_\Lambda]\,,
\\
%\end{eqnarray}
%\begin{eqnarray}
%
{\cal K}^{\Lambda}_{\;6_18_1} &=& - \frac{c_{\Lambda}^s E_6^2}{3 m_6^4} \!
\int_0^1 d\alpha \,\overline{\cal C}_1(\sigma_{\Lambda}^{u,6})
[M_u m_6^2[4m_6^2-m_\Lambda^2(1-\alpha)^2]+m_\Lambda^3[m_6^2+2m_\Lambda^2(1-\alpha)^2](1-\alpha)^2 \alpha]\,,\\
{\cal K}^{\Lambda}_{\;6_18_2} &=& \frac{c_{\Lambda}^s E_6^2 }{3 m_6^4} \!
\int_0^1 d\alpha \,\overline{\cal C}_1(\sigma_{\Lambda}^{u,6})
[-2m_6^4+m_6^2 m_\Lambda^2(1-\alpha)^2+m_\Lambda^4(1-\alpha)^4][M_u-\alpha m_\Lambda]\,, \\
\nonumber
{\cal K}^{\Lambda}_{\;6_28_1} & = & \frac{c_{\Lambda}^s E_6^2}{3 m_6^4} \!
\int_0^1 d\alpha \,\overline{\cal C}_1(\sigma_{\Lambda}^{u,6})
\bigg[
M_u [m_6^4-4m_6^2 m_\Lambda^2(1-\alpha)^2+6m_\Lambda^4(1-\alpha)^4]\\
&& \rule{15ex}{0ex}
 +m_\Lambda [-9m_6^4+10m_6^2 m_\Lambda^2(1-\alpha)^2+2m_\Lambda^4(1-\alpha)^4]\alpha
\bigg]\,,\\
{\cal K}^{\Lambda}_{\;6_28_2} & = & \frac{c_{\Lambda}^s E_6^2}{3 m_6^4} \!
\int_0^1 d\alpha \,\overline{\cal C}_1(\sigma_{\Lambda}^{u,6})
[5m_6^4-7m_6^2 m_\Lambda^2(1-\alpha)^2+2m_\Lambda^4(1-\alpha)^4][M_u-\alpha m_\Lambda]\,,
\end{eqnarray}}
\end{subequations}
\vspace*{-0.5\baselineskip}

\hspace*{-\parindent}with $M_R$ defined in connection with Eq.\,\eqref{KaonBSA} and
$\{E_i,F_i|i=1,2\}$, $E_6$ being canonically normalised Bethe-Salpeter amplitudes for diquarks corresponding to enumeration labels $i=1,2,6$ in Eq.\,\eqref{flavourarrays}.  This kernel was computed following the pattern in App.\,\ref{app:FEDelta}, using analogues of Eq.\,(\ref{replacementsDelta}).

\subsection{$\Sigma$ Baryon}
Given the flavour structure of the $\Sigma$, one can obtain its Faddeev equation directly from that of the proton by simply making the replacement $d\to u$.  However, we assumed isospin symmetry in writing the proton's Faddeev equation, so that replacement is not readily achieved.  We therefore report the complete structure here.

The Faddeev amplitude for the $\Sigma^+$ baryon is expressed in terms of
\begin{equation}
{\cal S}^\Sigma = s_\Sigma^2(P) \mathbf{I}_{\rm D}\,,\;
{\cal A}_\mu^{\Sigma i} = a_{\Sigma 1}^i(P) i \gamma_5\gamma_\mu + a_{\Sigma 2}^i(P) \gamma_5 \hat P_\mu\,,i=4,6\,,
\label{SigmaPlusFA}
\end{equation}
where the superscripts are diquark enumeration labels, Eq.\,\eqref{flavourarrays}.  The associated Faddeev equation is
\begin{equation}
\left(\begin{array}{c}
s_{\Sigma}^2(P)\\[0.7ex]
a_{\Sigma 1}^4(P)\\[0.7ex]
a_{\Sigma 1}^6(P)\\[0.7ex]
a_{\Sigma 2}^4(P)\\[0.7ex]
a_{\Sigma 2}^6(P)\end{array}\right)
=
\left( \begin{array}{ccccc}
{\cal K}^{\Sigma}_{\;22} & -\surd 2 \, {\cal K}^{\Sigma}_{\;24_1} & {\cal K}^{\Sigma}_{26_1} & -\surd 2\, {\cal K}^{\Sigma}_{\;24_2} & {\cal K}^{\Sigma}_{\;26_2}\\[0.7ex]
-\surd 2 \, {\cal K}^{\Sigma}_{\;4_12} & 0 & \surd 2 \, {\cal K}^{\Sigma}_{\;4_16_1} & 0 & \surd 2 \, {\cal K}^{\Sigma}_{\;4_16_2}\\[0.7ex]
{\cal K}^{\Sigma}_{\;6_12} & \surd 2 \, {\cal K}^{\Sigma}_{\;6_14_1} & {\cal K}^{\Sigma}_{\;6_16_1} & \surd 2 \, {\cal K}^{\Sigma}_{\;6_14_2} & {\cal K}^{\Sigma}_{6_16_2}\\[0.7ex]
-\surd 2 \, {\cal K}^{\Sigma}_{\;4_22} & 0 & \surd 2 \, {\cal K}^{\Sigma}_{\;4_26_1} & 0 & \surd 2 \, {\cal K}^{\Sigma}_{\;4_26_2}\\[0.7ex]
{\cal K}^{\Sigma}_{\;6_22} & \surd 2 \, {\cal K}^{\Sigma}_{\;6_24_1} & {\cal K}^{\Sigma}_{\;6_26_1} & \surd 2 \, {\cal K}^{\Sigma}_{\;6_24_2} & {\cal K}^{\Sigma}_{\;6_26_2}
\end{array}
\right)
\left(\begin{array}{c}
s_{\Sigma}^2(P)\\[0.7ex]
a_{\Sigma 1}^4(P)\\[0.7ex]
a_{\Sigma 1}^6(P)\\[0.7ex]
a_{\Sigma 2}^4(P)\\[0.7ex]
a_{\Sigma 2}^6(P)\end{array}
\right).
\label{SigmaPlusFE}
\end{equation}
In order to make each entry explicit, we define
\begin{equation}
c_{\Sigma}^f = \frac{g_8^2}{4 \pi^2 M_f}\,,\;
\sigma_{\Sigma}^{f,i} =
\sigma(\alpha,M_f^2,m_i^2,m_\Sigma^2)\,,
%(1-\alpha)\,M_I^2 + \alpha\,m_i^2 - \alpha(1-\alpha)m_{\Lambda}^2\,,
%
\end{equation}
where $\sigma(\alpha,x,y,z)$ was introduced in Eq.\,\eqref{definesigma}, $f=u,s$, $i=2,4,6$ is the diquark enumeration label in Eq.\,\eqref{flavourarrays}, so that $m_i$ is the mass of the associated correlation; and then
\begin{subequations}
{\allowdisplaybreaks
\begin{eqnarray}
{\cal K}^{\Sigma}_{\;22} &= & K^{\Sigma}_{22EE}+K^{\Sigma}_{22EF}+K^{\Sigma}_{22FF}\,,\\
K^{\Sigma}_{22EE} &=& c_{\Sigma}^s E_2^2 \!
\int_0^1 d\alpha \,\overline{\cal C}_1(\sigma_{\Sigma}^{u,2})
[M_u+\alpha m_{\Sigma}]\,,\\
K^{\Sigma}_{22EF} &=& -c_{\Sigma}^s E_2F_2 \frac{m_{\Sigma}}{M_R} \!
\int_0^1 d\alpha \,\overline{\cal C}_1(\sigma_{\Sigma}^{u,2})
(1-\alpha)[M_u+\alpha m_{\Sigma}]\,,\\
K^{\Sigma}_{22FF} &=& c_{\Sigma}^s F_2^2 \frac{m_2^2}{4M_R^2} \!
\int_0^1 d\alpha \,\overline{\cal C}_1(\sigma_{\Sigma}^{u,2})
(M_u+\alpha m_{\Sigma})\,,\\
{\cal K}^{\Sigma}_{\;24_1} &=& K^{\Sigma}_{24_1E}+K^{\Sigma}_{24_1F}\,,\\
K^{\Sigma}_{24_1E} &=& c_{\Sigma}^u \frac{E_2E_4}{m_4^2} \!
\int_0^1 d\alpha \,\overline{\cal C}_1(\sigma_{\Sigma}^{s,4})
[3M_s m_4^2 + \alpha m_{\Sigma}(m_4^2 + 2m_{\Sigma}^2(1-\alpha)^2)]\,,\\
K^{\Sigma}_{24_1F} &=& -c_{\Sigma}^u \frac{F_2E_4 }{m_4^2}\frac{m_{\Sigma}}{2M_R } \!
\int_0^1 d\alpha \,\overline{\cal C}_1(\sigma_{\Sigma}^{s,4})
(1-\alpha)[M_s[m_4^2+2m_{\Sigma}^2(1-\alpha)^2]+3\alpha m_4^2 ]\,,\\
{\cal K}^{\Sigma}_{\;24_2} &=& K^{\Sigma}_{24_2E} + K^{\Sigma}_{24_2F}\,,\\
K^{\Sigma}_{24_2E} &=& c_{\Sigma}^u  \frac{E_2E_4}{m_4^2} \!
\int_0^1 d\alpha \,\overline{\cal C}_1(\sigma_{\Sigma}^{s,4})
[m_{\Sigma}^2(1-\alpha)^2-m_4^2][\alpha m_{\Sigma}-M_s]\,,\\
K^{\Sigma}_{24_2F} &=& c_{\Sigma}^u  \frac{F_2E_4}{m_4^2} \frac{m_{\Sigma}}{2M_R }\!
\int_0^1 d\alpha \,\overline{\cal C}_1(\sigma_{\Sigma}^{s,4})
(1-\alpha) [m_{\Sigma}^2(1-\alpha)^2-m_4^2][\alpha m_{\Sigma}-M_s]\,,\\
{\cal K}^{\Sigma}_{\;26_1} &=& K^{\Sigma}_{26_1E}+K^{\Sigma}_{26_1F}\,,\\
K^{\Sigma}_{26_1E} &=& c_{\Sigma}^s  \frac{E_2E_6}{m_6^2} \!
\int_0^1 d\alpha \,\overline{\cal C}_1(\sigma_{\Sigma}^{u,6})
[3M_u m_6^2+ \alpha m_{\Sigma}(m_6^2+2m_{\Sigma}^2(1-\alpha)^2)]\,,\\
K^{\Sigma}_{26_1F} &=& -c_{\Sigma}^s  \frac{F_2E_6}{m_6^2}\frac{m_\Sigma}{2M_R } \!
\int_0^1 d\alpha \,\overline{\cal C}_1(\sigma_{\Sigma}^{u,6})
(1-\alpha)[M_u [m_6^2+2m_{\Sigma}^2(1-\alpha)^2]+3\alpha m_6^2 ]\,,\\
{\cal K}^{\Sigma}_{\;26_2} &=& K^{\Sigma}_{26_2E}+K^{\Sigma}_{26_2F}\,,\\
K^{\Sigma}_{26_2E} &=& c_{\Sigma}^s  \frac{E_2E_6}{m_6^2} \!
\int_0^1 d\alpha \,\overline{\cal C}_1(\sigma_{\Sigma}^{u,6})
[m_{\Sigma}^2(1-\alpha)^2-m_6^2][\alpha m_{\Sigma}-M_u]\,,\\
K^{\Sigma}_{26_2F} &=& c_{\Sigma}^s  \frac{F_2E_6}{m_6^2}\frac{m_\Sigma}{2M_R} \!
\int_0^1 d\alpha \,\overline{\cal C}_1(\sigma_{\Sigma}^{u,6}) (1-\alpha)
[m_{\Sigma}^2(1-\alpha)^2-m_6^2][\alpha m_{\Sigma}-M_u]\,,\\
{\cal K}^{\Sigma}_{\;4_12} &=& K^{\Sigma}_{4_12E}+K^{\Sigma}_{4_12F}\,,\\
K^{\Sigma}_{4_12E} &=& c_{\Sigma}^u  \frac{E_2E_4}{3m_4^2} \!
\int_0^1 d\alpha \,\overline{\cal C}_1(\sigma_{\Sigma}^{u,2})
[2m_4^2+m_{\Sigma}^2(1-\alpha)^2][M_u+\alpha m_{\Sigma}]\,,\\
K^{\Sigma}_{4_12F} &=& -c_{\Sigma}^u  \frac{F_2E_4}{4m_4^2}\frac{m_\Sigma}{2M_R}  \!
\int_0^1 d\alpha \,\overline{\cal C}_1(\sigma_{\Sigma}^{u,2})(1-\alpha)
[2m_4^2+m_{\Sigma}^2(1-\alpha)^2][M_u+\alpha m_{\Sigma}]\,,\\
{\cal K}^{\Sigma}_{\;4_22} &=& K^{\Sigma}_{4_22E}+K^{\Sigma}_{4_22F}\,,\\
K^{\Sigma}_{4_22E} &=& c_{\Sigma}^u  \frac{E_2E_4}{3m_4^2} \!
\int_0^1 d\alpha \,\overline{\cal C}_1(\sigma_{\Sigma}^{u,2})
[m_4^2-4m_{\Sigma}^2(1-\alpha)^2][M_u+\alpha m_\Sigma]\,,\\
K^{\Sigma}_{4_22F} &=& c_{\Sigma}^u  \frac{F_2E_4}{3 m_4^2}\frac{m_\Sigma}{2M_R} \!
\int_0^1 d\alpha \,\overline{\cal C}_1(\sigma_{\Sigma}^{u,2}) (1-\alpha)
(5m_4^2-2m_\Sigma^2(1-\alpha)^2)(M_u+\alpha m_\Sigma)\,,\\
\nonumber
{\cal K}^{\Sigma}_{\;4_16_1} &=&
-c_{\Sigma}^u  \frac{E_4E_6}{3m_4^2 m_6^2} \!
\int_0^1 d\alpha \,\overline{\cal C}_1(\sigma_{\Sigma}^{u,6})
\bigg[
M_u[m_4^2(4m_6^2-2m_\Sigma^2(1-\alpha)^2)+m_6^2 m_\Sigma^2(1-\alpha)^2]\\
&& \rule{15ex}{0ex}
+ \alpha(1-\alpha)^2 m_\Sigma^3[2m_4^2-m_6^2+2m_\Sigma^2(1-\alpha)^2]
\bigg]\,,\\
{\cal K}^{\Sigma}_{\;4_16_2} &=& c_{\Sigma}^u \frac{E_4E_6 }{3m_4^2 m_6^2} \!
\int_0^1 d\alpha \,\overline{\cal C}_1(\sigma_{\Sigma}^{u,6})
[m_6^2-m_\Sigma^2(1-\alpha)^2][2m_4^2+m_\Sigma^2(1-\alpha)^2][\alpha m_\Sigma-M_u]\,,\\
\nonumber
{\cal K}^{\Sigma}_{4_26_1} & = & c_{\Sigma}^u \frac{E_4E_6}{3m_4^2 m_6^2} \!
\int_0^1 d\alpha \,\overline{\cal C}_1(\sigma_{\Sigma}^{u,6})
\bigg[ M_u ( m_4^2 [ m_6^2-8m_\Sigma^2(1-\alpha)^2 ] +2 m_\Sigma^2 (1-\alpha)^2 [2m_6^2+3m_\Sigma^2(1-\alpha)^2] ) \\
&& \rule{15ex}{0ex}
+ \alpha m_\Sigma ( m_4^2[-9m_6^2+2m_\Sigma^2(1-\alpha)^2] +2m_\Sigma^2 (1-\alpha)^2 [4m_6^2+m_\Sigma^2(1-\alpha)^2]) \bigg] ,\\
{\cal K}^{\Sigma}_{\;4_26_2} &=& c_{\Sigma}^u \frac{E_4E_6 }{3m_4^2 m_6^2} \!
\int_0^1 d\alpha \,\overline{\cal C}_1(\sigma_{\Sigma}^{u,6})
[m_6^2 - m_\Sigma^2(1-\alpha)^2][5m_4^2 - 2m_\Sigma^2(1-\alpha)^2][M_u-\alpha m_\Sigma]\,,
\end{eqnarray}}
\label{SigmaPlusFEK1}
\end{subequations}
%\vspace*{-4ex}

\begin{subequations}
{\allowdisplaybreaks
\begin{eqnarray}
{\cal K}^{\Sigma}_{\;6_12} &=& K^{\Sigma}_{6_12E}+K^{\Sigma}_{6_12F}\,,\\
K^{\Sigma}_{6_12E} &=& c_{\Sigma}^s  \frac{E_2E_6}{3m_6^2} \!
\int_0^1 d\alpha \,\overline{\cal C}_1(\sigma_{\Sigma}^{u,2})
[2m_6^2+m_\Sigma^2(1-\alpha)^2][M_u+\alpha m_\Sigma]\,,\\
K^{\Sigma}_{6_12F} &=& -c_{\Sigma}^s  \frac{F_2E_6}{3m_6^2} \frac{m_\Sigma}{2M_R}\!
\int_0^1 d\alpha \,\overline{\cal C}_1(\sigma_{\Sigma}^{u,2})(1-\alpha)
[2m_6^2+m_\Sigma^2(1-\alpha)^2][M_u+\alpha m_\Sigma]\,,\\
{\cal K}^{\Sigma}_{\;6_22} &=& K^{\Sigma}_{6_22E}+K^{\Sigma}_{6_22F}\,,\\
K^{\Sigma}_{6_22E} &=& c_{\Sigma}^s \frac{E_2E_6 }{3m_6^2} \!
\int_0^1 d\alpha \,\overline{\cal C}_1(\sigma_{\Sigma}^{u,2})
[m_6^2-4m_\Sigma^2(1-\alpha)^2][M_u+\alpha m_\Sigma]\,,\\
K^{\Sigma}_{6_22E} &=& c_{\Sigma}^s \frac{F_2E_6}{3m_6^2} \frac{m_\Sigma}{2M_R}\!
\int_0^1 d\alpha \,\overline{\cal C}_1(\sigma_{\Sigma}^{u,2}) (1-\alpha)
[5m_6^2-2m_\Sigma^2(1-\alpha)^2][M_u+\alpha m_\Sigma]\,,\\
\nonumber
{\cal K}^{\Sigma}_{\;6_14_1} & = &-c_{\Sigma}^u \frac{E_4E_6 }{3m_4^2m_6^2} \!
\int_0^1 d\alpha \,\overline{\cal C}_1(\sigma_{\Sigma}^{s,4})
\bigg[
M_s [ m_4^2(4m_6^2+m_\Sigma^2(1-\alpha)^2)-2m_6^2m_\Sigma^2(1-\alpha)^2]\\
&& \rule{15ex}{0ex}
+ \alpha (1-\alpha)^2 m_\Sigma^3[2 m_6^2 -m_4^2 + 2m_\Sigma^2(1-\alpha)^2]
\bigg] , \\
{\cal K}^{\Sigma}_{\;6_14_2} &=& c_{\Sigma}^u  \frac{E_4E_6}{3m_4^2m_6^2} \!
\int_0^1 d\alpha \,\overline{\cal C}_1(\sigma_{\Sigma}^{s,4})
[m_4^2-m_\Sigma^2(1-\alpha)^2][2m_6^2+m_\Sigma^2(1-\alpha)^2][\alpha m_\Sigma-M_s]\,,\\
\nonumber
{\cal K}^{\Sigma}_{\;6_24_1} & = & c_{\Sigma}^u  \frac{E_4E_6}{3m_4^2m_6^2} \!
\int_0^1 d\alpha \,\overline{\cal C}_1(\sigma_{\Sigma}^{s,4})
\bigg[
M_s (m_4^2[m_6^2+4m_\Sigma^2(1-\alpha)^2] + 2 m_\Sigma^2 (1-\alpha)^2 [-4m_6^2+3m_\Sigma^2(1-\alpha)^2])\\
&& \rule{15ex}{0ex}
+ \alpha m_\Sigma (m_4^2 [-9m_6^2+8m_\Sigma^2(1-\alpha)^2] + 2 m_\Sigma^2 (1-\alpha)^2 [m_6^2+m_\Sigma^2(1-\alpha)^2])
\bigg]\,,\\
{\cal K}^{\Sigma}_{\;6_24_2} &=& c_{\Sigma}^u  \frac{E_4E_6}{3m_4^2m_6^2} \!
\int_0^1 d\alpha \,\overline{\cal C}_1(\sigma_{\Sigma}^{s,4})
[m_4^2 - m_\Sigma^2(1-\alpha)^2][5m_6^2 - 2m_\Sigma^2(1-\alpha)^2][M_s-\alpha M_s]\,,\\
{\cal K}^{\Sigma}_{\;6_16_1} &=& -c_{\Sigma}^u \frac{E_6^2}{3m_6^2} \!
\int_0^1 d\alpha \,\overline{\cal C}_1(\sigma_{\Sigma}^{u,6})
[M_u(2m_6^2+m_\Sigma^2(1-\alpha)^2) + \alpha m_\Sigma(-2m_6^2+5m_\Sigma^2(1-\alpha)^2)]\,,\\
{\cal K}^{\Sigma}_{\;6_16_2} &=& c_{\Sigma}^u  \frac{2 E_6^2}{3m_6^2} \!
\int_0^1 d\alpha \,\overline{\cal C}_1(\sigma_{\Sigma}^{u,6})
[m_6^2-m_\Sigma^2(1-\alpha)^2][\alpha m_\Sigma-M_u]\,,\\
{\cal K}^{\Sigma}_{\;6_26_1} &=& -c_{\Sigma}^u  \frac{E_6^2}{3m_6^2} \!
\int_0^1 d\alpha \,\overline{\cal C}_1(\sigma_{\Sigma}^{u,6})
[M_u(m_6^2-4m_\Sigma^2(1-\alpha)^2)+\alpha m_\Sigma(11m_6^2-14m_\Sigma^2(1-\alpha)^2)]\,,\\
{\cal K}^{\Sigma}_{\;6_16_2} &=& c_{\Sigma}^u  \frac{5 E_6^2}{3m_6^2} \!
\int_0^1 d\alpha \,\overline{\cal C}_1(\sigma_{\Sigma}^{u,6})
[m_6^2-m_\Sigma^2(1-\alpha)^2][M_u-\alpha m_\Sigma]\,.
\end{eqnarray}}
\label{SigmaPlusFEK2}
\end{subequations}
\vspace*{-0.5\baselineskip}

\hspace*{-\parindent}with $M_R$ defined in connection with Eq.\,\eqref{KaonBSA} and
$E_2$, $F_2$, $E_4$, $E_6$ being canonically normalised Bethe-Salpeter amplitudes for diquarks corresponding to enumeration labels $i=2,4,6$ in Eq.\,\eqref{flavourarrays}.  This kernel was computed following the pattern in App.\,\ref{app:FEDelta}, using analogues of Eq.\,(\ref{replacementsDelta}).

\subsection{$\Xi$ Baryon}
As apparent from Eqs.\,\eqref{uColumn}, the $\Xi^0$ baryon may be obtained from the $\Sigma^+$ by making the replacements $u\leftrightarrow s$.  It follows that the Faddeev equation for the $\Xi^0$ can simply be obtained from that for the $\Sigma^+$ by making the replacements $u\leftrightarrow s$ and $4\to 9$ in Eqs.\,\eqref{SigmaPlusFA}, \eqref{SigmaPlusFE}, \eqref{SigmaPlusFEK1}, \eqref{SigmaPlusFEK2}.

\subsection{$\Sigma^\ast$ Baryon}
\label{SigmaAstFE}
The Faddeev amplitude for the decuplet $\Sigma^\ast$ may be expressed as
\begin{equation}
{\cal D}^{\Sigma^\ast}_{\mu\rho} u^{\Sigma^\ast}_\rho(P) = \sum_{i=4,6} f^i_{\Sigma^\ast} \mathbf{I}_{\rm D} u^{\Sigma^\ast}_\mu(P)\,,
\end{equation}
so that the Faddeev equation can be written
\begin{equation}
\left(\begin{array}{c}
f_{\Sigma^\ast}^4(P)\\[0.7ex]
f_{\Sigma^\ast}^6(P)\end{array}\right)
=
\left( \begin{array}{cc}
0 & \surd 2 \, {\cal K}^{\Sigma^\ast}_{\;46}\\[0.7ex]
\surd 2 \, {\cal K}^{\Sigma^\ast}_{\;64} & {\cal K}^{\Sigma^\ast}_{\;66}
\end{array}
\right)
\left(\begin{array}{c}
f_{\Sigma^\ast}^4(P)\\[0.7ex]
f_{\Sigma^\ast}^6(P)\end{array}\right)\,.
\end{equation}
Defining
\begin{equation}
c_{\Sigma^\ast}^f = \frac{g_{10}^2}{4 \pi^2 M_f},\;
\sigma_{\Sigma^\ast}^{f,i}:=(1-\alpha)\,M_f^2 + \alpha\,m_i^2 - \alpha(1-\alpha)m_{\Sigma^\ast}^2\,,
\end{equation}
where $\sigma(\alpha,x,y,z)$ was introduced in Eq.\,\eqref{definesigma}, $f=u,s$, $i=4,6$ is the diquark enumeration label in Eq.\,\eqref{flavourarrays}, so that $m_i$ is the mass of the associated correlation; then
\begin{subequations}
{\allowdisplaybreaks
\begin{eqnarray}
\nonumber
{\cal K}^{\Sigma^\ast}_{\;46} &=& c_{\Sigma^\ast}^u \frac{E_4E_6 }{6 m_4^2m_6^2} \!
\int_0^1 d\alpha \,\overline{\cal C}_1(\sigma_{\Sigma^\ast}^{u,6})\,
\bigg[
m_4^2(5m_6^2 + 3m_{\Sigma^\ast}^2(1-\alpha)^2)  \\
&& \rule{15ex}{0ex}
+ (1-\alpha)^2 m_{\Sigma^\ast}^2(3m_6^2 + m_{\Sigma^\ast}^2(1-\alpha)^2)
\bigg] [M_u+\alpha m_{\Sigma^\ast}], \\
\nonumber
{\cal K}^{\Sigma^\ast}_{\;64} &=& c_{\Sigma^\ast}^u \frac{E_4E_6 }{6 m_4^2m_6^2} \!
\int_0^1 d\alpha \,\overline{\cal C}_1(\sigma_{\Sigma^\ast}^{s,4})\,
\bigg[
m_4^2(5m_6^2 + 3m_{\Sigma^\ast}^2 (1-\alpha)^2) \\
&& \rule{15ex}{0ex}
+ (1-\alpha)^2 m_{\Sigma^\ast}^2(3m_6^2+m_{\Sigma^\ast}^2 (1-\alpha)^2)\bigg]
[M_s+\alpha m_{\Sigma^\ast}]\,,\\
{\cal K}^{\Sigma^\ast}_{\;66} &=& c_{\Sigma^\ast}^s \frac{E_6^2}{m_6^2} \!
\int_0^1 d\alpha \,\overline{\cal C}_1(\sigma_{\Sigma^\ast}^{u,6})\,
[m_6^2+m_{\Sigma^\ast}^2(1-\alpha)^2][M_u+\alpha m_{\Sigma^\ast}]\,.
\end{eqnarray}}
\end{subequations}
\vspace*{-0.5\baselineskip}

\hspace*{-\parindent}with $E_4$, $E_6$ being canonically normalised Bethe-Salpeter amplitudes for diquarks corresponding to enumeration labels $i=4,6$ in Eq.\,\eqref{flavourarrays}.  This kernel was computed following the pattern in App.\,\ref{app:FEDelta}, using analogues of Eq.\,(\ref{replacementsDelta}).

The eigenvectors corresponding to the results in Table~\ref{OctetDecupletMasses} are
\begin{equation}
\begin{array}{l|c|c|c|c}
%GS:{-0.608283, -0.79372}
%RE:{-0.759854, -0.650094}
%PP:{-0.668717, -0.743517}
%PPRE:{-0.659808, -0.751434}
    & P=+, n=0 & P=+, n=1 & P=-, n=0 & P=-, n=0 \\\hline
f_{\Sigma^\ast}^4 & 0.61 & 0.76 & 0.67 & 0.66\\
f_{\Sigma^\ast}^6 & 0.79 & 0.65 & 0.74 & 0.75
\end{array}
\end{equation}
indicating that these states generally favour the $\{us\}u$ configuration over $\{uu\}s$.  This owes to the quark exchange character of the interaction in Fig.\,\ref{fig:FaddeevI}: both $\{uu\}s$ and $\{us\}u$ feed into $\{us\}u$ but only $\{us\}u$ can feed $\{uu\}s$.
The radial excitation of the positive-parity $\Sigma^\ast$ is an exception.  Its origin is dynamical, connected with the mass ordering $M_u < M_s <m_4<m_6$, with the switch in probability occurring for $1/d_{\cal F} = 0.64 (2 M^2)$, at which point the baryon's mass is $1.85\,$GeV.

\subsection{$\Xi^\ast$ Baryon}
As apparent from Eqs.\,\eqref{uColumn}, the $\Xi^{\ast 0}$ baryon may be obtained from the $\Sigma^{\ast+}$ by making the replacements $u\leftrightarrow s$.  It follows that the Faddeev equation for the $\Xi^{\ast 0}$ can simply be obtained from that for the $\Sigma^{\ast+}$ by making the replacements $u\leftrightarrow s$ and $4\to 9$ in the equations of App.\,\ref{SigmaAstFE}.

The eigenvectors corresponding to the results in Table~\ref{OctetDecupletMasses} are
\begin{equation}
\begin{array}{l|c|c|c|c}
%GS:{0.852904, 0.522067}
%RE:{0.900513, 0.434829}
%PP:0.872972, 0.4877
%PPRE:0.867305, 0.497776
    & P=+, n=0 & P=+, n=1 & P=-, n=0 & P=-, n=0 \\\hline
f_{\Xi^\ast}^6 & 0.85 & 0.90 & 0.87 & 0.87\\
f_{\Xi^\ast}^9 & 0.52 & 0.43 & 0.49 & 0.50
\end{array}
\end{equation}
indicating that these states favour the $\{us\}u$ configuration over $\{ss\}u$.  This owes again to the quark exchange character of the interaction in Fig.\,\ref{fig:FaddeevI}: both $\{ss\}u$ and $\{us\}s$ feed into $\{us\}s$ but only $\{us\}s$ can feed $\{ss\}u$.

\subsection{$\Omega$ Baryon}
We define
\begin{equation}
c_{\Omega} = \frac{g_{10}^2}{4\pi^2 M_s}\,,\;
\sigma_{\Omega}:=(1-\alpha)\,M_s^2 + \alpha\,m_9^2 - \alpha(1-\alpha)m_{\Omega}^2\,,
\end{equation}
so that the Faddeev equation for the $\Omega$ can be written
\begin{equation}
1 = 2 c_{\Omega} \frac{E_9^2}{m_9^2}
\int_0^1 d\alpha\, [m_9^2 + (1-\alpha)^2 m_\Omega^2][\alpha m_\Omega + M_s] \,
\overline{\cal C}^{\rm iu}_1(\sigma_\Omega^{9})\,.
\end{equation}
Of course, this is simply the Faddeev equation for the $\Delta^+$-resonance but for the replacement $u\to s$ or, equivalently, $4\to 9$.

%---\bibliographystyle{../../../../zProc/z10KITPC/h-physrev4}  %%%-- Use this for arXiv
%\bibliographystyle{spbasic} %%%-- Use this for FBS
%\bibliographystyle{cj} %%%-- Acceptable for FBS
%---\bibliography{../../../../CollectedBiB}  %%%-- Acceptable for arXiv

\end{document}